\documentclass[twocolumn,superscriptaddress,amsmath,amssymb,aps,prb,floatfix]{revtex4-2}

\usepackage{comment}
\usepackage{tensor}
\usepackage{amsmath,mathtools}
\usepackage[percent]{overpic}
\usepackage{color}
\usepackage{xcolor}
\usepackage{cancel}
\usepackage{ulem}
\usepackage{physics}
\usepackage{graphicx}
\usepackage{dcolumn}
\usepackage{bm}
\usepackage[breaklinks=true,colorlinks,citecolor=blue,linkcolor=blue,urlcolor=blue]{hyperref}
\usepackage[percent]{overpic}
\usepackage{braket}
\usepackage{bbold}
\usepackage{tabularx}

\begin{document}
\title{Amperean superconductivity cannot be induced by deep subwavelength cavities in a two-dimensional material}

\author{Gian Marcello Andolina}
\altaffiliation{These authors contributed equally to this work.}
\affiliation{ICFO-Institut de Ci\`{e}ncies Fot\`{o}niques, The Barcelona Institute of Science and Technology, Av. Carl Friedrich Gauss 3, 08860 Castelldefels (Barcelona),~Spain}
\author{Antonella De Pasquale}
\altaffiliation{These authors contributed equally to this work.}
\affiliation{Istituto Italiano di Tecnologia, Graphene Labs, Via Morego 30, I-16163 Genova,~Italy}
\email{Antonella.DePasquale@iit.it}
\author{Francesco Maria Dimitri Pellegrino}
\altaffiliation{These authors contributed equally to this work.}
\affiliation{Dipartimento di Fisica e Astronomia, Universit\`a di Catania,
Via S. Sofia, 64, I-95123 Catania, Italy}
\affiliation{INFN, Sez. Catania, I-95123 Catania, Italy}
\affiliation{CNR-IMM, Via S. Sofia 64, I-95123 Catania, Italy}
\author{Iacopo Torre}
\affiliation{ICFO-Institut de Ci\`{e}ncies Fot\`{o}niques, The Barcelona Institute of Science and Technology, Av. Carl Friedrich Gauss 3, 08860 Castelldefels (Barcelona),~Spain}
\author{Frank H. L. Koppens}
\affiliation{ICFO-Institut de Ci\`{e}ncies Fot\`{o}niques, The Barcelona Institute of Science and Technology, Av. Carl Friedrich Gauss 3, 08860 Castelldefels (Barcelona),~Spain}
\affiliation{ICREA-Instituci\'{o} Catalana de Recerca i Estudis Avan\c{c}ats, Passeig de Llu\'{i}s Companys 23, 08010 Barcelona,~Spain}
\author{Marco Polini}
\affiliation{Dipartimento di Fisica dell'Universit\`a di Pisa,
Largo B. Pontecorvo 3, I-56127 Pisa, Italy}
\affiliation{ICFO-Institut de Ci\`{e}ncies Fot\`{o}niques, The Barcelona Institute of Science and Technology, Av. Carl Friedrich Gauss 3, 08860 Castelldefels (Barcelona),~Spain}
\begin{abstract}
Amperean superconductivity is an exotic phenomenon stemming from attractive effective electron-electron interactions (EEEIs) mediated by a transverse gauge field. 
Originally introduced in the context of quantum spin liquids and high-$T_{\rm c}$ superconductors, Amperean superconductivity has been recently proposed to occur at temperatures on the order of $1$-$20~{\rm K}$ in two-dimensional, parabolic-band, electron gases embedded inside deep sub-wavelength optical cavities. 
In this work, we first generalize the microscopic theory of cavity-induced Amperean superconductivity to the case of graphene and then argue that this superconducting state cannot be achieved in the deep sub-wavelength regime. 
In the latter regime, indeed, a cavity induces only EEEIs between density fluctuations rather than the current-current interactions which are responsible for Amperean pairing.
\end{abstract}

\maketitle
\section{Introduction}
\label{sect:introduction}

The simplest microscopic model of superconductivity~\cite{Grosso}, namely the Bardeen-Cooper-Schrieffer (BCS) theory~\cite{BCS}, describes this phenomenon as arising from the condensation of Cooper pairs with zero total momentum, resulting in a spatially uniform order parameter.
Superconductors with ``pair density wave'' (PDW) order host, instead, a condensate where Cooper pairs have a non-zero center-of-mass momentum ${\bm Q}$~\cite{Casalbuoni_rmp_2004,Fradkin_rmp_2015}. Such exotic states of matter were first predicted to occur in metals in the presence of a sufficiently strong spin exchange field~\cite{Fulde_PhysRev_1964,Larkin_JEPT_1965}, leading to spin polarization and an imbalance between the populations of up- and down-spin electrons.  

Superconducting PDW order, however, can occur also in the presence of time-reversal symmetry. For example, it has been shown~\cite{Lee_prl_2007} that, in a gapless spin liquid state, a $U(1)$ gauge field mediates an effective interaction between spinons and that this is attractive when spinons have parallel momenta. This is akin to the Amperean attraction that occurs between two wires carrying parallel currents. The resulting superconducting instability has therefore been dubbed ``Amperean superconductivity''~\cite{Lee_prl_2007}.  Amperean superconductors are a special class of PDW superconductors, where the oscillatory nature of the pairing gap induced by the finite center-of-mass momentum ${\bm Q}$ occurs in the presence of time-reversal symmetry.
\begin{figure}[t]
\centering
 \begin{overpic}[width=0.95\columnwidth]{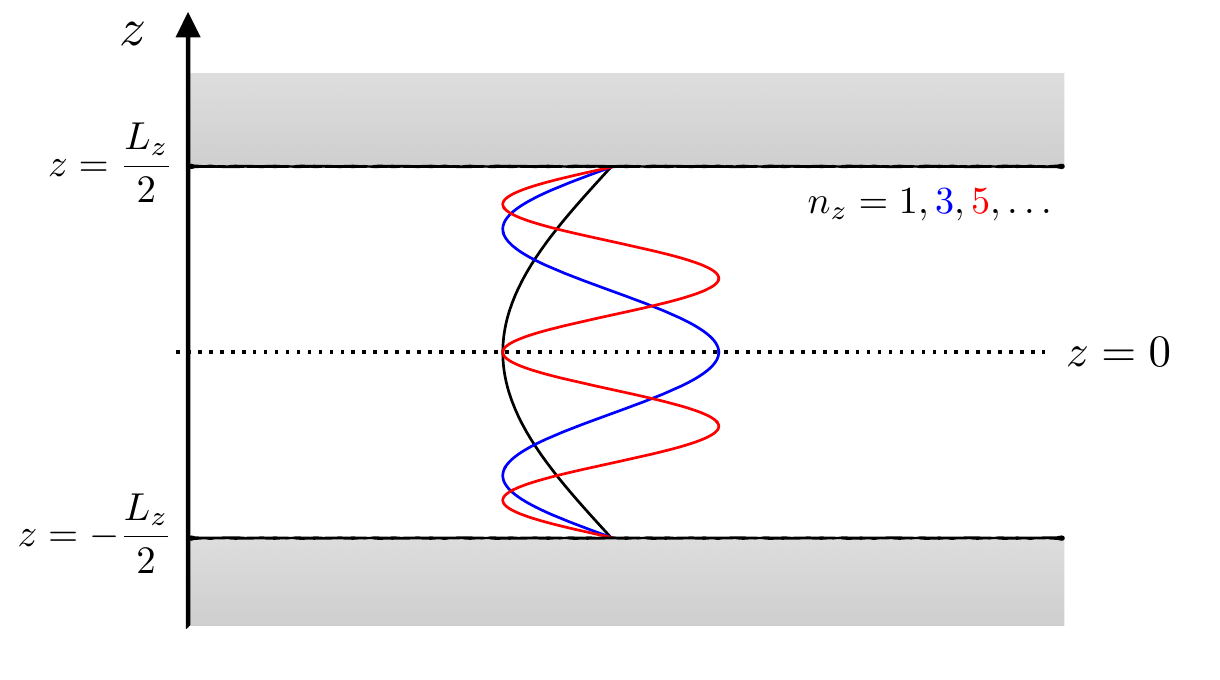}
 \end{overpic}
\caption{(Color online) A cartoon of the setup investigated in this Article. Side view of a planar Fabry-P\'erot cavity containing single-layer graphene (dashed line) at its center ($z=0$). The metallic mirrors at $z=\pm L_z/2$ are colored in gray. The distance between the two mirrors is therefore $L_z$. Only the odd cavity modes (labelled by $n_z=1,3,5, \ldots$) couple to the two-dimensional electron system.}\label{fig:cavity}
\end{figure}
More recently, the theory of Amperean superconductivity has been applied to the pseudogap phase of cuprate high-temperature superconductors~\cite{Lee_PRX_2014,Daniel_2020} and to the surface states of three-dimensional (3D) topological insulators (TIs)~\cite{Kargarian_2016,Subdo_2020_1,Subdo_2020_2}. 
In the latter case,  the interaction between fermions and gapless bosons that is needed to realize an effective fermion-gauge theory can be induced by depositing a ferromagnetic layer on the surface of a 3D TI. Finally, it has been shown~\cite{Baireuther_PRL_2015} that Andreev reflection between normal and superconducting regions in a three-terminal device is an extremely sensitive tool to detect Amperean PDW order.

Given this context, one may be tempted to conclude that Amperean superconductivity stems from effective electron-electron interactions (EEEIs) mediated by exotic gapless bosonic modes. In principle, however, ordinary transverse photons can mediate Amperean superconductivity. Indeed, current-current EEEIs mediated by the exchange of transverse photons have long been known to be responsible for exotic behavior, such as non-Fermi-liquid behavior of metals~\cite{Holstein_PhysRev_1973,Reizer_PRB_1989,Lee_PRL_1989,Khveshchenko_PRL_1993,Khveshchenko_PRB_1994}. In free space, Amperean superconductivity and non-Fermi-liquid behavior are essentially impossible to be observed since the dimensionless coupling constant is on the order of~\cite{Tsvelik} $v_{\rm F}/c \ll 1$, where $v_{\rm F}$ is the Fermi velocity and $c$ is the speed of light. To bypass this problem, Schlawin et al.~\cite{Jaksch_prl_2019} suggested an ingenious path to create a high-temperature Amperean superconducting state by coupling a two-dimensional (2D) parabolic-band electron gas~\cite{Giuliani_and_Vignale}, such as the one hosted by a GaAs quantum well, to an optical cavity. This is only one example of the many tantalizing phenomena that have been predicted to occur when quantum materials and cavity photons are brought to the strong-coupling regime. A largely incomplete list of other examples includes cavity-enhanced transport of excitons~\cite{Schachenmayer_prl_2015}, polaritonic-enhanced electron-phonon coupling and superconductivity~\cite{Sentef_ScienceAdvances_2018}, cavity-enhanced superconductivity in 2D superconducting films~\cite{Curtis_prl_2019}, cavity-controlled local interactions in strongly correlated materials~\cite{Kiffner_prb_2019,lede_jphysmater_2022},  photon condensation and magnetostatic instabilities~\cite{nataf_prl_2019,andolina_prb_2020,guerci_prl_2020}, ferroelectricity~\cite{Ashida_PRX_2020}, and topological bands~\cite{Kibis_prb_2011,Wang_prb_2019,Tokatly_PRB_2021}. For a broader view on the emergent research field dubbed ``cavity quantum electrodynamic control of matter'', we invite the reader to consult recent review articles~\cite{kockum_naturereviewsphysics_2019,Rubio_NatureMater_2021,Genet_PT_2021,GarciaVidal_Science_2021,Schlawin_APR_2022,Bloch_Nature_2022}. Experimentally, interactions between light and extended electron systems in the strong-coupling regime have been mainly studied in the quantum Hall regime, by using arrays of Terahertz (THz) split-ring resonators coupled to GaAs quantum wells~\cite{Scalari_science_2012,muravev_prb_2013,Maissen_prb_2014,smolka_science_2014,Keller_nanolett_2017,knuppel_nature_2019,Paravicini_nature_2019,Appugliese_science_2022}.  Recently, the authors of Ref.~\onlinecite{Jarc_arxiv_2022} have claimed that the critical temperature associated to the metal-to-insulator transition in ${\rm 1T}$-${\rm TaS}_2$ can be shifted by more than $70~{\rm K}$ by tuning the cavity resonant frequency. We finally stress that the role of genuine quantum effects in cavity QED of extended electron systems is subtle and a matter of debate (see, for example, Ref.~\onlinecite{Amelio_PRB_2021}).

Going back to cavity-induced superconductivity, Schlawin et al.~\cite{Jaksch_prl_2019} predicted record-high values of the critical temperature $T_{\rm c}$ for Amperean pairing, which was estimated to be in the range $T_{\rm c} \approx 1$-$20~{\rm K}$ for a parabolic-band 2D electron gas coupled to a THz cavity. Such high temperatures ultimately stem from a cavity compression (or ``mode volume confinement'') factor $A$ on the order of $10^{-5}$, which will be defined below and can be achieved e.g.~in nanoplasmonic THz cavities~\cite{Maissen_prb_2014,Keller_nanolett_2017}.

This Article is motivated by Ref.~\onlinecite{Jaksch_prl_2019} and goes beyond it in two main respects: i) On the one hand, we generalize the theory of Ref.~\onlinecite{Jaksch_prl_2019}  to the case of single-layer graphene, the most studied atomically-thin material~\cite{Geim2013,katsnelson,castroneto_rmp_2009}. The reason is easy to understand. Graphene can be doped either electrically (via a nearby metallic gate) or chemically (e.g.~via polymer electrolytes) to much larger values than GaAs and incredibly high carrier concentrations up to $5.5 \times 10^{14}~{\rm cm}^{-2}$ have been recently achieved~\cite{Starke}. Given the tremendous interest~\cite{Saito_NatRev_Mater_2017,Weitering_SST_2017,Li_Small_2021} in superconductivity in the 2D limit and the possibility to tune (e.g.~electrically) the critical temperature (e.g.~via a gate), we found it useful to generalize the theory of cavity-mediated EEEIs to the case of  graphene. ii) On the other hand, by working in the Coulomb gauge, we show that there is an intrinsic incompatibility between Amperean superconductivity, which stems from interactions mediated by a {\it transverse} gauge field ${\bm A}_{\rm cav}({\bm r},t)$, and the mode volume confinement of nanoplasmonic cavities, which is basically associated to a classical quasi-static electric field ${\bm E}_{\rm cav}({\bm r},t) = - \bm{\nabla}\phi_{\rm cav}({\bm r},t)$, which is {\it longitudinal}. 

Our manuscript is organized as follows. In Sect.~\ref{sec:Hamiltonian} we present our model Hamiltonian, describing electrons in single-layer graphene, coupled to the quantized electromagnetic modes of a planar optical cavity. In Sect.~\ref{sec:effectiveEEinteractions} we provide a microscopic derivation of the EEEI mediated by cavity photons---see Eq.~(\ref{eq:PQ6}). In Sect.~\ref{sec:AmpereanGap} we derive the linearized, Amperean gap equation---Eq.~(\ref{eq:strange_sign_second_term})---that needs to be solved while hunting for an Amperean superconducting instability. Finally, in Sect.~\ref{sec:results} we provide numerical evidence that no Amperean superconductivity occurs at measurable temperatures in a highly doped graphene sheet embedded inside a planar Fabry-P\'erot cavity. More importantly, in Sect.~\ref{sec:TransversVsLong} we present a Green's function approach to EEEI that highlights the profound difference between planar optical cavities and sub-wavelength cavities. This theory allows us to demonstrate that in the former physical setup, {\it no} ``mode-volume confinement" argument can be used for effectively boosting the value of the electron-photon coupling and related critical temperature $T_{\rm c}$. On the contrary, the same theory shows clearly that sub-wavelength nanoplasmonic cavities dramatically alter the simple Coulomb repulsion law, giving rise to density-density (rather than current-current) EEEIs. Appendices~\ref{sec:GrapheneH}-\ref{sec:Diamagnetic} contain a number of useful technical additions.

\section{Model Hamiltonian}
\label{sec:Hamiltonian}

\begin{table}[t]
    \centering
    \begin{tabularx}{\linewidth}{|c|X|}
        \hline
        & Cavity parameters: \\ \hline
        & Mirror distance: \( L_z\approx 0.4 - 2.0~\mu{\rm m} \) \\
        & Debye energy: \( \hbar\omega_{\rm D}  \approx 0.3 - 1.5~{\rm eV} \) \\ \hline\hline
        & Graphene parameters: \\ \hline
        & Hopping parameter: \( t =2.8~{\rm eV} \) \\
        & Bandwidth: \( W= 6~t \) \\
        & Chemical potential: \( \mu \approx 0.4 - 0.6~t \) \\
        & Electron density: \( n \approx 1-3\times 10^{14}~{\rm cm}^{-2} \) \\
        & C-C distance: \( a= 0.142~{\rm nm} \) \\
        \hline
    \end{tabularx}
    \caption{Physical parameters explored in this Article. The effective Debye energy is defined by \(\hbar\omega_{\rm D} \equiv  \pi \hbar c /(L_z\sqrt{\epsilon_{\rm r}})\). The value of \(\epsilon_{\rm r}\) has been fixed at \(\epsilon_{\rm r}=4\), which applies to the case of graphene encapsulated in hexagonal Boron Nitride (hBN), for example. The lower bound on \(L_z\) descends from the condition \(\hbar \omega_{\rm D} <2 \mu\) (no inter-band transitions).  The upper bound on \(L_z\), instead, stems from the recognition of the fact that in the limit \(L_z\to \infty\) the problem at hand reduces to that of electrons interacting with free-space electromagnetic radiation. Such interaction is known to yield interesting physics  at experimentally-unreachable ultra-low temperatures~\cite{Tsvelik}. Finally, we focus on large values of the chemical potential \(\mu\), since in this regime superconductivity is expected to be reached at high critical temperatures. Working under these conditions, we are able to establish an upper bound for the critical temperature for cavity-induced Amperean superconductivity. \label{table}}
\end{table}

We consider single-layer graphene (SLG) placed at the center ($z=0$) of a planar Fabry-P\'erot electromagnetic cavity, schematically depicted in Fig.~\ref{fig:cavity}. This is the exactly same cavity geometry used in Ref.~\onlinecite{Jaksch_prl_2019}. While the authors of Ref.~\onlinecite{Jaksch_prl_2019} have considered electrons roaming in a 2D square lattice, we here consider a 2D honeycomb lattice.   

The system is described  by the  following Hamiltonian:
\begin{equation}\label{eq:full_Hamiltonian}
{\cal \hat{H}} = \hat{\cal H}_{\rm e} + \hat{\cal H}_{\rm ph} +   \hat{\cal H}_{\rm para}~,
\end{equation} 
where $\hat{\cal H}_{\rm e}$ and $\hat{\cal H}_{\rm ph}$ are the free electron and the cavity Hamiltonian, respectively, and $\hat{\cal H}_{\rm para}$ is the electron-photon interaction term.  In writing Eq.~(\ref{eq:full_Hamiltonian}) we have neglected the {\it direct} repulsive Coulomb electron-electron interaction. The reason is twofold. On the one hand, this is a common starting point of any elementary (i.e.~BCS) theory of superconductivity~\cite{Grosso, Morel_PR_1962}. On the other hand, it would be important to analyze the role of the Coulomb repulsion if the critical temperature for the photon-mediated Amperean superconducting state was large, which, as we will see below is not the case.

In the next three Sections we will discuss the three terms in Eq.~(\ref{eq:full_Hamiltonian}).

\subsection{Free electron Hamiltonian}

We start by illustrating the free electron Hamiltonian $\hat{\cal H}_{\rm e}$, which describes electrons moving in SLG. Despite this is well-known textbook material~\cite{katsnelson}, we have decided to report in this Section a brief account of the main results and definitions in order for the Article to be self-contained. 

\begin{figure}[t]
\centering
\begin{overpic}[width=0.99\columnwidth]{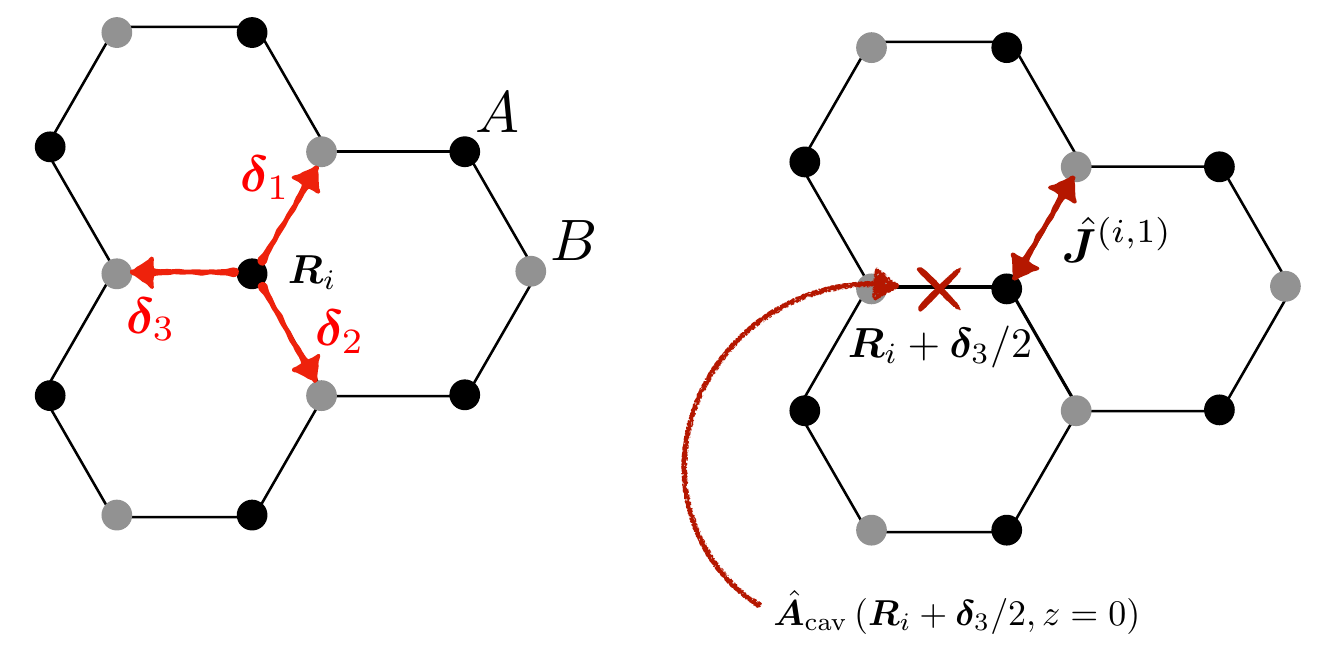}
\put(0,50){(a)}
\put(55,50){(b)}
\end{overpic}
\begin{overpic}[width=0.8\columnwidth]{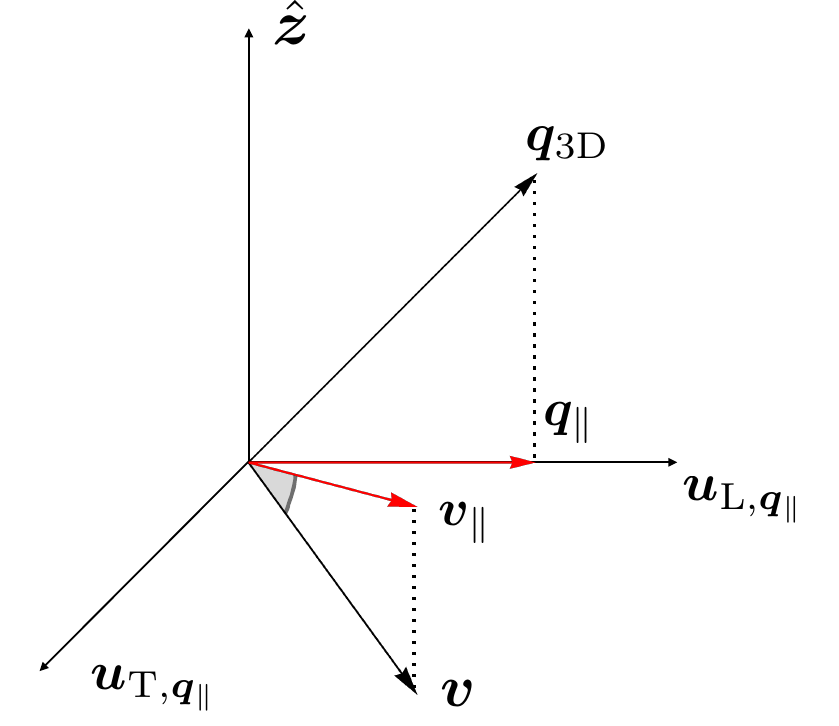}
\put(0,80){(c)}
\end{overpic}
\caption{(Color online) Two-dimensional honeycomb crystal structure of graphene, made out of two interpenetrating triangular sublattices, comprising $A$-type (black) and $B$-type (gray) atoms, respectively. Panel (a) shows the three nearest-neighbor vectors $\bm{\delta}_\ell$ introduced in the main text. Panel (b) shows the vector potential $\hat{\bm A}_{\rm cav}\left(\bm{R}_{i}  + {\bm{\delta}_\ell}/{2},z=0\right)$, with $\ell=3$, and the current $\hat{\bm J}^{(i,\ell)}$, with $\ell=1$. Panel (c) shows a vector $\bm{v}$ orthogonal to the 3D vector $\bm{q}_{\rm 3D}$, $\bm{v}\cdot\bm{q}_{\rm 3D}=0$, i.e.~$\bm{v}$ is {\it transverse} with respect to the 3D vector $\bm{q}_{\rm 3D}$. The projections of $\bm{q}_{\rm 3D}$ and $\bm{v}$ on the $\hat{\bm x}$-$\hat{\bm y}$ plane, i.e.~$\bm{q}_{\parallel}$ and $\bm{v}_\parallel$, respectively, are no longer orthogonal, i.e.~$\bm{v}_\parallel$ is {\it not} transverse with respect to $\bm{q}_{\parallel}$. For this reason, a 3D transverse field, such as $\hat{\bm A}_{\rm cav}(\bm{r}_{\parallel}, z)$, has both transverse and longitudinal components once it is projected onto a 2D plane and hence couples to both longitudinal and transverse components of the 2D graphene current.\label{fig:reticolo_gr}}
\end{figure}

Graphene's honeycomb lattice consists of a triangular lattice with a basis containing two atoms ($A$ and $B$) per unit cell. We introduce the two basis vectors, $\bm{a}_1=a(3, \sqrt{3})/2$ and $\bm{a}_2=a(3,- \sqrt{3})/2$, where $a$ is the Carbon-Carbon (C-C) distance---see Table~\ref{table}. Each site of the sublattice $A$ can be expressed as ${\bm R}_i=i_1 {\bm a}_1+i_2 {\bm a}_2$, where $i=1,\ldots,N_{\rm cell}$  ($N_{\rm cell}$ is the number of cells) identifies a pair of integers $i_1(i)$ and $i_2(i)$. The three nearest-neighbor sites of each site of the sublattice $A$ are identified by ${\bm R}_{(i,\ell)}={\bm R}_i+{\bm \delta}_\ell$, where $\ell=1\dots 3$ and $\bm{\delta}_1=a(1, \sqrt{3})/2$, $\bm{\delta}_2=a(1,- \sqrt{3})/2$, $\bm{\delta}_3=a (-1,0)$---see Fig.~\ref{fig:reticolo_gr}(a). In the real-space representation, the nearest-neighbor tight-binding Hamiltonian describing free electrons hopping on a honeycomb lattice reads as following~\cite{castroneto_rmp_2009}:
\begin{align}\label{eq:electronic_Hamiltonian_graphene_real_space}
{\cal \hat{H}}_{\rm e} &=  -  t\sum^{N_{\rm cell}}_{i=1}\sum_{\sigma=\uparrow, \downarrow}  \sum_{\ell=1}^3  \left(\hat c_{{\bm R}_i,\sigma,A}^{\dagger}  \hat c_{{\bm R}_{(i,\ell)}, \sigma, B} +{\rm H.c.} \right) \nonumber \\
&~ - \mu \hat{N}~,
\end{align}
where $t$ is the hopping parameter, $\hat{N}$ is the number operator,
\begin{align}
\hat{N}  =\sum^{N_{\rm cell}}_{i=1}\sum_{\sigma=\uparrow, \downarrow} 
&\left(\hat c_{{\bm R}_i,\sigma,A}^{\dagger}  \hat{c}^\dagger_{{\bm R}_i, \sigma, A}\right.   \nonumber\\
& + \left. \hat{c}^\dagger_{{\bm R}_{(i,\ell=3)}, \sigma, B}   \hat{c}_{{\bm R}_{(i,\ell=3)}, \sigma, B} \right)~,\end{align}
and $\mu$ is the chemical potential. Here, the operator $\hat c^\dagger_{{\bm R}_i, \sigma, A}$ ($\hat c_{{\bm R}_i, \sigma, A}$) creates (annihilates) an electron on the $A$ sublattice, at position ${\bm R}_i$ and with spin $\sigma$. The operator $\hat c^\dagger_{{\bm R}_{(i,\ell)}, \sigma, B}$ ($\hat c_{{\bm R}_{(i,\ell)}, \sigma, B}$) creates (annihilates) an electron on the $B$ sublattice at position ${\bm R}_{(i,\ell)}={\bm R}_i + {\bm \delta}_\ell$ and with spin $\sigma=\uparrow, \downarrow$---see Fig.~\ref{fig:reticolo_gr}{(a)}.

Eq.~\eqref{eq:electronic_Hamiltonian_graphene_real_space} can be easily transformed into the diagonal, band representation by going to momentum space (details are given in Appendix~\ref{sec:GrapheneH}):
\begin{equation}\label{eq:free_electron}
{\cal \hat H}_{\rm e}=\sum_{{\bm k} \in {\rm BZ}}\sum_{\sigma=\uparrow,\downarrow}  \sum_{\kappa= \pm} \xi_{\bm k,\kappa} \hat d_{{\bm k},\sigma,\kappa}^{\dagger} \hat d_{{\bm k },\sigma, \kappa}~,
\end{equation}
where $\hat d_{{\bm k},\sigma,\kappa}^{\dagger}$ ($\hat d_{{\bm k},\sigma,\kappa}$) creates (annihilates) an electron with wave vector $\bm k$ belonging to the first Brillouin zone (BZ), band index $\kappa=\pm$, and spin index $\sigma=\uparrow,\downarrow$. Furthermore, $\xi_{\bm k,\kappa} = \epsilon_{\bm k, \kappa}-\mu$ is the $\kappa$-th band energy, measured from the chemical potential $\mu$, where~\cite{castroneto_rmp_2009},
\begin{align}
\epsilon_{\bm k, \kappa} =\kappa t \sqrt{g({\bm k})}
\end{align}
and
\begin{align}
g({\bm k}) &= 3 + 2 \cos(\sqrt{3} k_y a) \nonumber\\
&+ 4 \cos \left(\frac{\sqrt{3}}{2} k_y a\right) \cos \left(\frac{3}{2} k_x a \right)~.
\end{align}
\subsection{Cavity Hamiltonian}

We now move on to discuss the cavity Hamiltonian $\hat{\cal H}_{\rm ph}$ and the fluctuating electromagnetic field. 

We describe the latter field in the Coulomb gauge, where $\bm{\nabla}\cdot \hat{\bm A}_{\rm cav}(\bm{r}_{\parallel}, z)=0$ and $\bm{\nabla}$ is the 3D nabla operator.  In this gauge, transverse radiation fields are given by the vector potential alone, while the instantaneous Coulomb potential contributes only to the near fields~\cite{Jackson}. In the Coulomb gauge, the quantized vector potential fulfilling the cavity boundary conditions is given by:
\begin{align}\label{vectorpot_2}
\hat{\bm A}_{\rm cav}(\bm{r}_{\parallel}, z)&=\sum_{\bm{q}_{\parallel},s, n_z} A^{\rm(2D)}_{\bm{q}_{\parallel} ,n_z}  \Big[{\bm e}_{\bm{q}_{\parallel},s,n_z }(z)\hat{a}_{\bm{q}_{\parallel},s ,n_z}e^{i\bm{q}_{\parallel} \cdot \bm{r}_{\parallel} }\nonumber\\&\qquad\qquad+ {\bm e}^\ast_{\bm{q}_{\parallel},s,n_z }(z)\hat{a}_{\bm{q}_{\parallel}, s ,n_z}^\dagger  
e^{-i\bm{q}_{\parallel} \cdot \bm{r}_{\parallel} }\Big]~.
\end{align}
Here, $\hat{a}_{\bm{q}_{\parallel},s ,n_z}$ ($\hat{a}_{\bm{q}_{\parallel}, s ,n_z}^\dagger$) annihilates (creates) a photon with momentum $\hbar\bm{q}_{\parallel}$, polarization $s=1,2$, and mode index $n_z$. The quantities ${\bm e}_{\bm{q}_{\parallel},s,n_z }(z)$ with $s=1,2$ denote the two polarization vectors~\cite{Kakazu_PRA_1994}
\begin{align}
{\bm e}_{\bm{q}_{\parallel},1,n_z } (z)&=  \frac{\hat{\bm{z}} \times {\bm q}_\parallel}{q_\parallel}   \sin\Big[\frac{\pi n_z}{L_z} \Big(z+\frac{L_z}{2} \Big) \Big]~,\label{polarization_1}\\ 
{\bm e}_{\bm{q}_{\parallel},2,n_z } (z)&=\frac{\bm q_\parallel}{q_\parallel} \sin\Big[\frac{\pi n_z}{L_z} \Big(z+\frac{L_z}{2} \Big) \Big] \label{polarization_2}
\frac{\pi n_z}{L_z \sqrt{q_\parallel^2+(\frac{\pi n_z}{L_z})^2}} \nonumber\\& \qquad+  \hat{\bm{z}} 
\cos\Big[\frac{\pi n_z}{L_z} \Big(z+\frac{L_z}{2} \Big)\Big] \frac{i q_\parallel}{ \sqrt{q_\parallel^2+(\frac{\pi n_z}{L_z})^2}}~,\nonumber\\
\end{align}
where $L_{z}$ is the length of the cavity in the $\hat{\bm z}$ direction (i.e.~the distance between the two mirrors) satisfying the quasi-2D condition $L_{z} \ll L_{x},L_{y}$, and $L_{x}$ ($L_{y}$) represents the lateral extension of the cavity in the $\hat{\bm x}$ ($\hat{\bm y}$) direction. Note that the 3D Coulomb gauge condition $\bm{\nabla}\cdot \hat{\bm A}_{\rm cav}(\bm{r}_{\parallel}, z)=0$ can be expressed in terms of the polarization vectors as following:
\begin{eqnarray}
\label{eq:CoulombC}
 [i\bm{q}_\parallel+\hat{\bm{z}}\partial_z]\cdot{\bm e}_{\bm{q}_{\parallel},s,n_z } (z) =0~.
\end{eqnarray}

In Eqs.~(\ref{vectorpot_2})-(\ref{polarization_2}), $n_z=1,2,3\dots$ is an integer, $\hbar{\bm q}_\parallel= (2\pi \hbar n_{x}/L_{x}, 2\pi \hbar n_{y}/L_{y})$ is the photon momentum in the $\hat{\bm x}$-$\hat{\bm y}$ plane, $n_{x}$, $n_{y}$ being relative integers, 
\begin{equation}
\label{eq:A2D}
A^{\rm(2D)}_{\bm{q}_{\parallel} ,n}= \sqrt{\frac{4\pi \hbar c^2}{V \omega_{\bm{q}_{\parallel} ,n_z} \epsilon_{\rm r}}}~,
\end{equation}
is the field amplitude, $V=L_z S$ is the cavity volume ($S = L_x L_y$ being the surface of the cavity in the $\hat{\bm x}$-$\hat{\bm y}$ plane),  $\epsilon_{\rm r}$ is the cavity relative dielectric constant, and
\begin{equation}
\label{eq:omega}
\hbar \omega_{\bm{q}_\parallel,n_z}= \frac{\hbar c}{\sqrt{\epsilon_{\rm r}}}\sqrt{ q_\parallel^2 + \left(\frac{\pi n_z}{L_z}\right)^2}~,
\end{equation}
is the cavity photon energy. We note that the lowest
cavity photon energy is
\begin{equation}
\label{eq:Denergy}
\hbar\omega_{\rm D} \equiv \hbar \omega_{{\bm 0},1} =  \frac{\pi \hbar c}{L_z \sqrt{\epsilon_{\rm r}}}~,
\end{equation}
which, as we will see below, plays the same role of the Debye phonon energy in the standard theory of phonon-mediated superconductivity~\cite{Grosso}. In other words, as we will show below, cavity photons introduce an effective attractive interaction between electrons lying near the Fermi surface, in an energy shell on the order of $\hbar\omega_{\rm D}$. It is possible to associate a Debye temperature $T_{\rm D}=\hbar \omega_{\rm D}/k_{\rm B}$ to the Debye energy, where $k_{\rm B}$ is the Boltzmann constant. For a distance $L_z$ between the two mirrors of $0.6~{\rm \mu m}$, the Debye temperature is on the order of $T_{\rm D}\approx 5.8 \times 10^3~{\rm K}$.

As highlighted in Fig.~\ref{fig:cavity},  SLG is placed on the $\hat{\bm x}$-$\hat{\bm y}$ plane, at $z=0$. In this case, only the {\it odd} cavity modes, i.e.~modes with $n_z=2 m_z+1$ ($m_z \in \mathbb{N}$), of both polarizations $s=1,2$ couple to the 2D electron system. The {\it even} cavity modes drop out of the problem. Indeed, for $n_z=2 m_z$ (with $m_z \in {\mathbb N}$) the $s=1$ polarization vector ${\bm e}_{\bm{q}_{\parallel},1,n_z } (0)$ evaluated at $z=0$ vanishes identically---see Eq.~\eqref{polarization_1}. The $s=2$ polarization vector ${\bm e}_{\bm{q}_{\parallel},2,n_z } (0)$ evaluated at $z=0$ is orthogonal to the $\hat{\bm x}$-$\hat{\bm y}$ plane and does not couple to the electronic orbital motion---see Eq.~\eqref{polarization_2}. In terms of $m_z$, the relevant polarization vectors evaluated at $z=0$ read as following:
\begin{eqnarray}
{\bm e}_{\bm{q}_{\parallel},1,2m_z+1 } (0)&=& \bm{u}_{{\rm T},\bm{q}_{\parallel}} (-1)^{m_z+1}~,\label{polarization_1z=0}\\ 
{\bm e}_{\bm{q}_{\parallel},2,2m_z+1 } (0)&=&\bm{u}_{{\rm L},\bm{q}_{\parallel}} 
\frac{(-1)^{m_z+1}\pi (2m_z+1)}{ \sqrt{(L_zq_\parallel)^2+[{\pi (2m_z+1)}]^2}}\label{polarization_2z=0} ~,\nonumber\\
\end{eqnarray}
where $\bm{u}_{{\rm T},\bm{q}_{\parallel}}={\hat{\bm{z}} \times {\bm q}_\parallel}/{q_\parallel} $ ($\bm{u}_{{\rm L},\bm{q}_{\parallel}}={\bm q}_\parallel/{q_\parallel} $) is a unit vector orthogonal (parallel) to the photon wave vector $\bm{q}_{\parallel}$; $\bm{u}_{{\rm T},\bm{q}_{\parallel}}$ ($\bm{u}_{{\rm L},\bm{q}_{\parallel}}={\bm q}_\parallel/{q_\parallel}$) is therefore transverse (longitudinal) with respect to ${\bm q}_{\parallel}$. We emphasize that, here, the notions of ``transverse'' and ``longitudinal'' are defined with respect to the 2D photon wave vector $\bm{q}_{\parallel}$ and not with respect to the 3D wave vector $\bm{q}_{\rm 3D}$. As emphasized above, the vector field $\hat{\bm A}_{\rm cav}(\bm{r}_{\parallel}, z)$ is a 3D transverse field due to the Coulomb condition in Eq.~\eqref{eq:CoulombC}, i.e.~it is orthogonal to ${\bm q}_{\rm 3D}$. However, once it is projected on the 2D $\hat{\bm x}$-$\hat{\bm y}$ plane where SLG is lying, it acquires both a longitudinal and a transverse component with respect to $\bm{q}_\parallel$---see Fig.~\ref{fig:reticolo_gr}(c).
 
Finally, the cavity Hamiltonian reads
\begin{equation}\label{eq:free_boson}
\hat{\cal H}_{\rm ph}= \sum_{\bm{q}_\parallel, s, n_z}  \hbar\omega_{\bm{q}_{\parallel},n_z}\hat{a}_{\bm{q}_{\parallel},s,n_z}^\dagger\hat{a}_{\bm{q}_{\parallel},s,n_z}~.
\end{equation}
\subsection{Electron-photon coupling}

The orbital motion of electrons roaming in SLG is coupled to the cavity modes by means of the Peierls substitution:
\begin{eqnarray}\label{eq:Peierls_factor}
&~&\hat c^{\dagger}_{{\bm R}_{(i,\ell)}, \sigma, B}\hat c_{{\bm R}_i,\sigma,A}  \to \hat c^{\dagger}_{{\bm R}_{(i,\ell)}, \sigma, B}\hat c_{{\bm R}_i,\sigma,A} \times \nonumber \\ &\times&\exp[{- \frac{i e}{c \hbar} \int_{\bm{R}_{i} }^{\bm{R}_{(i,\ell)} } \hat{\bm A}_{\rm cav}(\bm{r}_\parallel,z=0) \cdot d \bm{r}_\parallel}]~,
\end{eqnarray}
where $-e$ is the elementary electron's charge, ${\bm R}_i$ is the position of a given site of the sublattice $A$ in the cell labeled by the index $i$,  and $\bm{R}_{(i,\ell)} = {\bm R}_i + {\bm \delta}_\ell$ is the position of a given site of the sublattice $B$, which is linked by the nearest-neighbour vector ${\bm \delta}_\ell$ to the site ${\bm R}_i$ of the sublattice $A$. In passing, we note that, in quantum mechanics, exponentiating operators is dangerous. For example, time-ordering is needed in time-dependent perturbation theory~\cite{Sakurai}. In the case of Eq.~(\ref{eq:Peierls_factor}), however, no problems arise since $[\hat{\bm A}_{\rm cav}(\bm{r}_\parallel,z=0), \hat{\bm A}_{\rm cav}(\bm{r}^\prime_\parallel,z=0)]=0$.

Assuming that the vector potential changes slowly with respect to the atomic scales, the line integral in Eq.~(\ref{eq:Peierls_factor}) can be simplified as following:
\begin{align}
\label{eq:Peierls_factor1}
&\int_{\bm{R}_{i} }^{\bm{R}_{(i,\ell)} } \hat{\bm A}_{\rm cav}(\bm{r}_\parallel, z=0) \cdot d \bm{r}_\parallel \nonumber\\
&\simeq \hat{\bm A}_{\rm cav}\left(\bm{R}_{i}  + {\bm{\delta}_\ell}/{2},z=0\right) \cdot \bm{\delta}_\ell~.
\end{align}
The vector potential is evaluated at the middle of a link connecting an $A$-type site at position $\bm{R}_{i}$ with a $B$-type site at position ${\bm R}_i + {\bm \delta}_\ell$---see Fig.~\ref{fig:reticolo_gr}(b).

The Peierls replacement in Eq.~\eqref{eq:Peierls_factor1} couples the electronic motion to the cavity modes at all orders in the elementary electron charge $e$. If we expand the exponential phase factor in Eq.~(\ref{eq:Peierls_factor}) up to {\it second order} in the vector potential $\hat{\bm A}_{\rm cav}$ we get
\begin{eqnarray}
{\cal \hat{H}} &=&  \hat{\cal H}_{\rm ph} + \hat{\cal H}_{\rm e} +  \hat{\cal H}_{\rm para} + \hat{\cal H}_{\rm diam}+\dots~,
\end{eqnarray}
where ``$\dots$'' represent terms in the expansion of the Peierls factor of order higher than quadratic in the elementary charge $e$. We will show momentarily that these terms can be safely neglected. 

The linear (``paramagnetic''~\cite{Giuliani_and_Vignale}) term $\hat{\cal H}_{\rm para}$ reads as following:
\begin{equation}
\label{eq:para}
\hat{\cal H}_{\rm para}=\sum^{N_{\rm cell}}_{i=1}  \sum_{\ell=1}^3 \frac{e}{c} \hat{\bm J}^{(i,\ell)} \cdot \hat{\bm A}_{\rm cav}\big(\bm{R}_{i} + \bm{\delta}_\ell/2,z=0\big)~,
\end{equation}
where the paramagnetic current operator is given by:
\begin{equation}
\label{eq:current}
\hat{\bm J}^{(i,\ell)}=  \sum_{\sigma=\uparrow, \downarrow} \frac{i t}{\hbar}\; \bm{\delta}_\ell    \Big[\hat c_{{\bm R}_i,\sigma,A}^{\dagger} \hat c_{{\bm R}_{(i,\ell)},\sigma, B} - \hat c_{{\bm R}_{(i,\ell)},\sigma, B} ^{\dagger}  \hat c_{{\bm R}_i,\sigma,A}\Big]~.
\end{equation}
For a given site ${\bm R}_i$ belonging to the $A$ sublattice, there are three current operators $\hat{\bm J}^{(i,\ell)}$, flowing parallel to $\bm{\delta}_\ell$ and between the site ${\bm R}_i$ and the site ${\bm R}_i+{\bm \delta}_\ell$ belonging to the $B$ sublattice---see Fig.~\ref{fig:reticolo_gr}(b).
  
It is extremely important to understand the scaling of the light-matter interaction Hamiltonian $\hat{\cal H}_{\rm para} $ in Eq.~\eqref{eq:para} with respect to physical parameters. The paramagnetic current $\hat{\bm J}^{(i,\ell)}$---see Eq.~\eqref{eq:current}---is on the order of $\sim at/\hbar$ while the vector potential $\hat{\bm A}_{\rm cav}\left(\bm{R}_{i}  + \bm{\delta}_\ell/2,z=0\right)$---see Eq.~\eqref{eq:A2D}---can be estimated as
\begin{equation} 
A^{\rm(2D)}_{\bm{q}_{\parallel} ,n} \sim \epsilon_{\rm r}^{-1/2} \hbar c \frac{1}{\sqrt{V}} \frac{1}{\sqrt{\hbar\omega_{\rm D}}}~,
\end{equation}
where $V=S L_z$ is the cavity volume and the Debye energy $\hbar\omega_{\rm D}$ has been introduced in Eq.~(\ref{eq:Denergy}).  Finally, in the paramagnetic Hamiltonian, there is a factor $(e/c)$ we have to take care of. We therefore conclude that $\hat{\mathcal{H}}_{\hat{\bm{A}}}$ is controlled by the following quantity,
\begin{equation}\label{eq:coupling_constant_of_the_whole_theory}
\widetilde{g} \equiv \epsilon_{\rm r}^{-1/2}\frac{a}{\sqrt{V}} \frac{1}{\sqrt{\hbar\omega_{\rm D}}}te~.
\end{equation}
We now assume that the 2D electron system area, which is $\propto N_{\rm cell}a^2$, coincides with the cavity surface $S$, i.e.~we take $S \sim N_{\rm cell}a^2$.  Replacing this result into Eq.~(\ref{eq:coupling_constant_of_the_whole_theory}), we are led to introduce the ``reduced''  coupling constant $g$ defined in such a way that $\widetilde{g} = N^{-1/2}_{\rm cell} g$. We get
\begin{equation}\label{eq:coupling_constant_of_the_whole_theory2}
g = \epsilon_{\rm r}^{-1/2}\frac{1}{\sqrt{L_z}} \frac{1}{\sqrt{\hbar\omega_{\rm D}}}te~.
\end{equation}
Replacing the explicit expression of $\hbar\omega_{\rm D}$ in Eq.~(\ref{eq:coupling_constant_of_the_whole_theory2}), we conclude that, for a Fabry-P\'erot cavity, $g$ reduces to
\begin{equation}\label{eq:coupling_constant_of_the_whole_theory_FB}
g_{\rm FP} \equiv \frac{te}{\sqrt{\hbar c}} \epsilon_{\rm r}^{-1/4} = t \epsilon_{\rm r}^{-1/4}\sqrt{\alpha_{\rm QED}}~,
\end{equation}
where $\alpha_{\rm QED}=e^2/(\hbar c)\approx 1/137$ is the fine structure constant. The quantity $g_{\rm FP}$ is the coupling constant of our theory. Notice that $g_{\rm FP}$ is independent of $L_{z}$ because of the cancellation that occurs between $V \propto L_z$ and $\hbar\omega_{\rm D}\propto 1/L_{z}$. The fact that the light-matter coupling is independent of $L_z$ is typical of Fabry-P\'erot cavities and, more in general, of all cavities where genuine transverse photons are confined.

The paramagnetic Hamiltonian $\hat{\cal H}_{\rm para}$ in Eq.~\eqref{eq:para} can be conveniently expressed in momentum space. As detailed in Appendix~\ref{sec:GrapheneH}, the current $\hat{\bm J}^{(i,\ell)}$ can be expressed as
\begin{align}\label{eq:currentFourier}
\hat{\bm J}^{(i,\ell)}= 
\frac{1}{N_{\rm cell}}\sum_{\bm{q}_{\parallel}}   e^{i \bm{q}_{\parallel}\cdot \bm{R}_{i}}  
\hat{\bm{J}}_{\ell,\bm{q}_{\parallel}}~,
\end{align}
where the Fourier transform of the current $\hat{\bm J}_{\ell,\bm{q}_{\parallel}}$ is given by
\begin{align}\label{eq:Fourier_link_current}
 \hat{\bm{J}}_{\ell,\bm{q}_{\parallel}} \equiv \sum_{{\bm k},\bm{k}^\prime \in {\rm BZ}}   \left(
 \hat{\bm \jmath}_{\ell, \bm{k}^\prime, \bm k} + 
 \hat{\bm \jmath}^\dagger_{\ell,\bm k, \bm{k}^\prime} 
 \right)\delta_{\bm{k}-\bm{k}^\prime,\bm{q}_{\parallel}}~,
\end{align}
with
\begin{align}
\label{eq:componentsJ}
\hat{\bm \jmath}_{\ell,\bm k^\prime, \bm k}  \equiv  \frac{i t}{\hbar} \sum_{\sigma=\uparrow, \downarrow} \bm{\delta}_\ell e^{i {\bm k}\cdot {\bm \delta}_\ell} \frac{\varphi^\ast_{\bm{k}}}{2} \sum_{\kappa,\kappa^\prime} \kappa^\prime \hat d_{{\bm k^\prime},\sigma,\kappa^\prime}^{\dagger} \hat d_{{\bm k},\sigma,\kappa} ~.
\end{align}
Here, $\varphi_{\bm{k}}=- \sum_{\ell=1}^3 e^{i {\bm k } \cdot  {\bm \delta}_\ell}/|\sum_{\ell=1}^3 e^{i {\bm k } \cdot  {\bm \delta}_\ell}|$.

Similarly, the vector potential $\hat{{\bm A}}_{\rm cav} \left(\bm{R}_{i}  + \bm{\delta}_\ell/2,z=0\right)$ in Eq.~\eqref{vectorpot_2} can be rewritten in terms of its Fourier components $\hat{\bm{A}}_{\bm{q}_\parallel,\ell}$:
\begin{eqnarray}\label{eq:Fourier_middle_link_A}
\hat{\bm A}_{\rm cav} \big(\bm{R}_{i}  + \bm{\delta}_\ell/2,z=0\big)=\sum_{\bm{q}_{\parallel}} \sum_{\ell=1}^3  e^{i\bm{q}_{\parallel} \cdot \bm{R}_{i}} \hat{\bm{A}}_{\bm{q}_\parallel,\ell}~, 
\end{eqnarray}
with
\begin{align}
\label{eq:componentsA}
\hat{\bm{A}}_{\bm{q}_\parallel,\ell}& \equiv \sum_{s, n_z} A^{\rm(2D)}_{\bm{q}_{\parallel} ,n_z}e^{i   \frac{\bm{q}_{\parallel} \cdot \bm{\delta}_\ell}{2}} \big[ {\bm e}_{\bm{q}_{\parallel},s,n_z } (0)\hat{a}_{\bm{q}_{\parallel},s ,n_z} +\nonumber
\\ &+{\bm e}^*_{-\bm{q}_{\parallel},s,n_z } (0)\hat{a}_{-\bm{q}_{\parallel}, s ,n_z}^\dagger  \big]~. 
\end{align}
Notice that there is a Fourier component of the vector potential $\hat{\bm{A}}_{\bm{q}_\parallel,\ell}$ for each $\ell$, since the vector potential that couples with the current $\hat{\bm J}^{(i,\ell)}$ is  $\hat{\bm A}_{\rm cav}\left(\bm{R}_{i}  + {\bm{\delta}_\ell}/{2},z=0\right)$, which, as stated earlier, is evaluated at the middle of a link connecting an atom $A$ at position $\bm{R}_{i}$ with an atom $B$ at position  ${\bm R}_i + {\bm \delta}_\ell$---see Fig.~\ref{fig:reticolo_gr}(b).

Replacing Eqs.~(\ref{eq:currentFourier}) and~(\ref{eq:Fourier_middle_link_A}) in Eq.~\eqref{eq:para} we find:
\begin{align}\label{eq:para1}
\hat{\cal H}_{\rm para}
&=\frac{e}{c}\frac{1}{N_{\rm cell}} \sum^{N_{\rm cell}}_{i=1} \sum_{\bm{q}_{\parallel},\bm{q}_{\parallel}^\prime}\sum_{\ell=1}^3   e^{i(\bm{q}_{\parallel}+\bm{q}^\prime_\parallel) \bm{R}_{i} }\times \nonumber\\  &\times\hat{\bm{A}}_{\bm{q}_\parallel,\ell} \cdot  \hat{\bm{J}}_{\ell,\bm{q}^\prime_{\parallel}} ~.
\end{align}
Performing the sum over the unit cell index $i$ we find:
\begin{equation}
\label{eq:para2}
\hat{\cal H}_{\rm para}=
\frac{e}{c} \sum_{\ell=1}^3 \sum_{\bm{q}_{\parallel},\bm{q}_{\parallel}^\prime} \delta_{\bm{q}_\parallel, -\bm{q}^\prime_\parallel+\bm G} \hat{\bm{A}}_{\bm{q}_\parallel,\ell} \cdot   \hat{\bm{J}}_{\ell,\bm{q}^\prime_{\parallel}}~.
\end{equation}
Summing over $\bm{q}^\prime_{\parallel}$ by using the Kronecker delta and neglecting {\it Umklapp} processes \cite{footnote1} we finally find:
\begin{align}
\label{eq:para4}
\hat{\cal H}_{\rm para}&=
\frac{e}{c} \sum_{\ell=1}^3\sum_{\bm{q}_{\parallel}} \hat{\bm{A}}_{\bm{q}_{\parallel},\ell} \cdot \hat{\bm{J}}_{\ell,-\bm{q}_{\parallel}}~.
\end{align}
In order to get a clear physical interpretation, it is convenient to further express Eq.~\eqref{eq:para4} in terms of $\hat{\bm \jmath}_{\ell, \bm k, {\bm k}^\prime}$ and 
$ \hat{\bm{A}}_{\bm{q}_\parallel,\ell}$. By using Eqs.~(\ref{eq:componentsJ}) and~(\ref{eq:componentsA}) we find:
\begin{widetext}
\begin{align}
\label{eq:para5}
\hat{\cal H}_{\rm para}=
\frac{e}{c} \sum_{\ell=1}^3\sum_{\bm{q}_{\parallel}}\sum_{s, n_z} A^{\rm(2D)}_{\bm{q}_{\parallel} ,n_z}e^{i   \frac{\bm{q}_{\parallel} \cdot \bm{\delta}_\ell}{2}} \big[ {\bm e}_{\bm{q}_{\parallel},s,n_z } (0)\hat{a}_{\bm{q}_{\parallel},s ,n_z} +{\bm e}^*_{-\bm{q}_{\parallel},s,n_z } (0)\hat{a}_{-\bm{q}_{\parallel}, s ,n_z}^\dagger  \big] \cdot
    \sum_{{\bm k},\bm{k}^\prime \in {\rm BZ}}   \left(
 \hat{\bm \jmath}_{\ell, \bm{k}^\prime, \bm k} + 
 \hat{\bm \jmath}^\dagger_{\ell,\bm k, \bm{k}^\prime} 
 \right)\delta_{\bm{k}-\bm{k}^\prime,-\bm{q}_{\parallel}}~.
\end{align}
\end{widetext}
We therefore see that this Hamiltonian describes processes whereby an electron with momentum ${\bm k}$ scatters with the photonic field, either emitting a photon with momentum $-\bm{q}_\parallel$ 
or absorbing a photon with momentum $\bm{q}_\parallel$, and  ends up in a state with momentum $\bm{k}+\bm{q}_\parallel$.

The second-order ``diamagnetic'' $\hat{\cal H}_{\rm diam}$ term is reported in Appendix~\ref{sec:Diamagnetic}. For the problem at hand, it is enough to stop the expansion at first order in the electrical charge $e$, retaining only the paramagnetic term. Indeed, as we proceed to show, the diamagnetic term is parametrically smaller than the paramagnetic one. The energy scale of the bare electronic Hamiltonian is the
 hopping parameter $t$, while, as we have seen in Eq.~(\ref{eq:coupling_constant_of_the_whole_theory}), the energy scale of paramagnetic term is $g_{\rm FP} = t \epsilon_{\rm r}^{-1/4}\sqrt{\alpha_{\rm QED}}$. The strength of ${\cal \hat{H}}_{\rm para}$ compared to the band energy is therefore on the order of $g_{\rm FP}/t =\epsilon_{\rm r}^{-1/4} \sqrt{\alpha_{\rm QED}}\ll 1$. Suppose now that we expand the Peierls factor to next order in perturbation theory. In this case, we find a light-matter coupling at second order in the electrical charge, ${\cal \hat{H}}_{\rm diam}$. The order of magnitude of the diamagnetic term, relative to the band energy, is $g^2_{\rm FP}/t^2=\epsilon_{\rm r}^{-1/2} \alpha_{\rm QED}$. We therefore conclude that ${\cal \hat{H}}_{\rm diam}$ is smaller than ${\cal \hat{H}}_{\rm para}$ by a factor $g_{\rm FP}/t \propto \epsilon^{-1/4}_{\rm r}\sqrt{\alpha_{\rm QED}}\ll 1$.  

In the next Section (Sect.~\ref{sec:effectiveEEinteractions}) we will retain ${\cal \hat{H}}_{\rm para}$, neglect ${\cal \hat{H}}_{\rm diam}$ (and higher-order terms), and integrate out the photonic degrees of freedom. In this manner we will obtain an EEEI whose magnitude is given by the second-order perturbation theory energy scale $g^2_{\rm FP}/(\hbar\omega_{\rm D})$. Such EEEI is therefore $\sim \alpha_{\rm QED}$, i.e.~it is on the same order of magnitude of the expectation value $\tensor[_{\rm ph}]{\langle}{}0|\hat{\cal H}_{\rm diam}|0 \tensor[]{\rangle}{_{\rm ph}}$ of the diamagnetic term ${\cal \hat{H}}_{\rm diam}$ over the bare vacuum field $|0 \tensor[]{\rangle}{_{\rm ph}}$. However, this expectation value is {\it quadratic} (and not {\it quartic}) in the electronic operators and thus represents a renormalization of the electronic bands, which is not the focus of this Article. Our interest here is indeed ``limited'' to the possibility that EEEIs mediated by cavity photons induce a superconducting instability at the Fermi surface. Besides the vacuum contribution $\tensor[_{\rm ph}]{\langle}{}0|\hat{\cal H}_{\rm diam}|0 \tensor[]{\rangle}{_{\rm ph}}$, the diamagnetic term can give rise to an EEEI (once it is integrated out), but this is suppressed by a factor $\alpha_{\rm QED}\ll 1$ with respect to the EEEI induced by the paramagnetic coupling ${\cal \hat{H}}_{\rm para}$. On the basis of these arguments, we neglect from now on ${\cal \hat{H}}_{\rm diam}$ and all other higher-order terms deriving from the expansion of the Peierls phase factor in powers of the elementary charge $e$.
 
Before concluding this Section a comment is in order. It is by now established (see e.g. Ref.~\onlinecite{andolina_prb_2019,Eckhardt_commphys_2022} and references therein) that discarding the diamagnetic term can be dangerous for some physical phenomena induced by light-matter interactions in cavity QED. For example, if one considers only the paramagnetic coupling ${\cal \hat{H}}_{\rm para}$ (disregarding the fact that retaining ${\cal \hat{H}}_{\rm para}$ {\it alone} is a weak-coupling approximation), the ground state is unstable~\cite{andolina_prb_2019,Eckhardt_commphys_2022} with respect to a superradiant quantum phase transition (a.k.a.~photon condensation) when the light-matter coupling reaches a critical value $g_{\rm c}$. This threshold is attained when the coupling is comparable with the bare frequency of the system, a regime that is well beyond the applicability of perturbation theory and inconsistent with the aforementioned weak-coupling assumption. The inclusion of the diamagnetic term restores the stability of the ground state~\cite{andolina_prb_2019,nataf_prl_2019, andolina_prb_2020,guerci_prl_2020}. Here, we consider light-matter interactions described by ${\cal \hat{H}}_{\rm para}$ only in the weak-coupling $g_{\rm FP}\ll g_{\rm c}$ regime, where the diamagnetic can be safely discarded.

\section{Microscopic derivation of the cavity-induced effective current-current interactions}
\label{sec:effectiveEEinteractions}

In this Section we trace out the photonic degrees of freedom in the total Hamiltonian $\hat{\cal{H}}$ in Eq.~(\ref{eq:full_Hamiltonian}), in order to obtain an {\it effective} Hamiltonian for the electronic degrees of freedom only. This procedure naturally yields a photon-mediated EEEI. We remind that the paramagnetic coupling $\hat{\cal H}_{\hat{\bm A}}$, which couples electrons to photons, scales as $g_{\rm FP} = t \epsilon_{\rm r}^{-1/4} \sqrt{\alpha_{\rm QED}}$, which is assumed to be small. In the rest of this Section, we perform a perturbative expansion in $g_{\rm FP}$. 

We start by defining the eigenstates $\ket{\psi_n}_{\rm e}$ of the bare electronic Hamiltonian $\hat{\cal H}_{\rm e}$, i.e.~$\hat{\cal H}_{\rm e}\ket{\psi_n}_{\rm e}=E_n\ket{\psi_n}_{\rm e}$.
The global ground state $\ket{\Psi}$ of the system defined by the full Hamiltonian $\hat{\cal H}$ in Eq.~\eqref{eq:full_Hamiltonian} satisfies the Schr\"odinger equation: $(\hat{\cal H}-E)\ket{\Psi}=0$. 

In the absence of the paramagnetic term, i.e.~for $g_{\rm FP}=0$, the ground state $\ket{\Psi_0}$ of the full Hamiltonian is the product of the bare electronic ground state $\ket{\psi_0}_{\rm e}$ and the photonic vacuum $\ket{0}_{\rm ph}$, i.e.~$\ket{\Psi_0}=\ket{\psi_0}_{\rm e}\ket{0}_{\rm ph}$. For finite but small $g_{\rm FP}$, the true ground state will have a large overlap with $\ket{\Psi_0}$. Hence, for small $g_{\rm FP}$, it makes sense to perform a perturbative expansion around the bare photonic vacuum.
Following this idea, we separate the total Hilbert space into two subspaces~\cite{Grosso}, $P$ and $Q$. States in $P$ have no constrains on the electronic wave-function but contain zero photons. Conversely, $Q$ is spanned by states having one or more photons.
Let $\hat{\cal P}$ be the projection operator on the subspace $P$. This operator leaves the electronic wave-function unaffected and projects the electromagnetic field onto its vacuum state $\ket{0}_{\rm ph}$, 
\begin{equation}
\hat{\cal P}=\mathbb{1}_{\rm e}|{0}\tensor[]{\rangle}{_{\rm ph}}{}\tensor[_{\rm ph}]{\langle}{}{0}|~,
\end{equation}
where $\mathbb{1}_{\rm e}=\sum_n|{\psi_n}\tensor[]{\rangle}{_{\rm e}} \tensor[_{\rm e}]{\langle}{}{\psi_n}|$ is the identity acting on the electronic degrees of freedom.
Similarly, $\hat{\cal Q}=\mathbb{1}-\hat{\cal P}$ is the projector operator on the subspace $Q$, characterized by non-zero photons. The projectors have the following properties: $\hat{\cal P}+\hat{\cal Q} = \mathbb{1}$, $\hat{\cal P}^2=\hat{\cal P}$, $\hat{\cal Q}^2=\hat{\cal Q}$, and $\hat{\cal P}\hat{\cal Q}=\hat{\cal Q}\hat{\cal P}=0$.

Since $\hat{\cal P}+\hat{\cal Q} = \mathbb{1}$ we have:
\begin{equation}
\ket{\Psi} = (\hat{\cal P}+\hat{\cal Q})\ket{\Psi} = \hat{\cal P}\ket{\Psi} +\hat{\cal Q}\ket{\Psi}~.
\end{equation}
Replacing this result into the left-hand side of the eigenvalue problem $\hat{\cal H}\ket{\Psi}= E\ket{\Psi}$ we find
\begin{equation}\label{eq:intermediate_step}
 \hat{\cal H}\hat{\cal P}\ket{\Psi} + \hat{\cal H}\hat{\cal Q}\ket{\Psi}= E\ket{\Psi}~.
\end{equation}
In order to find two coupled eigenvalue equations for $\hat{\cal P}\ket{\Psi}$ and $\hat{\cal Q}\ket{\Psi}$ we simply need to multiply Eq.~(\ref{eq:intermediate_step}) on the left, one time for $\hat{\cal P}$ and one time for $\hat{\cal Q}$. We find:
\begin{align}\label{eq:proj1}
(\hat{\cal P} \hat{\cal H} \hat{\cal P} + \hat{\cal P} \hat{\cal H} \hat{\cal Q} ) \ket{\Psi} =E \hat{\cal P}  \ket{\Psi}~,
\end{align}
and
\begin{align}\label{eq:proj2}
(\hat{\cal Q} \hat{\cal H} \hat{\cal P} + \hat{\cal Q} \hat{\cal H} \hat{\cal Q} ) \ket{\Psi} = E \hat{\cal Q}  \ket{\Psi}~.
\end{align}
We now use in Eq.~(\ref{eq:proj2}) that $\hat{\cal Q}$ is idempotent, i.e.~ $\hat{\cal Q}^2=\hat{\cal Q}$:
\begin{align}
&(\hat{\cal Q} \hat{\cal H} \hat{\cal P} + \hat{\cal Q} \hat{\cal H} \hat{\cal Q}^2) \ket{\Psi} = 
\hat{\cal Q} \hat{\cal H} \hat{\cal P} \ket{\Psi} + \hat{\cal Q} \hat{\cal H} \hat{\cal Q}( \hat{\cal Q} \ket{\Psi})=\nonumber\\
&= E \hat{\cal Q}  \ket{\Psi}~.
\end{align}
Solving the previous equation for $\hat{\cal Q} \ket{\Psi}$ we find:
\begin{align}\label{eq:proj22}
\hat{\cal Q} \ket{\Psi} = (E - \hat{\cal Q}\hat{\cal H}\hat{\cal Q})^{-1}\hat{\cal Q}\hat{\cal H}\hat{\cal P}\ket{\Psi}~.
\end{align}
We now rewrite Eq.~(\ref{eq:proj1}) as following:
\begin{align}\label{eq:proj1_bis}
&(\hat{\cal P} \hat{\cal H} \hat{\cal P}^2 + \hat{\cal P} \hat{\cal H} \hat{\cal Q}^2) \ket{\Psi} =E \hat{\cal P}\ket{\Psi} \nonumber\\
&\implies \hat{\cal P} \hat{\cal H} \hat{\cal P}(\hat{\cal P}\ket{\Psi}) + \hat{\cal P} \hat{\cal H} \hat{\cal Q}(\hat{\cal Q} \ket{\Psi}) =E \hat{\cal P}\ket{\Psi}~,
\end{align}
where we have again used $\hat{\cal P}^2=\hat{\cal P}$ and $\hat{\cal Q}^2=\hat{\cal Q}$. We now replace Eq.~(\ref{eq:proj22}) into Eq.~(\ref{eq:proj1_bis}) and find:
\begin{align}\label{eq:proj1_tris}
&\hat{\cal P} \hat{\cal H} \hat{\cal P}(\hat{\cal P}\ket{\Psi}) + \hat{\cal P} \hat{\cal H} \hat{\cal Q}(E - \hat{\cal Q}\hat{\cal H}\hat{\cal Q})^{-1}\hat{\cal Q}\hat{\cal H}\hat{\cal P}\ket{\Psi} \nonumber\\
&=E \hat{\cal P}\ket{\Psi}~,
\end{align}
or, equivalently, 
\begin{align}\label{eq:proj1_quater}
&\hat{\cal P} \hat{\cal H} \hat{\cal P}(\hat{\cal P}\ket{\Psi}) + \hat{\cal P} \hat{\cal H} \hat{\cal Q}(E - \hat{\cal Q}\hat{\cal H}\hat{\cal Q})^{-1}\hat{\cal Q}\hat{\cal H}\hat{\cal P}^2\ket{\Psi} \nonumber\\
&=E \hat{\cal P}\ket{\Psi} \nonumber\\
&\implies \hat{\cal P} \hat{\cal H} \hat{\cal P}(\hat{\cal P}\ket{\Psi}) + \hat{\cal P} \hat{\cal H} \hat{\cal Q}(E - \hat{\cal Q}\hat{\cal H}\hat{\cal Q})^{-1}\hat{\cal Q}\hat{\cal H}\hat{\cal P}(\hat{\cal P}\ket{\Psi})\nonumber\\
&=E \hat{\cal P}\ket{\Psi}
\end{align}
We have therefore demonstrated the following important result:
\begin{align}
\label{eq:PQ}
&\left[\hat{\cal P} \hat{\cal H} \hat{\cal P} + \hat{\cal P} \hat{\cal H} \hat{\cal Q}(  E -\hat{\cal Q} \hat{\cal H}\hat{\cal Q})^{-1}
\hat{\cal Q} \hat{\cal H} \hat{\cal P}
\right]\hat{\cal P} \ket{\Psi} \nonumber\\
&= E \hat{\cal P}   \ket{\Psi}~. 
\end{align} 
Eq.~(\ref{eq:PQ}) represents an effective Hamiltonian for $\hat{\cal P}\ket{\Psi}$, which acts on states of the subspace $P$ with no photons.

We note that $\hat{\cal P} \hat{\cal H} \hat{\cal P}=\hat{\cal H}_{\rm e}$ since the two photon-related terms $\hat{\cal H}_{\rm ph}$ and $\hat{\cal H}_{\rm para}$ vanish, once projected onto the vacuum, i.e.~$\hat{\cal P}\hat{\cal H}_{\rm ph}\hat{\cal P}=0$ and $\hat{\cal P}\hat{\cal H}_{\rm para}\hat{\cal P}=0$.  Also, $\hat{\cal P} \hat{\cal H} \hat{\cal Q}=\hat{\cal P}\hat{\cal H}_{\rm para}\hat{\cal Q}$ due to the fact that the paramagnetic term $\hat{\cal H}_{\rm para}$ is the only term in the total Hamiltonian $\hat{\cal H}$ which changes the number of photons from zero to one, connecting the two subspaces $P$ and $Q$. Eq.~\eqref{eq:PQ} therefore naturally defines a retarded (i.e.~energy dependent) self-energy operator $\hat {\Sigma}(E)$,
\begin{equation}
\label{eq:PQ1}
\hat{\Sigma}(E)=\hat{\cal P}\hat{\cal H}_{\rm para} \hat{\cal Q}(  E -\hat{\cal Q} \hat{\cal H}\hat{\cal Q})^{-1}
\hat{\cal Q} \hat{\cal H}_{\rm para} \hat{\cal P}~,
\end{equation}
and an effective Hamiltonian
\begin{equation}\label{eq:abstract_effective_Hamil}
\hat{\mathcal{H}}_{\rm eff} (E)=\hat{\mathcal{H}_{\rm e}}+\hat{\Sigma}(E)~.
\end{equation}
Up to this point, Eqs.~(\ref{eq:PQ1}) and (\ref{eq:abstract_effective_Hamil}) are exact as we have made no approximations. We now use the perturbative expansion in $g_{\rm FP}$ in order to eliminate the energy dependence of $\hat{\Sigma}(E)$. 

First, we split the subspace $Q$ into two subspaces, $Q_1$, which contains states with one photon, and $Q_>$, which is spanned by states with more than one photon. We then introduce the projector onto the subspace $Q_1$ ($Q_>$), which we denote by the symbol $\hat{\cal Q}_1$ ($\hat{\cal Q}_>$). The explicit form of $\hat{\cal Q}_1$ is:
\begin{equation}
\hat{\cal Q}_1=\mathbb{1}_{\rm e}\sum_{\bm{q}_{\parallel},s ,n_z}|{\bm{q}_{\parallel},s ,n_z}\tensor[]{\rangle}{_{\rm ph}}{\tensor[_{\rm ph}]{\langle}{}{\bm{q}_{\parallel},s ,n_z}|}~, 
\end{equation}
where $\ket{{\bm{q}_{\parallel},s ,n_z}}_{\rm ph}=\hat{a}^\dagger_{\bm{q}_{\parallel},s ,n_z}\ket{0}_{\rm ph}$ and $\hat{\cal Q}_>=\hat{\cal Q}-\hat{\cal Q}_1$. Since $\hat{\cal H}_{\rm para}$ is linear in the photonic creation and annihilation operators, it can create or destroy only {\it one} photon. Hence $\hat{\cal H}_{\rm para}$ connects the space $P$ with the space $Q_1$, while $Q_>$ remains decoupled, i.e.~$\hat{\cal P}\hat{\cal H}_{\rm para} \hat{\cal Q}=\hat{\cal P}\hat{\cal H}_{\rm para} \hat{\cal Q}_1$. Using this result in Eq.~(\ref{eq:PQ1}) we find:
\begin{equation}
\label{eq:PQ11stepAlt}
\hat{\Sigma}(E)=\hat{\cal P}\hat{\cal H}_{\rm para} \hat{\cal Q}_1(  E -\hat{\cal Q}\hat{\cal H}\hat{\cal Q})^{-1}
\hat{\cal Q}_1 \hat{\cal H}_{\rm para} \hat{\cal P}~.
\end{equation}
We are now in the position to perform the perturbative expansion. 

Since $\hat{\cal H}_{\rm para}$ is linear in $g_{\rm FP}$, the perturbative expansion of the self-energy $\hat{\Sigma}(E)$ in Eq.~\eqref{eq:PQ1} starts at order $g^2_{\rm FP}$. In this Article we are interested only in the contribution to $\hat{\Sigma}(E)$ of ${\cal O}(g^2_{\rm FP})$ and will therefore discard higher-order contributions. Hence, in  the term $ (  E - {\hat{\cal Q} \hat{\cal H}\hat{\cal Q}})^{-1}$ we can discard the paramagnetic term, which is linear in $g_{\rm FP}$, replacing the full Hamiltonian $\hat{\cal H}$ with the sum of $\hat{\cal H}_{\rm e}$ and $\hat{\cal H}_{\rm ph}$, i.e.~${\hat{\cal Q} \hat{\cal H}\hat{\cal Q}} \to {\hat{\cal Q}(\hat{\cal H}_{\rm e}+\hat{\cal H}_{\rm ph})\hat{\cal Q}}$. The latter term does not change the number of photons and it is therefore not able to connect ${Q}_1$ and ${Q}_>$, i.e.
\begin{equation}\label{eq:useful_property}
{\hat{\cal Q}_1(\hat{\cal H}_{\rm e}+\hat{\cal H}_{\rm ph})\hat{\cal Q}}_>= {\hat{\cal Q}_> (\hat{\cal H}_{\rm e}+\hat{\cal H}_{\rm ph})\hat{\cal Q}}_1=0~.
\end{equation}
For the sake of brevity, we introduce in what follows the shorthand $\hat{\cal H}_{\rm d} = \hat{\cal H}_{\rm e}+\hat{\cal H}_{\rm ph}$.

We now expand the denominator $(E - {\hat{\cal Q} \hat{\cal H}_{\rm d}\hat{\cal Q}})^{-1}$ by using the following identity:
\begin{equation}\label{eq:expansion}
\frac{1}{E-\hat{\cal{O}}}=\frac{1}{E}\Big[\sum_{n=1}^\infty \Big(\frac{\hat{\cal{O}}}{E}\Big)^n+1\Big]~,
\end{equation}
where $\hat{\cal{O}}$ is a generic operator. Using Eqs.~(\ref{eq:useful_property}) and~(\ref{eq:expansion}) we can express the energy denominator in Eq.~\eqref{eq:PQ11stepAlt} as following:
\begin{align}
\label{eq:PQ11stepAlt1}
&[ E -(\hat{\cal Q}_1+\hat{\cal Q}_>) \hat{\cal H}_{\rm d}(\hat{\cal Q}_1+\hat{\cal Q}_>)]^{-1}= \nonumber\\
&=\frac{1}{E}\left\{\sum_{n=1}^\infty \Big[ \Big(\frac{\hat{\cal Q}_1 \hat{\cal H}_{\rm d}\hat{\cal Q}_1}{E}\Big)^n+ \Big(\frac{\hat{\cal Q}_> \hat{\cal H}_{\rm d}\hat{\cal Q}_>}{E}\Big)^n\Big]+1\right\}~.
\end{align}
Recalling that in Eq.~(\ref{eq:PQ11stepAlt}) we need the quantity $\hat{\cal Q}_1(  E -\hat{\cal Q}\hat{\cal H}_{\rm d}\hat{\cal Q})^{-1}
\hat{\cal Q}_1$ and using that $\hat{\cal Q}_1(\hat{\cal Q}_> \hat{\cal H}\hat{\cal Q}_>)^n\hat{\cal Q}_1=0$, we get:

\begin{align}
\label{eq:PQ11stepAlt1_bord}
&\hat{\cal Q}_1[ E -(\hat{\cal Q}_1+\hat{\cal Q}_>) \hat{\cal H}_{\rm d}(\hat{\cal Q}_1+\hat{\cal Q}_>)]^{-1}\hat{\cal Q}_1= \nonumber\\
&=\frac{\hat{\cal Q}_1}{E}\left\{\sum_{n=1}^\infty \Big(\frac{\hat{\cal Q}_1 \hat{\cal H}_{\rm d}\hat{\cal Q}_1}{E}\Big)^n+1\right\}\hat{\cal Q}_1=\nonumber\\
&= \hat{\cal Q}_1\frac{1}{E - \hat{\cal Q}_1 \hat{\cal H}_{\rm d}\hat{\cal Q}_1}\hat{\cal Q}_1~.
\end{align}
Replacing this result in Eq.~(\ref{eq:PQ11stepAlt}) we find
\begin{equation}
\label{eq:PQ11}
\hat{\Sigma}(E)=\hat{\cal P}\hat{\cal H}_{\rm para} \hat{\cal Q}_1(  E -\hat{\cal Q}_1 \hat{\cal H}_{\rm d}\hat{\cal Q}_1)^{-1}
\hat{\cal Q}_1 \hat{\cal H}_{\rm para} \hat{\cal P}~.
\end{equation}
The last step is to expand $E$ in powers of $g_{\rm FP}$ by assuming that the energy is close to that of the non-interacting ground state, i.e.~$E=E_0+\delta E$, where $\delta E \sim g_{\rm FP}$. Hence, in the term $ (  E -\hat{\cal Q}_1 \hat{\cal H}\hat{\cal Q}_1)^{-1}$ we can approximate $E$ with $E_0$, obtaining a non-retarded self-energy, $\hat{\Sigma}(E)\simeq\hat{\Sigma}(E_0)$. After all these simplifications, we obtain
\begin{align}
&(E_0 -\hat{\cal Q}_1 \hat{\cal H}_{\rm d}\hat{\cal Q}_1)^{-1} \simeq \sum_n\sum_{\bm{q}_{\parallel},s ,n_z} |{\psi_n} \tensor[]{\rangle}{_{\rm e}} |{\bm{q}_{\parallel},s ,n_z}\tensor[]{\rangle}{_{\rm ph}} \times \nonumber \\ &\times \frac{1}{(  E_0 -E_n-\hbar\omega_{\bm{q}_{\parallel},s ,n_z} )} {\tensor[_{\rm ph}]{\langle}{}{\bm{q}_{\parallel},s ,n_z}|}\tensor[_{\rm e}]{\langle}{} \psi_n|~.
\end{align} 
We now define the EEEI Hamiltonian as the self-energy evaluated at  energy $E_0$ {\it and} projected onto the bare photon vacuum, i.e.
\begin{equation}\label{eq:profound_definition_EEEI}
\hat{\mathcal{H}}_{\rm EEEI} \equiv \tensor[_{\rm ph}]{\langle}{}0|\hat {\Sigma}(E_0)|0\tensor[]{\rangle}{_{\rm ph}}~.
\end{equation} 
Explicitly, we have:
\begin{align}
\label{eq:PQ2}
\hat{\mathcal{H}}_{\rm EEEI}&= \sum_n\sum_{\bm{q}_{\parallel},s ,n_z} \tensor[_{\rm ph }]{\langle}{} {0}|\hat{\cal H}_{\rm para}|{\bm{q}_{\parallel},s ,n_z}\tensor[]{\rangle}{_{\rm ph}}\ket{\psi_n}_{\rm e} \times\nonumber \\ &\times \frac{1}{(  E_0 -E_n-\hbar\omega_{\bm{q}_{\parallel},s ,n_z} )}  \times \nonumber\\
&\times \prescript{}{\rm e }{\langle} \psi_n|\tensor[_{\rm ph }]{\langle}{}{\bm{q}_{\parallel},s ,n_z}|\hat{\cal H}_{\rm para} |{0}\tensor[]{\rangle}{_{\rm ph}}~. \nonumber\\
\end{align}
This is an Hamiltonian acting only on the electronic sector.

We can further simplify Eq.~(\ref{eq:PQ2}) by performing the {\it adiabatic} approximation, i.e.~by assuming that the following inequality is satisfied: $E_n -E_0 \ll \hbar\omega_{\bm{q}_{\parallel},s ,n_z} $. In our case this assumption is safe since the energy difference between electrons in the conduction band (which are responsible for superconductivity) is much smaller that the
photon energy. By using the {\it adiabatic} assumption and the completeness relation, $\sum_n|{\psi_n}\tensor[]{\rangle}{_{\rm e}} \tensor[_{\rm e}]{\langle}{}{\psi_n}|=\mathbb{1}_{\rm e}$, we finally get:
\begin{align}
\label{eq:PQ3}
\hat{\mathcal{H}}_{\rm EEEI}\simeq  -\sum_{\bm{q}_{\parallel},s ,n_z}\frac{1}{\hbar\omega_{\bm{q}_{\parallel},s ,n_z} }  |\tensor[_{\rm ph} ]{\langle}{} {0}|\hat{\cal H}_{\rm para}|{\bm{q}_{\parallel},s ,n_z}\tensor[]{\rangle}{_{\rm ph}}|^2~.
\end{align}
Following this projection approach we have integrated out the photonic degrees of freedom and found an EEEI. Notice that $ |\tensor[_{\rm ph}]{\langle}{}{0}|\hat{\cal H}_{\rm para}|{\bm{q}_{\parallel},s ,n_z} \tensor[]{\rangle}{_{\rm ph}}|^2$ is still an {\it operator} acting on the electronic degrees of freedom---$|\hat{O}|^2=\hat{O}^\dagger\hat{O}$ is the modulus of the operator $\hat{O}$. Eq.~(\ref{eq:PQ3}) is the most important result of this Section.

We can write Eq.~(\ref{eq:PQ3}) more explicitly by employing the expression for the paramagnetic term $\hat{\cal H}_{\rm para}$ worked out in Eq.~\eqref{eq:para4}. We find,
\begin{widetext}
\begin{equation}
\label{eq:PQ4}
\hat{\mathcal{H}}_{\rm EEEI}= -\Big(\frac{e}{c}\Big)^2\sum_{\bm{q}_{\parallel},s ,n_z}\frac{1}{\hbar\omega_{\bm{q}_{\parallel},s ,n_z} } \big| \sum_{\bm{q}^\prime_{\parallel},\ell}  \hat{\bm J}_{\ell, -\bm{q}^\prime_{\parallel}} \cdot \tensor[_{\rm ph}]{\langle}{}{0|\hat{\bm{A}}_{\bm{q}^\prime_\parallel,\ell} |\bm{q}_{\parallel},s ,n_z}\tensor[]{\rangle}{_{\rm ph}}\big|^2~. 
\end{equation}
\end{widetext}
The matrix element $\tensor[_{\rm ph}]{\langle}{}{0|\hat{\bm{A}}_{\bm{q}^\prime_\parallel,\ell} |\bm{q}_{\parallel},s ,n_z}\tensor[]{\rangle}{_{\rm ph}}$ can be calculated from Eq.~\eqref{eq:componentsA}.  By noticing that $\tensor[_{\rm ph}]{\langle}{}{0|\hat{a}_{\bm{q}^\prime_{\parallel}, s^\prime ,n^\prime_z} |\bm{q}_{\parallel},s ,n_z}\tensor[]{\rangle}{_{\rm ph}}=\delta_{\bm{q}_{\parallel},\bm{q}_{\parallel}^\prime}\delta_{s,s^\prime}\delta_{n_z,n_z^\prime}$ and $\tensor[_{\rm ph }]{\langle}{}{{0|\hat{a}^\dagger_{\bm{q}^\prime_{\parallel}, s^\prime ,n^\prime_z} |\bm{q}_{\parallel},s ,n_z}}\tensor[]{\rangle}{_{\rm ph}}=0$, we get
\begin{align}
\tensor[_{\rm ph }]{\langle}{} {0|\hat{\bm{A}}_{\bm{q}^\prime_\parallel,\ell} |\bm{q}_{\parallel},s ,n_z}\tensor[]{\rangle}{_{\rm ph}}=\delta_{\bm{q}^\prime_{\parallel},\bm{q}_{\parallel}} A^{\rm(2D)}_{\bm{q}_{\parallel} ,n_z} e^{i   \frac{\bm{q}_{\parallel} \cdot \bm{\delta}_\ell}{2}} {\bm e}_{\bm{q}_{\parallel},s,n_z } (0)~.
\end{align}
Using this result in Eq.~\eqref{eq:PQ4} we find an effective long-range interaction between currents,
\begin{equation}
\label{eq:PQ5}
\hat{\mathcal{H}}_{\rm EEEI}= -\Big(\frac{e}{c}\Big)^2\sum_{\ell,{\ell}^\prime,\bm{q}_{\parallel}}  \hat{\bm{J}}_{\ell,-\bm{q}_{\parallel}}\cdot \mathcal{F}_{\ell,{\ell}^\prime,\bm{q}_{\parallel}}\cdot
 \hat{\bm{J}}_{\ell^\prime, -\bm{q}_{\parallel}}^\dagger~,
\end{equation}
where,
\begin{align}
\label{eq:Ftensor}
\mathcal{F}_{\ell,{\ell}^\prime,\bm{q}_{\parallel}}= \sum_{s ,n_z}{\bm e}_{\bm{q}_{\parallel},s,n_z} (0) \frac{ e^{\frac{i\bm{q}_{\parallel} \cdot (\bm{\delta}_\ell - \bm{\delta}_{\ell^\prime}) }{2}}\big( A^{\rm(2D)}_{\bm{q}_{\parallel} ,n_z}\big)^2}{\hbar\omega_{\bm{q}_{\parallel},s ,n_z} }  {\bm e}^*_{\bm{q}_{\parallel},s,n_z} (0)~.
\end{align}
This term contains only quantities associated with the vector potential, specifically the photonic energy $\hbar\omega_{\bm{q}_{\parallel},s ,n_z}$, the photonic polarization ${\bm e}_{\bm{q}_{\parallel},s,n_z} (0)$, the vector potential amplitude $A^{\rm(2D)}_{\bm{q}_{\parallel} ,n_z}$, and a phase factor, $ \exp[{i{\bm{q}_{\parallel} \cdot (\bm{\delta}_\ell - \bm{\delta}_{\ell^\prime} })/{2}}]$, corresponding to the position at which the field is evaluated---see Fig.~\ref{fig:reticolo_gr}(b). 

By using the explicit expressions for the field amplitude $A^{\rm(2D)}_{\bm{q}_{\parallel} ,n_z}$ and the photonic energy $\hbar\omega_{\bm{q}_{\parallel},s ,n_z}$ reported in Eqs.~(\ref{eq:A2D})-(\ref{eq:omega}) we get,
\begin{align}
\label{eq:F}
 \mathcal{F}_{\ell,{\ell}^\prime,\bm{q}_{\parallel}}
 &=\frac{4 L_z}{ \pi S} e^{\frac{i\bm{q}_{\parallel} \cdot (\bm{\delta}_\ell - \bm{\delta}_{\ell^\prime}) }{2}}\times  \nonumber\\ 
 &\times  \sum_{s ,n_z}{\bm e}_{\bm{q}_{\parallel},s,n_z} (0) \frac{1}{(L_zq_\parallel/\pi)^2 + n^2_z} {\bm e}^*_{\bm{q}_{\parallel},s,n_z} (0)~.
 \end{align}
We remind the reader that the sum over $n_z$ in Eq.~(\ref{eq:F}) runs only over the odd modes, i.e.~$n_z=2m_z+1$ with $m_z\in \mathbb{N}$.

Using the explicit expressions for the polarization vectors reported in Eqs.~(\ref{polarization_1z=0})-(\ref{polarization_2z=0}),  we can decompose $\mathcal{F}_{\ell,{\ell}^\prime,\bm{q}_{\parallel}}$ in Eq.~\eqref{eq:F} into a transverse and a longitudinal contribution (with respect to $\bm{q}_{\parallel}$):
\begin{align}
\label{eq:F1}
\mathcal{F}_{\ell,{\ell}^\prime,\bm{q}_{\parallel}}&= \mathcal{F}^{\rm T}_{\ell,{\ell}^\prime,\bm{q}_{\parallel}}\bm{u}_{{\rm T},\bm{q}_{\parallel}}  \bm{u}_{{\rm T},\bm{q}_{\parallel}} +
  \mathcal{F}^{\rm L}_{\ell,{\ell}^\prime,\bm{q}_{\parallel}} \bm{u}_{{\rm L},\bm{q}_{\parallel}}  \bm{u}_{{\rm L},\bm{q}_{\parallel}}~, 
\end{align}
where
\begin{align}\label{eq:F2}
 \mathcal{F}^{\rm T}_{\ell,{\ell}^\prime,\bm{q}_{\parallel}}  &=\frac{4 L_z}{\pi S} e^{\frac{i\bm{q}_{\parallel} \cdot (\bm{\delta}_\ell - \bm{\delta}_{\ell^\prime}) }{2}} \nonumber\\
 &\times \sum_{m_z=0}^\infty\frac{1}{(L_zq_\parallel/\pi)^2 + (2m_z+1)^2}~,
\end{align}
and
\begin{align}\label{eq:F2_2}
\mathcal{F}^{\rm L}_{\ell,{\ell}^\prime,\bm{q}_{\parallel}}  &=\frac{4 L_z}{\pi S} e^{\frac{i\bm{q}_{\parallel} \cdot (\bm{\delta}_\ell - \bm{\delta}_{\ell^\prime}) }{2}} \nonumber\\
&\times \sum_{m_z=0}^{\infty} \frac{(2m_z+1)^2}{[(L_z q_\parallel/\pi)^2 + (2m_z+1)^2]^2}~. 
\end{align} 
The two sums over the mode index $m_z$ can be performed analytically. We find:
\begin{equation}\label{eq:ST_tanh}
{\cal S}_{\rm T}(x) \equiv \sum^\infty_{m_z=0} \frac{1}{(2m_z+1)^2+x^2} =  \frac{\pi}{4x}\tanh\left(\frac{\pi x}{2}\right)~,
\end{equation}
and
\begin{align}\label{eq:SL_sinh_cosh}
{\cal S}_{\rm L}(x) &\equiv \sum^\infty_{m_z=0}\frac{(2m_z+1)^2}{[(2m_z+1)^2+x^2]^2}  \nonumber\\
&=  \frac{\pi  [\sinh (\pi  x)+\pi x]}{8 x [\cosh (\pi  x)+1]}~.
\end{align}
Using these results we can rewrite Eqs.~(\ref{eq:F2})-(\ref{eq:F2_2}) as following: 
\begin{equation}\label{eq:F3}
\mathcal{F}^{\rm T}_{\ell,{\ell}^\prime,\bm{q}_{\parallel}}=\frac{4 L_z}{\pi S} e^{\frac{i\bm{q}_{\parallel} \cdot (\bm{\delta}_\ell - \bm{\delta}_{\ell^\prime}) }{2}}  {\cal S}_{\rm T}\left(\frac{L_z q_\parallel}{\pi}\right)~,
\end{equation}
and
\begin{equation}\label{eq:F3_3}
\mathcal{F}^{\rm L}_{\ell,{\ell}^\prime,\bm{q}_{\parallel}}=\frac{4 L_z}{\pi S} e^{\frac{i\bm{q}_{\parallel} \cdot (\bm{\delta}_\ell - \bm{\delta}_{\ell^\prime}) }{2}}  {\cal S}_{\rm L}\left(\frac{L_z q_\parallel}{\pi}\right)  ~.
\end{equation} 

Finally, the EEEI Hamiltonian $\hat{\mathcal{H}}_{\rm EEEI}$ in Eq.~\eqref{eq:PQ5} can be rewritten as,
\begin{equation}\label{eq:PQ6}
\hat{\mathcal{H}}_{\rm EEEI} = -\left(\frac{e}{c}\right)^2\sum_{\alpha={\rm T},{\rm L}}\sum_{\ell,{\ell}^\prime,\bm{q}_{\parallel}}  \hat{{{J}}}_{\ell,-\bm{q}_{\parallel},\alpha} \mathcal{F}^\alpha_{\ell,{\ell}^\prime,\bm{q}_{\parallel}}
 \hat{{{J}}}_{\ell^\prime,-\bm{q}_{\parallel},\alpha}^\dagger~,
\end{equation}
where we introduced the transverse and longitudinal currents (with respect to $\bm{q}_\parallel$) as:
\begin{align}\label{eq:TLcurrents}
\hat{J}_{\ell,\bm{q}_{\parallel},{\rm T}}&\equiv \bm{u}_{{\rm T},\bm{q}_{\parallel}} \cdot \hat{\bm J}_{\ell,\bm{q}_{\parallel}}~,\\
\hat{J}_{\ell,\bm{q}_{\parallel},{\rm L}}&\equiv  \bm{u}_{{\rm L},\bm{q}_{\parallel}} \cdot \hat{\bm J}_{\ell,\bm{q}_{\parallel}}~.
\end{align}

We now estimate the strength of the effective current-current interaction $\hat{\mathcal{H}}_{\rm EEEI}$ in Eq.~\eqref{eq:PQ6}. The product $\hat{J}_{\ell,-\bm{q}_{\parallel},\alpha}\hat{J}^\dagger_{\ell^\prime,-\bm{q}_{\parallel},\alpha}$ is on the order of $(ta/\hbar)^2$. The quantity
$\mathcal{F}_{\ell,{\ell}^\prime,\bm{q}_{\parallel}}$ in Eq.~\eqref{eq:F} is on the order of $L_z/S$. Provided that 2D electron system area and the surface $S$ of the cavity in the $\hat{\bm x}$-$\hat{\bm y}$ plane coincide, i.e.~provided that $S\sim N_{\rm cell}a^2$, and recalling the prefactor $(e/c)^2$ in Eq.~\eqref{eq:PQ6}, we immediately see that  $\hat{\cal{H}}_{\rm EEEI}$  scales like
\begin{align}
\label{eq:mangetostaticEEEI}
\hat{\cal{H}}_{\rm EEEI}\sim \frac{g^2_{\rm FP}}{\hbar \omega_{\rm D}} = t \alpha_{\rm QED} \frac{t L_z}{\pi \hbar c}~.
\end{align} 
Notice that the relative dielectric constant $\epsilon_{\rm r}$ drops out of the problem. This stems from the fact that, in the {\it adiabatic} limit, the photon-mediated EEEI corresponds to a magnetostatic interaction between currents, which does not feel the influence of a dielectric. We conclude that it is not possible to tune current-current EEEI by altering the electrostatic environment provided e.g. by a substrate.

Since we are interested in the possibility that the EEEI in Eq.~(\ref{eq:PQ6}) induces a superconducting instability, we can perform a further approximation onto $\hat{\cal{H}}_{\rm EEEI}$. Assuming to have an electron-doped graphene sheet, we can limit our analysis to the conduction band and set $\kappa= \kappa^\prime = +$.  Indeed, superconductivity emerges from attractive interactions between electrons lying in a thin energy shell around the Fermi energy. Also, we focus on a high doping regime where superconductivity is expected to be strongly favoured and where cavity photons cannot induce inter-band transitions. Mathematically, the absence of inter-band transitions is ensured by the {\it Pauli-blocking} inequality $\hbar \omega_{\rm D} < 2\mu$, the effective Debye energy $\hbar \omega_{\rm D}$ representing half of the interested energy shell around the chemical potential $\mu$. In this intra-band approximation we can express the current-current interaction in terms of the conduction-band fermionic operators $ \hat d_{{\bm k} ,\sigma }\equiv\hat d_{{\bm k} ,\sigma,+}$ as
\begin{widetext}
\begin{eqnarray}
\hat{\mathcal{H}}_{\rm EEEI} &=&-\frac{g^2_{\rm FP}}{\hbar \omega_{\rm D}} \frac{a^2}{S}\sum_{\bm{q}_\parallel}\sum_{\alpha={\rm T,L}}  {\cal S}_{\alpha}\Big(\frac{L_z q_\parallel}{\pi}\Big)\sum_{\bm{k},\ell,\sigma}\Big[\frac{ \bm{\delta}_{\ell}}{a}\cdot \bm{u}_{\alpha,\bm{q}_{\parallel}} \Big(e^{i(\bm{k}+\bm{q}_\parallel/2)\cdot \bm{\delta}_{\ell}} {\varphi^\ast_{\bm k}} {-}e^{-i(\bm{k}+\bm{q}_\parallel/2)\cdot  \bm{\delta}_{\ell}} {\varphi_{\bm{k}+\bm{q}_{\parallel}}}\Big)\hat d^\dagger_{\bm{k}+\bm{q}_\parallel ,\sigma }\hat d_{{\bm k} ,\sigma }\Big]
\times \nonumber\\&\times &\sum_{\bm{k}^\prime,\ell^\prime,\sigma^\prime} \Big[ \frac{\bm{\delta}_{\ell^\prime}}{a}\cdot \bm{u}_{\alpha,\bm{q}_{\parallel}} \Big(e^{-i(\bm{k}^\prime-\bm{q}_\parallel/2)\cdot \bm{\delta}_{\ell^\prime}} {\varphi_{{\bm{k}^\prime}-\bm{q}_{\parallel}}} {-}e^{i(\bm{k}^\prime-\bm{q}_\parallel/2)\cdot \bm{\delta}_{\ell^\prime}} {\varphi^\ast_{\bm{k}^\prime}} \Big)\hat d^\dagger_{\bm{k}^\prime-\bm{q}_\parallel ,\sigma^\prime }\hat d_{{\bm k}^\prime ,\sigma^\prime }\Big]~.
\end{eqnarray}
\end{widetext}
In writing the previous equation we have used the explicit expression of the current operator in terms of the fermionic creation and annihilation operators---see Eqs.~(\ref{eq:Fourier_link_current}) and~(\ref{eq:componentsJ}).

We now perform a change of variables introducing the three vectors $\bm Q$, $\bm p$, and $\bm p'$ in such a way that ${\bm k} = {\bm Q} + {\bm p}^\prime$,  
${\bm k}^\prime = {\bm Q} -  {\bm p}^\prime$, and  ${\bm q}_\parallel = {\bm p} - {\bm p}^\prime$. This will enable us to express the EEEI in terms of the center-of-mass momentum $\bm Q$ and the relative momentum $\bm p^\prime$ of the two interacting electrons. This choice of variables is motivated by our interest in establishing whether or not it is possible to induce an Amperean superconducting instability in a 2D electron system (in our case SLG) placed inside a cavity. As we will see below in Sect.~\ref{sec:AmpereanGap}, in an Amperean superconductor $\bm Q$ is allowed to hold a finite value. This is in stark contrast with an ordinary BCS superconductor where the center-of-mass momentum ${\bm Q}$ of a Cooper pair is ${\bm Q}_{\rm BCS}={\bm 0}$.

Taking $S= 3\sqrt{3}N_{\rm cell} a^2 /2$ and performing the aforementioned change of variables, the EEEI becomes:
\begin{align}\label{eq:Veff_f2}
\hat{\mathcal{H}}_{\rm EEEI} &=
 \frac{1}{2 N_{\rm cell}}  \sum_{ {\bm p}, {\bm p}', \bm Q} V_{\bm p, \bm p', \bm Q} \sum_{\sigma,\sigma' = \uparrow,\downarrow} \Big( 
\hat d_{{\bm Q} + {\bm p},\sigma }^{\dagger} 
\hat d_{{\bm Q}-{\bm p},\sigma' }^{\dagger} \nonumber\\
&\times
\hat d_{{\bm Q }-{\bm p}',\sigma'}
\hat d_{{\bm Q }+{\bm p'},\sigma} 
+
\hat d_{{\bm Q} + {\bm p},\sigma }^{\dagger} \hat d_{{\bm Q} + {\bm p},\sigma }  \delta_{\bm p',-\bm p} \delta_{\sigma',\sigma}\Big)~.
\end{align}
Here, $ V_{\bm p, \bm p', \bm Q}$ is the electron-electron interaction potential defined as $V_{\bm p, \bm p', \bm Q}= [g^2_{\rm FP}/(\hbar\omega_{\rm D})] \tilde{V}_{\bm p, \bm p', \bm Q}$, where $\tilde{V}_{\bm p, \bm p', \bm Q}$ is the following dimensionless quantity:

\begin{widetext}
\begin{align}
\label{eq:potential}
\tilde{V}_{\bm p, \bm p', \bm Q}&= -\frac{4}{3\sqrt{3}} \Big(\bm{\eta}_{\bm{Q}+(\bm{p}+\bm{p}^\prime)/2} {\varphi^\ast_{\bm{Q}+\bm{p}^\prime}} -\bm{\eta}^\ast_{\bm{Q}+(\bm{p}+\bm{p}^\prime)/2} {\varphi_{\bm{Q}+\bm{p}}}\Big) \cdot\Big[ \sum_{\alpha={\rm T,L}} \bm{u}_{\alpha,\bm{p}-\bm{p}^\prime} {\cal S}_{\alpha}\Big(\frac{L_z |\bm{p}-\bm{p}^\prime|}{\pi}\Big)  \bm{u}_{\alpha,\bm{p}-\bm{p}^\prime} \nonumber\Big]\cdot\\ &\cdot
\Big(\bm{\eta}^\ast_{\bm{Q}-(\bm{p}+\bm{p}^\prime)/2} {\varphi_{\bm{Q}-\bm{p}}} -\bm{\eta}_{\bm{Q}-(\bm{p}+\bm{p}^\prime)/2} {\varphi^\ast_{\bm{Q}-\bm{p}^\prime}} \Big)~,\nonumber\\
\end{align}
\end{widetext}
where the vector $\bm{\eta}_{\bm{k}}$ is
\begin{equation}
\bm{\eta}_{\bm{k}} \equiv \sum_{\ell=1}^3 \bm{\delta}_{\ell} e^{i\bm{k}\cdot \bm{\delta}_{\ell}}~.
\end{equation}

We now discuss a few important properties of $\tilde{V}_{\bm p, \bm p', \bm Q}$: 

i) The Hermitian nature of $\hat{\mathcal{H}}_{\rm EEEI}$ as defined in Eq.~(\ref{eq:profound_definition_EEEI}) implies that $\tilde{V}_{\bm p, \bm p', \bm Q}$ satisfies the following property: $\tilde{V}_{\bm p, \bm p', \bm Q}=\tilde{V}^\ast_{\bm p^\prime, \bm p, \bm Q}$. 

ii) We can see by direct inspection that $\tilde{V}_{\bm p, \bm p', \bm Q}$ does not depend on the cavity length $L_z$ if one takes $\bm{p}=\bm{p}^\prime$. This happens because, in the function $ {\cal S}_{\alpha}(x)$, the length $L_z$ always appears in the product $L_z(\bm{p}-\bm{p}^\prime)$. 

iii) We note that the quantity $g^2_{\rm FP}/(\hbar\omega_{\rm D}) \propto L_z$ in the definition of $V_{\bm p, \bm p', \bm Q}$ diverges in free space, i.e.~in the limit $L_z\to \infty$. One may therefore be tempted to conclude (erroneously) that the effective current-current interaction potential diverges in free space. We now show that this not the case. In the limit $x\to \infty$, the functions ${\cal S}_\alpha(x)$ for $\alpha={\rm T}$ and $\alpha={\rm L}$ behave as following:
\begin{align}
\label{eq:asympt1}
{\cal S}_{\rm T}(x\to \infty)&= \frac{ \pi}{4 x}~,\\
\label{eq:asympt2}
{\cal S}_{\rm L}(x\to \infty)&= \frac{ \pi}{8 x}~.
\end{align}
These results imply that $\tilde{V}_{\bm p, \bm p', \bm Q}$ scales as $1/(L_z|\bm{p}-\bm{p}^\prime|)$ in the free-space $L_z\to \infty$ limit. Hence, globally ${V}_{\bm p, \bm p', \bm Q}$ tends to an $L_z$-independent quantity for  $L_z\to \infty$.

iv)
Notice that in the same limit, the electromagnetic functions $ \mathcal{F}^{\alpha}_{\ell,{\ell}^\prime,\bm{q}_{\parallel}}$ for $\alpha={\rm T}$ and $\alpha={\rm L}$ tend to the following asymptotical results:
\begin{align}\label{eq:F2asy}
\lim_{L_z\to \infty} \mathcal{F}^{\rm T}_{\ell,{\ell}^\prime,\bm{q}_{\parallel}}&=\frac{\pi}{ q_\parallel S} e^{\frac{i\bm{q}_{\parallel} \cdot (\bm{\delta}_\ell - \bm{\delta}_{\ell^\prime}) }{2}}  
 ~, \\
\lim_{L_z\to \infty} \mathcal{F}^{\rm L}_{\ell,{\ell}^\prime,\bm{q}_{\parallel}}&=\frac{\pi }{2 q_\parallel S} e^{\frac{i\bm{q}_{\parallel} \cdot (\bm{\delta}_\ell - \bm{\delta}_{\ell^\prime}) }{2}}  ~,
\end{align}
which are the correct results in the absence of a cavity. Notice that, in this limit, the quantities $ \mathcal{F}^{\alpha}_{\ell,{\ell}^\prime,\bm{q}_{\parallel}}$ do not depend on $L_z$, as expected for the free-space case.

v) Finally, we comment on the range of values of $L_z$ we have explored in the numerical calculations that are discussed below in  Sect.~\ref{sec:AmpereanGap}. A {\it lower bound} on the distance $L_{z}$ between the mirrors is provided by the fact that we have decided to operate in the regime of Pauli blocking, i.e.~in the regime of chemical potentials $\mu$ obeying the inequality $\hbar \omega_{\rm D} <2 \mu$. By expressing the Debye energy scale $\hbar \omega_{\rm D}$ in terms of $L_z$ as in Eq.~\eqref{eq:Denergy} we get an expression for the lower bound $L_z^{\rm min}$ on $L_z$:
\begin{equation}
L_z^{\rm min} \equiv \frac{\pi\hbar c}{2\mu\sqrt{\epsilon_{\rm r}}}~.
\end{equation}

Using $\epsilon_{\rm r}=4$, which is appropriate to an hBN spacer separating the two metallic mirrors (see Table~\ref{table}), we obtain
\begin{equation}
L_z^{\rm min} \simeq \frac{0.15}{\mu[{\rm eV}]}~{\rm \mu m} = \frac{150}{\mu[{\rm eV}]}~{\rm nm}~.
\end{equation}
For $\mu= 0.5 t$ we get $L_z^{\rm min}\simeq 0.11~{\rm \mu m} = 110~{\rm nm}$.

\begin{figure*}[t]
  \centering
  \vspace{1.em}
  \begin{overpic}[width=0.68\columnwidth]{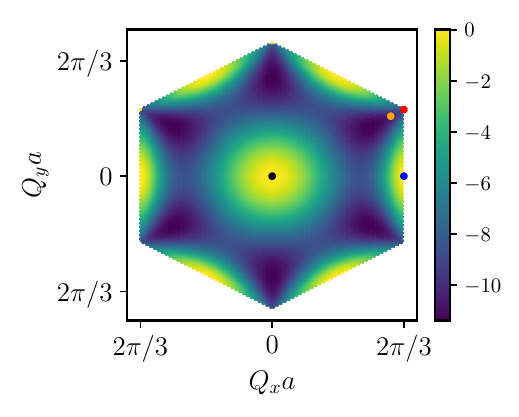}\put(2,80){\normalsize (a)}\end{overpic} 
  \begin{overpic}[width=0.68\columnwidth]{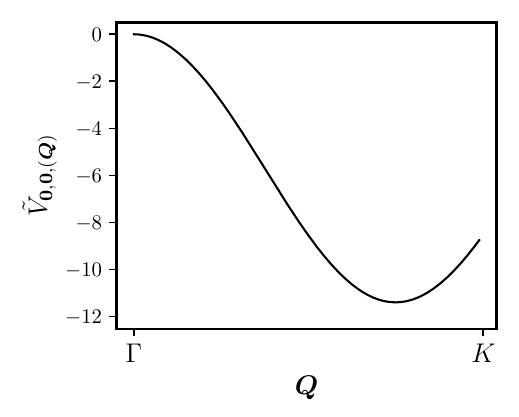}\put(2,80){\normalsize (b)}\end{overpic}
  \begin{overpic}[width=0.68\columnwidth]{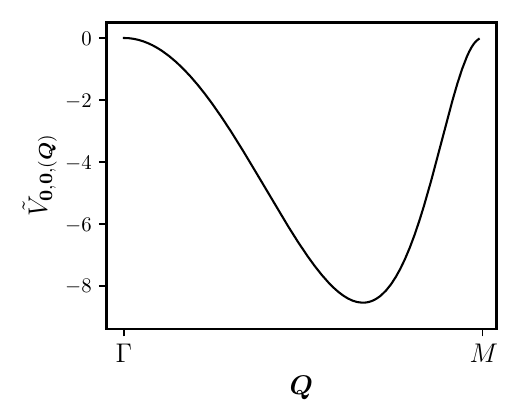}\put(2,80){\normalsize (c)}\end{overpic}\\
    \begin{overpic}[width=0.68\columnwidth]{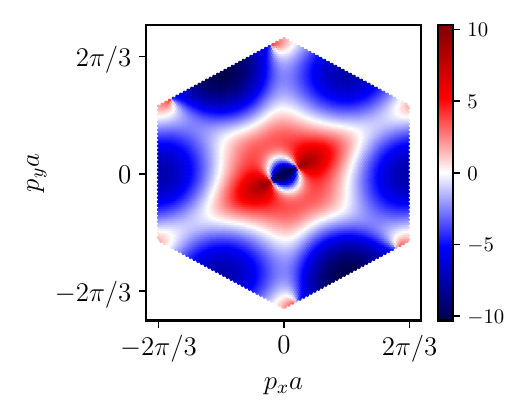}\put(2,80){\normalsize (d)}\end{overpic} 
  \begin{overpic}[width=0.68\columnwidth]{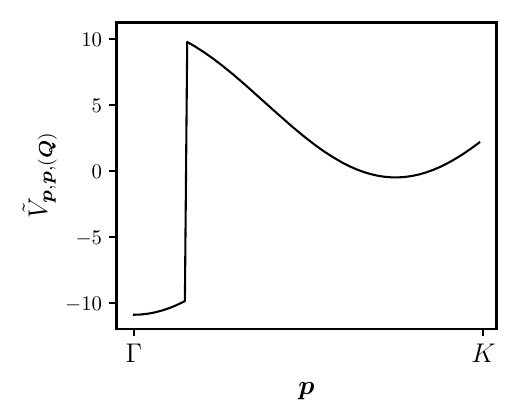}\put(2,80){\normalsize (e)}\end{overpic}
  \begin{overpic}[width=0.68\columnwidth]{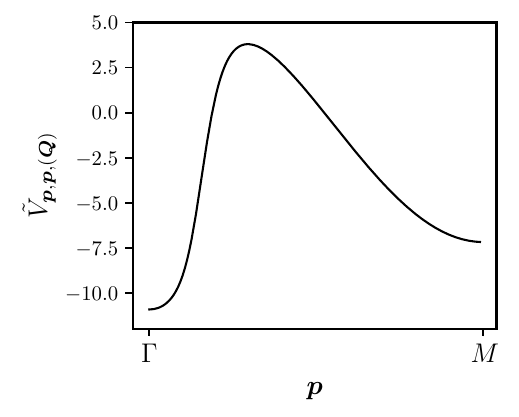}\put(2,80){\normalsize (f)}\end{overpic}\\
    \begin{overpic}[width=0.68\columnwidth]{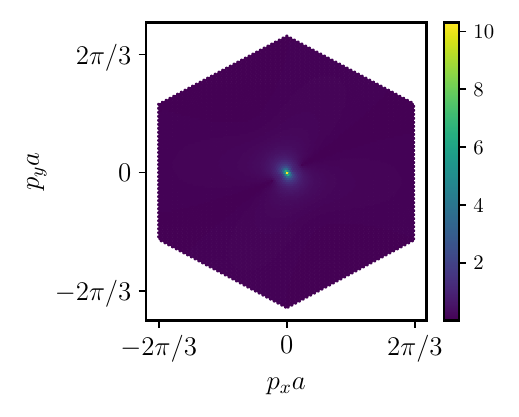}\put(2,80){\normalsize (g)}\end{overpic} 
  \begin{overpic}[width=0.68\columnwidth]{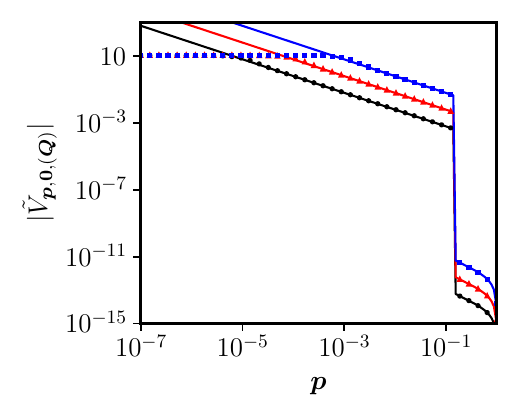}\put(2,80){\normalsize (h)}\end{overpic}
  \begin{overpic}[width=0.68\columnwidth]{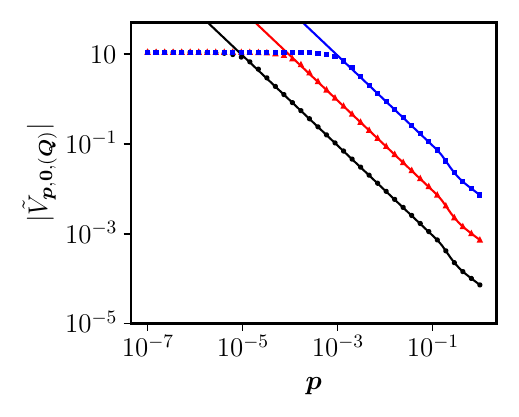}\put(2,80){\normalsize (i)}\end{overpic}\\
  \caption{(Color online) Panel (a) Two-dimensional color map of the dimensionless EEEI potential $\tilde{V}_{\bm{p},{\bm{p}'},\bm{Q}}$ as a function of $\bm{Q}$ in the first BZ and for $\bm{p}=\bm{p'}=\bm{0}$. The filled black, blue, and red filled circles denote the $\Gamma$, $M$, and $K$ points of the BZ, respectively. We recall that $\bm{\Gamma}=(0,0)$, $\bm{K}=2\pi/(3a)(1,1/\sqrt{3})$, and $\bm{M}=2\pi/(3a)(1,0)$.  The orange filled circle corresponds to the point $0.9\bm{K}$. In our numerical calculations of the Amperean superconducting gap we have taken ${\bm{Q}}=0.9\bm{K}$.
Panels (b)-(c) show $\tilde{V}_{\bm{0},{\bm{0}},\bm{Q}}$ as a function of $\bm{Q}$, which is varied along a high-symmetry path in the BZ. In panel (b) the potential is evaluated along the path $\Gamma-K$ while in panel (c) along the path $\Gamma-M$. It is clear from panel (b) that the minimum of the potential is reached for $\bm{Q}\approx(2/3)\bm{K}$.
Panel (d) shows a color map of $\tilde{V}_{\bm{p},{\bm{p}'},\bm{Q}}$  as a function of $\bm{p}$ in the first BZ, for $\bm p=\bm p'$ and $\bm{Q}=0.9\bm{K}$. Panels (e) and (f) show $\tilde{V}_{\bm{p},\bm{p},0.9\bm{K}}$  evaluated along the path $\Gamma-K$ and $\Gamma-M$, respectively. Panels (a)-(f) of this figure show quantities that are {\it independent} of the cavity length $L_z$.
Panel (g) shows a color map of $|\tilde{V}_{\bm{p},{\bm{p}'},\bm{Q}}|$ as a function of $\bm{p}$ in the first BZ, and evaluated for ${\bm p}^\prime = {\bm 0}$ and $\bm{Q}=0.9\bm{K}$. Results in panel (g) have been obtained by setting $L_z=10~{\rm nm}$. This value of $L_z$, which is below the lower bound $L_z^{\rm min}$, has been chosen only to show the potential structure, since for higher values of $L_z$ the potential is highly peaked around $\bm{p}=\bm{p}^\prime$. The symbols in panels (h) and (i) show $|\tilde{V}_{\bm{p},\bm{0},0.9\bm{K}}|$ evaluated along the paths $\Gamma-K$ and  $\Gamma-M$, respectively. Notice that the horizontal axis (${\bm p}$) is in logaritmic scale. The solid lines represent the asymptotic expansion of $|\tilde{V}_{\bm{p},\bm{0},0.9\bm{K}}|$, which can be easily obtained by using Eqs.~(\ref{eq:asympt1}) and~(\ref{eq:asympt2}).  Black, red, and blue colors in panels (h) and (i) refer to three different values of $L_z$, i.e.~$L_z=10~{\rm \mu m}$, $L_z = 1~{\rm \mu m}$, and $L_z= 100~{\rm nm}$, respectively. \label{fig:Vp_pp_Q}}
\end{figure*}

We now provide some illustration of the electron-electron interaction potential $V_{\bm p,\bm p', \bm Q}$. In Fig.~\ref{fig:Vp_pp_Q}, panels (a)-(c), we plot the dimensionless quantity $\tilde{V}_{\bm p,\bm p', \bm Q}$ defined in Eq.~(\ref{eq:potential}) as a function of $\bm{Q}=(Q_x,Q_y)$ and for~\cite{Jaksch_prl_2019} $\bm p=\bm p'=\bm 0$. Note that the resulting potential $\tilde{V}_{\bm 0,\bm 0, \bm Q}$ is {\it attractive} and the minimum is reached for ${\bm Q} = {\bm Q}^\star \simeq 2/3\bm{K}$, where $\bm{K}=2\pi/(3a)(1,1/\sqrt{3})$ is an high-symmetry point of the honeycomb-lattice BZ. However, at ${\bm Q} = {\bm Q}^\star$ the density of states corresponding to the symmetrized dispersion ${\bar\epsilon}_{\bm{p}}$ 
\begin{eqnarray}\label{eq:Sdispersion}
\bar{\epsilon}_{\bm p} \equiv \frac{\xi_{\bm{p}+\bm{Q}}+ \xi_{-\bm{p}+\bm{Q}}}{2},~
\end{eqnarray}
vanishes for $\mu \lesssim 0.65 t$. As we will see below, 
$\bar{\epsilon}_{\bm p}$ plays an important role in the theory of the Amperean superconductivity that will be discussed in the next Section. A better choice of $\bm{Q}$ is ${\bm Q} \simeq 0.9\bm{K}$ for which $\tilde{V}_{\bm 0,\bm 0, \bm Q}$ is attractive {\it and} there is a finite density of states for high but experimentally accessible~\cite{Starke} values of $\mu$ such as $\mu\approx 0.5 t$. In what follows we will focus on this choice for ${\bm Q}$. 

In Fig.~\ref{fig:Vp_pp_Q}, panels~(d)-(f), we show that, for ${\bm Q} \simeq 0.9\bm{K}$, the minimum of $\tilde{V}_{\bm p,\bm p', \bm Q}$ is reached for $\bm p = \bm p'=\bm 0$, justifying the choice we made above. In Fig.~\ref{fig:Vp_pp_Q}, panels (g)-(i), we finally show that the potential is peaked around $\bm p = \bm p'=\bm 0$. Results in panels (h) and (i) confirm that, in most of the BZ, the asymptotic scalings of ${\cal S}_\alpha(x)$ reported in Eqs.~(\ref{eq:asympt1})-(\ref{eq:asympt2}) well approximate the full functions ${\cal S}_\alpha(x)$ in Eqs.~(\ref{eq:ST_tanh})-(\ref{eq:SL_sinh_cosh}).


%
\section{Amperean superconductivity}
\label{sec:AmpereanGap}
The Hamiltonian~\eqref{eq:Veff_f2} is an EEEI mediated the cavity gauge field, akin to the microscopic phonon-mediated EEEI Hamiltonian on which the BCS theory relies~\cite{Grosso,BCS}. The main difference between Eq.~\eqref{eq:Veff_f2} and the conventional BCS Hamiltonian is that in the latter electrons on {\it opposite} sides of the Fermi surface (i.e.~electrons with ${\bm k^\prime}=-{\bm k}$ and center-of-mass momentum ${\bm Q}_{\rm BCS}={\bm 0}$) become partners in a Cooper pair. On the contrary, the Hamiltonian~\eqref{eq:Veff_f2} describes a pairing mechanism whereby electrons traveling in the same direction (${\bm Q} \neq {\bm 0}$) feel an attractive force. Since this is analogous to the Amperean attraction between parallel current wires that appears in classical electromagnetism, such pairing mechanism has also been dubbed Amperean~\cite{Lee_prl_2007}. 

Amperean superconductivity occurs when the Amperean gap $\Delta^{\bm{Q},\bm{p}}_{\sigma, \sigma^\prime}$, i.e. the expectation value of the operator $\hat{d}_{-\bm{p}^\prime+\bm{Q},\sigma^\prime} \hat{d}_{\bm{p}^\prime+ \bm{Q},\sigma}$, is finite. More precisely, we have
\begin{eqnarray}
\label{eq:AmpereanGap}
\Delta^{\bm{Q},\bm{p}}_{\sigma, \sigma^\prime} \equiv -\frac{1}{N_{\rm cell}} \sum_{\bm{p}^\prime}   V_{\bm{p},\bm{p}^\prime,\bm{Q}} \braket{\hat{d}_{-\bm{p}^\prime+\bm{Q},\sigma^\prime} \hat{d}_{\bm{p}^\prime+ \bm{Q},\sigma}}_{T}\,,
\end{eqnarray}
where the average is performed on the thermal state at a finite temperature $T$. 

In order to develop a mean-field theory of Amperean superconductivity, we follow the usual path~\cite{Giuliani_and_Vignale} and apply a Hartree-Fock (HF) decoupling onto the two-body interaction term in~\eqref{eq:Veff_f2}, i.e.~we approximate the right-hand side of Eq.~\eqref{eq:Veff_f2} as following (see, e.g., Chapter~2 in Ref.~\onlinecite{Giuliani_and_Vignale}):
\begin{align}
\label{eq:HF}
&\hat d_{{\bm Q} + {\bm p},\sigma }^{\dagger} 
\hat d_{-{\bm p}+{\bm Q},\sigma' }^{\dagger}
\hat d_{-{\bm p}'+{\bm Q },\sigma'}
\hat d_{{\bm Q }+{\bm p'},\sigma}\nonumber\\
&\simeq  \braket{\hat d_{{\bm Q} + {\bm p},\sigma }^{\dagger} 
\hat d_{-{\bm p}+{\bm Q},\sigma' }^{\dagger}}_T
\hat d_{-{\bm p}'+{\bm Q },\sigma'}
\hat d_{{\bm Q }+{\bm p'},\sigma} \nonumber\\
&+\hat d_{{\bm Q} + {\bm p},\sigma }^{\dagger} 
\hat d_{-{\bm p}+{\bm Q},\sigma' }^{\dagger}
\braket{\hat d_{-{\bm p}'+{\bm Q },\sigma'}
\hat d_{{\bm Q }+{\bm p'},\sigma}}_T \nonumber\\
&-\braket{\hat d_{{\bm Q} + {\bm p},\sigma }^{\dagger} 
\hat d_{-{\bm p}+{\bm Q},\sigma' }^{\dagger}}_T
\braket{\hat d_{-{\bm p}'+{\bm Q },\sigma'}
\hat d_{{\bm Q }+{\bm p'},\sigma}}_T ~,
\end{align}
which yields to the following mean-field HF Hamiltonian:
\begin{align}\label{eq:HIAmperean2}
\hat{\cal H}^{\rm HF}_{\rm EEEI}&=\frac{1}{2}\sum_{\bm{p},\bm{Q}} \sum_{\sigma,\sigma^\prime}  \Big[ \braket{\hat{d}^\dagger_{\bm{p}+\bm{Q},\sigma} \hat{d}^\dagger_{-\bm{p}+\bm{Q},\sigma^\prime}}_T \Delta^{\bm{Q},\bm{p}}_{\sigma, \sigma^\prime}\nonumber\\
&-  \hat{d}^\dagger_{\bm{p}+\bm{Q},\sigma} \hat{d}^\dagger_{-\bm{p}+\bm{Q},\sigma^\prime}
 \Delta^{\bm{Q},\bm{p}}_{\sigma, \sigma^\prime}\nonumber\\
&-\hat{d}_{-\bm{p}+\bm{Q},\sigma^\prime} \hat{d}_{\bm{p}+ \bm{Q},\sigma}  (\Delta^{\bm{Q},\bm{p}}_{\sigma, \sigma^\prime} )^*\Big] ~.
\end{align}

Following extensive earlier literature on PDW superconducting states \cite{Subdo_2020_1,Subdo_2020_2,Lee_prl_2007,Jaksch_prl_2019}, we assume that the momentum $\bm Q$ is {\it fixed} (by the analysis carried out in the previous Section, where it was concluded that the best choice is ${\bm Q}\simeq 0.9 {\bm K}$) and drop the sum over it. 

If we add  the free electron Hamiltonian to Eq.~\eqref{eq:HIAmperean2}, and fix the spin degrees of freedom $\sigma,\sigma^\prime$ so as to focus only on triplet or singlet superconductivity, the mean-field electronic Hartree-Fock Hamiltonian reads,
\begin{widetext}
\begin{eqnarray}\label{eq:eff-H-MF-bis1}
\hat{\cal H}^{\rm HF} = \frac{1}{2}\!\sum_{\bm{p}}\!\!
 \begin{pmatrix}
\hat{d}^\dagger_{\bm{p}+\bm{Q},\sigma}& \hat{d}_{-\bm{p}+\bm{Q},\sigma^\prime}
\end{pmatrix}
 \begin{pmatrix}
\xi_{\bm{p}+\bm{Q}} &  -\Delta^{\bm{Q},\bm{p}}_{\sigma, \sigma^\prime}  \\
 - \Delta^{\bm{Q},\bm{p}}_{\sigma^\prime, \sigma} &- \xi_{-\bm{p}+\bm{Q}}  \\
\end{pmatrix}
\begin{pmatrix}
\hat{d}_{\bm{p}+\bm{Q},\sigma} \nonumber\\ \hat{d}^\dagger_{-\bm{p}+\bm{Q},\sigma^\prime}
\end{pmatrix} +\frac{1}{2}\sum_{\bm{p}}  \braket{\hat{d}^\dagger_{\bm{p}+\bm{Q},\sigma} \hat{d}^\dagger_{-\bm{p}+\bm{Q},\sigma^\prime}}_T \Delta^{\bm{Q},\bm{p}}_{\sigma, \sigma^\prime} ~.
\end{eqnarray}
\end{widetext}
This Hamiltonian can be diagonalized by introducing the following Bogoliubov transformation:
\begin{eqnarray}
\hat{\gamma}_{{\bm{p}+\bm{Q}} ,\sigma}&=&u_{\bm p} \hat{d}_{{\bm{p}+\bm{Q}} ,\sigma}+v^*_{\bm p} \hat{d}^\dagger_{{-\bm{p}+\bm{Q}} ,\sigma^\prime}~,\\
\hat{\gamma}^\dagger_{{-\bm{p}+\bm{Q}} ,\sigma^\prime}&=&v_{\bm p} \hat{d}_{{\bm{p}+\bm{Q}} ,\sigma}-u_{\bm p}  \hat{d}^\dagger_{{-\bm{p}+\bm{Q}} ,\sigma^\prime}~,
\end{eqnarray}
where $u_{\bm{p}} =\cos(\theta_{\bm{p}}/2)$ and $v_{\bm{p}}=\sin(\theta_{\bm{p}}/2)e^{i\phi_{\bm{p}}}$, with 
\begin{equation}
\cos(\theta_{\bm{p}})=\frac{\bar{\epsilon}_{\bm p}}{\epsilon_{\bm{p}}}~,
\end{equation}
\begin{equation}
\sin(\theta_{\bm{p}})=\frac{\sqrt{h^2_{1}(\bm{p})+h^2_{2}(\bm{p})}}{\epsilon_{\bm{p}}}~,
\end{equation}
\begin{equation}
e^{i\phi_{\bm{p}}}=\frac{{h_{1}(\bm{p})+i h_{2}(\bm{p})}} {\sqrt{h^2_{1}(\bm{p})+h^2_{2}(\bm{p})}}~,
\end{equation}
and
\begin{equation}
\epsilon_{\bm{p}}=\sqrt{h^2_{1}(\bm{p})+h^2_{2}(\bm{p})+\bar{\epsilon}^2_{\bm p}}~.
\end{equation}
Here, $h_1(\bm{p})\equiv-{\rm Re}( \Delta^{\bm{Q},\bm{p}}_{\sigma, \sigma^\prime})$, $h_2(\bm{p})\equiv{\rm Im}(\Delta^{\bm{Q},\bm{p}}_{\sigma, \sigma^\prime})$, and $\bar{\epsilon}_{\bm p}$ has been defined above in Eq.~(\ref{eq:Sdispersion}). 

The above diagonalization procedure brings us to the following expression for the total Hamiltonian:
\begin{eqnarray}\label{eq:Bogoliubov_Hamiltonianbis}
\hat{\cal H}^{\rm HF} &=&\sum_{\bm{p}} E_{+}(\bm{p}) \hat{\gamma}^\dagger_{{\bm{p}+\bm{Q}} ,\sigma} \hat{\gamma}_{{\bm{p}+\bm{Q}} ,\sigma}+\nonumber\\
&-& \sum_{\bm{p}} E_{-}(\bm{p})\hat{\gamma}^\dagger_{{\bm{p}+\bm{Q}} ,\sigma^\prime} \hat{\gamma}_{{\bm{p}+\bm{Q}} ,\sigma^\prime} + {\cal C}~,
\end{eqnarray}
where
${\cal C}\equiv \sum_{\bm{p}}  \braket{\hat{d}^\dagger_{\bm{p}+\bm{Q},\sigma} \hat{d}^\dagger_{-\bm{p}+\bm{Q},\sigma^\prime}} \Delta^{\bm{Q},\bm{p}}_{\sigma, \sigma^\prime}/2 $ and 
\begin{eqnarray}
E_{\pm}(\bm{p})&=& \frac{\delta_{\bm p}\pm \epsilon_{\bm{p}}}{2}~,
\end{eqnarray}
the particle-hole asymmetry parameter $\delta_{\bm p}$ being defined by:
\begin{equation}
\delta_{\bm p} \equiv  \frac{\xi_{\bm{p}+\bm{Q}}-\xi_{-\bm{p}+\bm{Q}}}{2}~.
\end{equation}
Note that $\delta_{-\bm p}=-\delta_{\bm p}$, which, in turn, leads to $E_{-}(\bm{p})={-}E_{+}(-\bm{p})$.

The Bogoliubov quasiparticles follow the standard Fermi-Dirac distribution $n_{\rm F}(x)=1/(1+e^x)$, i.e.~$\braket{\hat{\gamma}_{{\bm{p}+\bm{Q}} ,\sigma} \hat{\gamma}_{{\bm{p}+\bm{Q}} ,\sigma} }_T=\braket{\hat{\gamma}_{{\bm{p}+\bm{Q}} ,\sigma^\prime} \hat{\gamma}_{{\bm{p}+\bm{Q}} ,\sigma^\prime} }_T=0$, 
 $\braket{ \hat{\gamma}^\dagger_{{\bm{p}+\bm{Q}} ,\sigma} \hat{\gamma}_{{\bm{p}+\bm{Q}},\sigma} }_T=n_{\rm F}[\beta E_+(\bm{p})]$, 
 and $\braket{ \hat{\gamma}^\dagger_{{\bm{p}+\bm{Q}} ,\sigma^\prime} \hat{\gamma}_{{\bm{p}+\bm{Q}},\sigma^\prime} }_T=n_{\rm F}[\beta E_-(\bm{p})]$,  where $\beta=1/(k_{\rm B} T)$, $k_{\rm B}$ being the Boltzmann constant. These relations imply:
\begin{align}
&\braket{ \hat{d}_{\bm{p}+\bm{Q},\sigma^\prime} \hat{d}_{\bm{p}+\bm{Q},\sigma}}_{T}=-v^*_{\bm{p}}u_{\bm{p}}  \Big\{1-n_{\rm F}[\beta E_+(\bm{p})]\nonumber\\
&-n_{\rm F}[\beta E_+(-\bm{p})]\Big\}~.
\end{align}
Noticing that $1/2-n_{\rm F}[x]=\tanh(x/2)/2$ we have,
\begin{align}
&\braket{ \hat{d}_{\bm{p}+\bm{Q},\sigma^\prime} \hat{d}_{\bm{p}+\bm{Q},\sigma}}_{T}=-\frac{v^*_{\bm{p}}u_{\bm{p}}}{2}  \Big\{ \tanh(\beta E_+(\bm{p})/2)+\nonumber\\
&+\tanh(\beta E_+(-\bm{p})/2)\Big\}~.
\end{align}
Finally, recalling, on the one hand, the definition of the order parameter in Eq.~\eqref{eq:AmpereanGap}, and, on the other hand, the definitions of $u_{\bm p}$ and $v_{\bm p}$, we get a self-consistency condition known as gap equation:
\begin{align}\label{gapeq-II-Bis}
\Delta^{\bm{Q},\bm{p}}_{\sigma, \sigma^\prime}&= -\frac{1}{N_{\rm cell}}  \sum_{\bm{p}^\prime} V_{\bm{p},\bm{p}^\prime,\bm{Q}} \frac{ \Delta^{\bm{Q},\bm{p}^\prime}_{\sigma, \sigma^\prime} }{4{ \epsilon}_{\bm p^\prime}}\nonumber\\ & \times  \Big\{ \tanh(\beta E_+(\bm{p}^\prime)/2)+\tanh(\beta E_+(-\bm{p}^\prime)/2)\Big\}~.
\end{align}
\subsection{Linearization of the gap equation}

\begin{figure}[t]
\centering
\vspace{1.em}
\begin{overpic}[width=0.90\columnwidth]{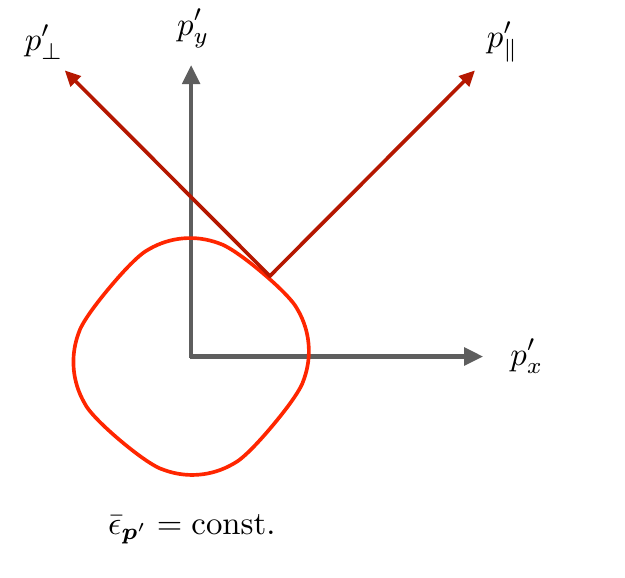}\put(2,95){\normalsize (a)}\end{overpic}
\begin{overpic}[width=0.95\columnwidth]{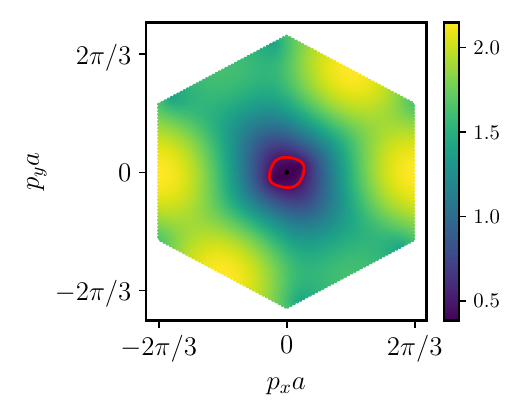}\put(2,80){\normalsize (b)}\end{overpic} 
\caption{(Color online) Panel (a) shows the change of variables $(p_x^\prime, p_y^\prime) \to (p^\prime_\perp,  p^\prime_\parallel)$, where $p^\prime_\perp$ moves on a line with fixed energy $\bar{\epsilon}_{\bm{p}^\prime}={\rm constant}$, while $p^\prime_\parallel$, being parallel to $\nabla_{{\bm p}^\prime}\bar{\epsilon}_{\bm{p}^\prime}$, moves in the direction of maximum variation of $\bar{\epsilon}_{\bm{p}^\prime}$.   Panel (b) shows the symmetrized dispersion ${\bar\epsilon}_{\bm{p}}$ defined in Eq.~(\ref{eq:Sdispersion}) (in units of the hopping energy $t$), the red line denotes the FS, ${\bar\epsilon}_{\bm{p}}=0$, with $\mu=0.5t$.}
\label{fig:FS}
\end{figure}

Our aim is to determine the inverse critical temperature $\beta_{\rm c}$ at which the order parameter vanishes. To this end, we linearize the gap equation (\ref{gapeq-II-Bis}) in the proximity of $\beta_{\rm c}$, by replacing the quantities ${ E}_+(\pm\bm{p})$ and $\epsilon_{\bm p}$ in the right-hand side of Eq.~(\ref{gapeq-II-Bis}) with the corresponding expressions calculated for $\Delta^{\bm{Q},\bm{p}}_{\sigma, \sigma^\prime}=0$.  For $\Delta^{\bm{Q},\bm{p}}_{\sigma, \sigma^\prime}=0$, we have: $h_1(\bm{p})=h_2(\bm{p})=0$, $\epsilon_{\bm p}= \bar{\epsilon}_{\bm p}$, and $E_{+}(\bm{p}) = \xi_{\bm{p}+\bm{Q}}$.  Eq.~(\ref{gapeq-II-Bis}) then becomes:
\begin{align}\label{gapeq-II-Tris}
\Delta^{\bm{Q},\bm{p}}_{\sigma, \sigma^\prime}&= -\frac{1}{N_{\rm cell}}  \sum_{\bm{p}^\prime} V_{\bm{p},\bm{p}^\prime,\bm{Q}} \frac{ \Delta^{\bm{Q},\bm{p}^\prime}_{\sigma, \sigma^\prime} }{4\bar{ \epsilon}_{\bm p^\prime}}\nonumber\\ & \times \Big\{\tanh[\beta_{\rm c} (\bar{\epsilon}_{\bm{p}^\prime}+\delta_{\bm{p}^\prime})/2]+\tanh[\beta_{\rm c} (\bar{\epsilon}_{\bm{p}^\prime}-\delta_{\bm{p}^\prime})/2]\Big\}~.
\end{align}
We can rewrite the second line of the previous equation as following:
\begin{align}\label{eq:hyperbolic_identity}
 &\frac{\tanh[\beta_{\rm c} (\bar{\epsilon}_{\bm{p}^\prime}+\delta_{\bm{p}^\prime})/2]+\tanh[\beta_{\rm c} (\bar{\epsilon}_{\bm{p}^\prime}-\delta_{\bm{p}^\prime})/2]}{2\bar{\epsilon}_{\bm{p}^\prime}}=\nonumber\\
 & = \frac{\sinh{(\beta_{\rm c}\bar{\epsilon}_{\bm{p}^\prime})}}{\bar{\epsilon}_{\bm{p}^\prime}}\frac{1}{{\cosh{(\beta_{\rm c}\bar{\epsilon}_{\bm{p}^\prime})}}+\cosh{(\beta_{\rm c}{\delta}_{\bm{p}^\prime})}}~.
\end{align}
We then take the continuum limit through the customary replacement
\begin{equation}\label{eq:continuum_limit_lattice}
\frac{1}{N_{\rm cell}}\sum_{\bm{p} \in {\rm BZ}}\to  S_{\rm c} \int_{{\rm BZ}} \frac{d^2  \bm{p}}{(2\pi)^2}~, 
\end{equation}
where $S_{\rm c}= 3\sqrt{3} a^2 /2$ is the unit cell area. 
Replacing Eqs.~(\ref{eq:hyperbolic_identity}) and~(\ref{eq:continuum_limit_lattice}) inside Eq.~(\ref{gapeq-II-Tris}) we find:
\begin{align}\label{eq:strange_sign_second_term}
\Delta_{\sigma,\sigma^\prime}^{\bm{Q},\bm{p}}&=- \frac{S_{\rm c}}{2}  \int_{ {\rm BZ}} \frac{ d^2 \bm{p}^\prime}{(2\pi)^2} V_{\bm{p},\bm{p}^\prime,\bm{Q}} { \Delta^{\bm{Q},\bm{p}^\prime}_{\sigma, \sigma^\prime} }\times \nonumber\\ &\times\Bigg\{ \frac{\sinh{(\beta_{\rm c}\bar{\epsilon}_{\bm{p}^\prime})}}{\bar{\epsilon}_{\bm{p}^\prime}}\frac{1}{{\cosh{(\beta_{\rm c}\bar{\epsilon}_{\bm{p}^\prime})}}+\cosh{(\beta_{\rm c}{\delta}_{\bm{p}^\prime})}}\Bigg\}~.
\end{align}
Superconductivity is a Fermi surface (FS) effect. Below, we will therefore need to restrict ourselves to a  neighborhood of the FS, which is defined by $\bar{\epsilon}_{\bm{p}^\prime}=0$. We therefore consider the iso-energy contour lines defined by the condition $\bar{\epsilon}_{\bm{p}^\prime}= {\rm const}$ and perform a change of variables---see Fig.~\ref{fig:FS}(a)---$(p_x^\prime, p_y^\prime) \to (p^\prime_\parallel, p^\prime_\perp)$. Here, $p^\prime_\parallel$ refers to the component of ${\bm p}^\prime$ which is parallel to $\nabla_{{\bm p}^\prime} \bar{\epsilon}_{\bm{p}^\prime}$. On the other hand, $p^\prime_\perp$ refers to the component of ${\bm p}^\prime$ which is orthogonal to $\nabla_{{\bm p}^\prime} \bar{\epsilon}_{\bm{p}^\prime}$. By definition, we therefore have that $\bar{\epsilon}_{\bm{p}^\prime}$ depends only on $p^\prime_\parallel$, i.e.~$\bar{\epsilon}_{\bm{p}^\prime}=\bar{\epsilon}_{p^\prime_\parallel}$. Performing this change of variables in Eq.~(\ref{eq:strange_sign_second_term}) we find:
\begin{align}\label{eq:geq}
&\Delta_{\sigma,\sigma^\prime}^{\bm{Q},{p}_\perp,{p}_\parallel}=- \frac{S_{\rm c}}{2} \int_{{\rm BZ}} \frac{d {p}^\prime_\perp dp^\prime_\parallel}{(2\pi)^2} V_{{{p}_\perp,{p}_\parallel,{p}^\prime_\perp,{p}^\prime_\parallel,{\bm Q}}} {\Delta_{\sigma,\sigma^\prime}^{\bm{Q},{p}^\prime_\perp,{p}^\prime_\parallel} }\times \nonumber\\ &\times\Bigg \{ \frac{\sinh{(\beta_{\rm c}\bar{\epsilon}_{p_\parallel^\prime})}}{\bar{\epsilon}_{p^\prime_\parallel}}\frac{1}{{\cosh{(\beta_{\rm c}\bar{\epsilon}_{p_\parallel^\prime})}}+\cosh{(\beta_{\rm c}{\delta}_{p^\prime_\perp,p^\prime_\parallel})}} \Bigg \}~.
\end{align}

Since the superconducting instability refers to a small energy region centered on the FS and the potential $V_{{{p}_\perp,{p}_\parallel,{p}^\prime_\perp,{p}^\prime_\parallel,{\bm Q}}}$ is peaked at $p_\parallel=p_\parallel^\prime$ and $p_\perp=p_\perp^\prime$ over a length scale on the order of $1/L_z$, the integration over $p_\parallel^\prime$ and $p_\perp^\prime$ in Eq.~\eqref{eq:geq} can be approximated by choosing  $p_\parallel=p_\parallel^{\rm F}$ lying on the FS (see Fig.~\ref{fig:FS}(a)) and $p_\perp=p_\perp^\prime$. Hence, the integral over $p_\parallel^\prime,p_\perp^\prime$ can be estimated as following:
 \begin{eqnarray}\label{eq:approxInt}
 \int dp_{\parallel}^\prime dp_{\perp}^\prime V_{{{p}_\perp,{p}_\parallel^{\rm F},{p}^\prime_\perp,{p}^\prime_\parallel,{\bm Q}}}\simeq \frac{1}{L^2_z}V_{{{p}_\perp,{p}_\parallel^{\rm F},{p}_\perp,{p}^{\rm F}_\parallel,{\bm Q}}}~.
\end{eqnarray}
Moreover, since the EEEI potential is approximated as a delta function, the last line of Eq.~\eqref{eq:geq} can be evaluated on the FS, where ${\sinh{(\beta_{\rm c}\bar{\epsilon}_{p_\parallel})}}/{\bar{\epsilon}_{p_\parallel}}\simeq \beta_{\rm c} $ and ${\cosh{(\beta_{\rm c}\bar{\epsilon}_{p_\parallel})}}\simeq 1$.
Within these approximations, Eq.~\eqref{eq:geq} reduces to
 \begin{eqnarray}\label{eq:geq1}
   \Delta_{\sigma,\sigma^\prime}^{\bm{Q},{p}_\perp,{p}^{\rm F}_\parallel}&=&- \frac{S_c   }{2L^2_z(2\pi)^2} V_{{{p}_\perp,{p}^{\rm F}_\parallel,{p}_\perp,{p}^{\rm F}_\parallel,{\bm Q}}} {\Delta_{\sigma,\sigma^\prime}^{\bm{Q},{p}_\perp,{p}_\parallel} } \nonumber\\ &\times& \Bigg\{ \frac{\beta_{\rm c}}{1+\cosh{(\beta_{\rm c}{\delta}_{p_\perp,p^{\rm F}_\parallel})}}\Bigg\}~.
\end{eqnarray}

Since this equation is diagonal in ${p}_\perp$, an expression for the critical temperature reads,
\begin{eqnarray}\label{eq:Tc}
k_{\rm B}T_{\rm c}={\rm max}_{{p}_\perp}\nu_{{p}_\perp,{p}^{\rm F}_\parallel}~,
\end{eqnarray}
where
 \begin{eqnarray}\label{eq:geq2}
\nu_{{p}_\perp,{p}^{\rm F}_\parallel}&=&- \frac{S_c   }{2L^2_z(2\pi)^2} V_{{{p}_\perp,{p}^{\rm F}_\parallel,{p}_\perp,{p}^{\rm F}_\parallel,{\bm Q}}}  \nonumber\\ &\times& \Bigg\{ \frac{1}{1+\cosh{(\nu_{{p}_\perp,{p}^{\rm F}_\parallel}^{-1}{\delta}_{p_\perp,p^{\rm F}_\parallel})}}\Bigg\}~.
\end{eqnarray}
We can further express the gap equation in terms of the EEEI $\tilde{V}_{{{p}_\perp,{p}^{\rm F}_\parallel,{p}_\perp,{p}^{\rm F}_\parallel,{\bm Q}}}$ defined in Eq.~\eqref{eq:potential}:
 \begin{eqnarray}\label{eq:geq3}
\nu_{{p}_\perp,{p}^{\rm F}_\parallel}&=&- \frac{3 \sqrt{3}   }{4(2\pi)^2}\Big(\frac{a}{L_z}\Big)^2 \frac{g^2_{\rm FP}}{\hbar \omega_{\rm D}} \tilde{V}_{{{p}_\perp,{p}^{\rm F}_\parallel,{p}_\perp,{p}^{\rm F}_\parallel,{\bm Q}}}  \nonumber\\ &\times& \Bigg\{ \frac{1}{1+\cosh{(\nu_{{p}_\perp,{p}^{\rm F}_\parallel}^{-1}{\delta}_{p_\perp,p^{\rm F}_\parallel})}}\Bigg\}~.
\end{eqnarray}

\subsection{Results and estimation of the critical temperature}
\begin{figure}[t]
\centering
\includegraphics[width=0.9\columnwidth]{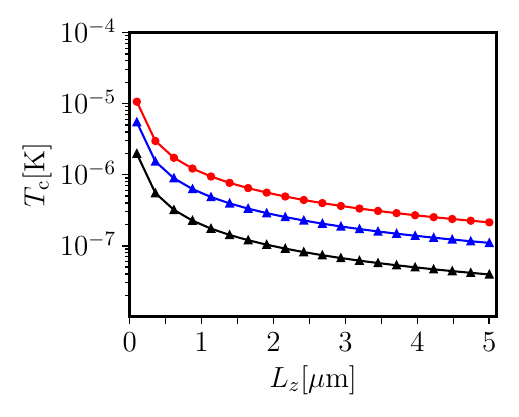}
\caption{(Color online) Critical temperature $T_{\rm c}$ as a function of the distance $L_z$ between the cavity mirrors. Different sets of data refer to different values of the chemical potential $\mu$. Namely, red, black, and blue colors refer to $\mu=0.4 t$, $\mu=0.45 t$, and $\mu=0.5 t$, respectively.
\label{fig:nu_vs_Lz}}
\end{figure}

Eq.~\eqref{eq:geq3} has been solved numerically and a summary of our main results is reported in Fig.~\ref{fig:nu_vs_Lz}. Obtained critical temperatures are on the order of $T_{\rm c}\sim 10^{-6}~{\rm K}$ for a distance between the two mirrors of $L_z=1~{\rm \mu m}$. 

The numerical solution of Eq.~\eqref{eq:geq3} shows that the maximal eigenvalue $\nu_{{p}_\perp,{p}^{\rm F}_\parallel}$ is reached for ${\tilde{p}_\perp,\tilde{p}^{\rm F}_\parallel}$ such that ${\delta}_{\tilde{p}_\perp,\tilde{p}^{\rm F}_\parallel}=0$.
It is therefore possible to estimate the critical temperature by approximating the term $\cosh{(\nu_{\tilde{p}_\perp,\tilde{p}^{\rm F}_\parallel}^{-1}{\delta}_{\tilde{p}_\perp,\tilde{p}^{\rm F}_\parallel})}\simeq1$. The equation for the critical temperature can be expressed as,
\begin{equation}
\label{eq:nu1}
k_{\rm B}T_{\rm c} = - \frac{3\sqrt{3}}{8(2\pi)^2}\Big(\frac{a}{L_z}\Big)^2 \frac{g^2_{\rm FP}}{\hbar \omega_{\rm D}} \tilde{V}_{\tilde{p}_\perp,\tilde{p}^{\rm F}_{\parallel},\tilde{p}_\perp,\tilde{p}^{\rm F}_{\parallel}\bm{Q}}~.
\end{equation}
The algebraic (rather than exponential) relation between the critical temperature and the coupling constant $g_{\rm FP}$ of the theory is peculiar of light-mediated superconductivity~\cite{Sentef_ScienceAdvances_2018,Piazza21} or, more in general, of superconductivity mediated by bosons carrying a small momentum, when forward scattering is dominant~\cite{Rademaker_NJP_2016}.  In conventional BCS superconductors, electron-phonon interactions are essentially local, allowing electrons to exchange arbitrary values $\hbar \bm{q}$ of momentum. This leads to the famous exponential BCS relationship for the critical temperature. 
Conversely, in systems characterized by non-local interactions that are peaked at ${\bm q} = {\bm 0}$---such as those mediated by photons---the critical temperature shows a distinctive power-law dependence. This behavior can also be found in some high-temperature superconductors~\cite{Aperis},  in stark contrast to conventional BCS-type phonon-mediated superconductors.

We can estimate $\tilde{V}_{\tilde{p}_\perp,\tilde{p}^{\rm F}_{\parallel},\tilde{p}_\perp,\tilde{p}^{\rm F}_{\parallel}\bm{Q}}\simeq -10$  (see Fig.~\ref{fig:Vp_pp_Q}). The other factors are $(a/L_z)^2\simeq 2\times 10^{-8} L^{-2}_z[{\rm \mu m}] $ and $g^2_{\rm FP}/(\hbar \omega_{\rm D})\simeq 3.2 \times 10^{-2} L_z[{\rm \mu m}] t$. Finally, we obtain,
\begin{equation}\label{eq:nu2}
T_{\rm c} =\frac{3.4 \times 10^{-6}~{\rm K}}{L_z[{\rm \mu m}]}~,
\end{equation}
which, in terms of orders of magnitude, is in good agreement with our numerical findings shown in Fig.~\ref{fig:nu_vs_Lz}. The analytical estimate reported in Eq.~(\ref{eq:nu1}) allows us to conclude that the smallness of the critical temperature in our setup is mainly due to the dimensionless parameter $(a/L_z)^2\ll 1$. We envision two possible ways to greatly increase this parameter: i) On the one hand, one could increase the lattice spacing $a$ by resorting e.g.~to artificial graphene realized in semiconducting or cold-atom platforms~\cite{Polini13} or 2D moir{\'e} super-lattices~\cite{Bistritzer2011,Cao2018a,Cao2018b,Andrei2021,Kennes2021}; ii) on the other hand, one could greatly decrease the field extension, which in the Fabry-P\'erot setup is on the order of $L_z$, by means of nano-plasmonic cavities (see, however, the following discussion).
\section{Discussion}
\label{sec:results}

Given the disappointing results reported in Fig.~\ref{fig:nu_vs_Lz}, it is interesting to think about the possibility of boosting the critical temperature $T_{\rm c}$ for light-induced Amperean superconductivity to higher values. We stress that, so far, we have established the absence of an Amperean superconducting instability in graphene embedded in a Fabry-P\'erot cavity. We remind the reader that such cavities harbor {\it transverse} electromagnetic modes.

For the case of a parabolic-band 2D electron gas roaming in a GaAs quantum well, critical temperatures in the range of $1$-$20~{\rm K}$ have been obtained~\cite{Jaksch_prl_2019} by placing the system into a nanoplasmonic cavity. It is well known~\cite{Reserbat_ACS_2021,Basov_Nanophotonics_2021} that such cavities operate in the sub-wavelength regime, where the plasmon-induced confinement length scale $\lambda_{\rm p}$ is much smaller than the lengthscale $\lambda_{\rm d}$ imposed by the free-space diffraction limit, i.e.~$\lambda_{\rm d}\equiv \lambda_0/2$. 

With the aim of deepening the comparison between our disappointing results and the spectacular predictions for GaAs quantum wells~\cite{Jaksch_prl_2019}, we therefore follow the authors of Ref.~\onlinecite{Jaksch_prl_2019} and introduce the cavity {\it compression factor} as the ratio between the plasmonic mode volume $\lambda^3_{\rm p}$ and the free-space mode volume $\lambda^3_{\rm d}$:
\begin{equation}
A \equiv \left(\frac{\lambda_{\rm p}}{\lambda_{\rm d}}\right)^3~.
\end{equation}

The starting point of the theory developed by the authors of Ref.~\onlinecite{Jaksch_prl_2019} is the same as ours, i.e.~it is a theory of Amperean superconductivity induced in a 2D electron system by the vacuum of a  Fabry-P{\'e}rot cavity. The authors, however, go beyond by generalizing such theory to a nanoplasmonic cavity. Heuristically, they argue that the light-matter coupling defined in Eq.~(\ref{eq:coupling_constant_of_the_whole_theory}) scales like $1/\sqrt{V}$, where $V$ is the cavity volume. Hence, a drastic reduction of the volume occupied by light should result in a dramatic enhancement of the critical temperature $T_{\rm c}$.  The key idea proposed by the authors of Ref.~\onlinecite{Jaksch_prl_2019} is therefore to: i) replace the cavity volume $V=SL_z$ with the mode volume $V_{\rm mode} \equiv \lambda_{\rm p}^3=A\lambda_{\rm d}^3$; ii) estimate the volume imposed by the diffraction limit, i.e.~$\lambda^3_{\rm d}$,  with the cavity volume $V$; iii) leave the Debye energy $\hbar\omega_{\rm D}$ unaffected by the rescaling $V \to V_{\rm mode}$.  Following this logic, we therefore see that the compression factor $A$ linearly relates the  effective mode volume with the bare Fabry-P\'erot cavity volume, i.e.~$V_{\rm mode}=A V$. Replacing this result in Eq.~(\ref{eq:coupling_constant_of_the_whole_theory}) we finally find the following light-matter coupling constant for a nanoplasmonic cavity:
\begin{equation}\label{eq:gnano_vs_gFP}
g_{\rm nano} = \frac{g_{\rm FP}}{\sqrt{A}}~.
\end{equation}
Since $A\ll 1$, we have that $g_{\rm nano} \gg g_{\rm FP}$: nanoplasmonic cavities enable the reach of ultra-strong light-matter interactions~\cite{Reserbat_ACS_2021}. Employing values of $A$ on the order of $10^{-5}$, the authors of Ref.~\onlinecite{Jaksch_prl_2019} find a critical temperature for Amperean superconductivity on the order of $1$-$20~{\rm K}$.

In terms of the analysis performed in this Article, the proposed replacement~\cite{Jaksch_prl_2019} $g_{\rm FP} \to g_{\rm nano}$ implies that the EEEI potential $V_{\bm p, \bm p', \bm Q}$ in Eq.~\eqref{eq:Veff_f2} needs to be multiplied by $1/A$. Hence, we find a critical temperature $T_{\rm c}(A)$ at $A\neq 1$ that reads as following:
\begin{eqnarray}
T_{\rm c}(A)=  \frac{T_{\rm c}(A=1)}{A}~,
\end{eqnarray}
where $T_{\rm c}(A=1) \simeq 10^{-6}~{\rm K}$ is the critical temperature evaluated with the diffraction-limited light-matter coupling constant $g_{\rm FP}$ defined in Eq.~(\ref{eq:coupling_constant_of_the_whole_theory_FB}). Recent experimental progress~\cite{Lundeberg_Science_2017,David_Science_2018,Epstein_Science_2020} in the fabrication of nanoplasmonic cavities using graphene plasmons combined with an engineered metal-dielectric environment has enabled the reach of ultra-strong compression factors, with record-high values~\cite{Epstein_Science_2020} on the order of $A\simeq 5 \times 10^{-10}$. Using this value of $A$, one would get a renormalized critical temperature for Amperean superconductivity $T_{\rm c}(A)\simeq 2 \times 10^3~{\rm K}$, which is clearly too optimistic.

The key point we wish to emphasize is that, unfortunately, the formal replacement $g_{\rm FP} \to g_{\rm nano}$  is not justified in the context of developing a theory of Amperean superconductivity. Indeed, the latter stems, microscopically, from the coupling between the electronic current (\ref{eq:current}) and the cavity vector potential~\eqref{vectorpot_2}, which is achieved by placing the 2D electron system inside a planar Fabry-P{\'e}rot cavity. As emphasized in this Article, such cavities host genuine 3D transverse electromagnetic modes. Conversely, in a sub-wavelength nanoplasmonic cavity satisfying the condition $\lambda_{\rm p}\ll \lambda_{\rm d}$, the field is purely {\it quasi-static}~\cite{Torre_ArXiv_2021}, i.e.~it is a field that can be described by an electric scalar potential $\phi_{\rm cav}(\bm{r},t)$. Such scalar potential couples to matter through the density (rather than the current) operator~\cite{Giuliani_and_Vignale}. Tracing out the nanoplasmonic-cavity degrees of freedom would therefore result in an effective {\it density-density} interaction, rather than an Amperean current-current one, as we will show momentarily in Sect.~\ref{sec:TransversVsLong}. Employing ``mode volume confinement'' arguments in the theory of Amperean superconductivity is therefore fundamentally inconsistent and one is forced to take $A=1$ in Eq.~(\ref{eq:gnano_vs_gFP}).

\subsection{Dyadic Green's function and EEEIs}
\label{sec:TransversVsLong}

In this Section, we use a Green's function approach to demonstrate that, in sub-wavelength cavities, the EEEI (see Eq.~\eqref{eq:profound_definition_EEEI}) is  dominated by a density-density contribution. 

We begin by reminding the reader about Maxwell's equations~\cite{Jackson}:
\begin{align}
\partial_\alpha D_\alpha(\bm r, \omega)& = 4\pi \rho^{\rm ext} (\bm r, \omega)~,\label{eq:divD}\\
\partial_\alpha B_\alpha (\bm r, \omega)& = 0~,\label{eq:divB}\\
\varepsilon_{\alpha \beta \gamma} \partial_\beta E_\gamma (\bm r, \omega) &= \frac{i\omega}{c}  B_\alpha (\bm r, \omega)~,\label{eq:rotE}\\
\varepsilon_{\alpha \beta  \gamma} \partial_\beta H_\gamma (\bm r, \omega)& = \frac{4\pi}{c} { J}_\alpha^{\rm ext}(\bm r, \omega)-\frac{i\omega}{c} { D}_\alpha (\bm r, \omega)~.\label{eq:rotH}
\end{align}
Here, $\rho^{\rm ext} (\bm r, \omega)$ and ${ J}_\alpha^{\rm ext}(\bm r, \omega)$ are the external charge and current densities, evaluated at position $\bm{r}$ and frequency $\omega$ and $\alpha=1,2,3$ is a Cartesian index that specifies the component of vectorial quantities. Throughout this Section, we adopt the Einstein summation convention. Specifically, for two vectors $ v_\alpha $ and $w_\alpha$, repeated indices imply summation, i.e.~$v_\alpha w_\alpha\equiv \sum_\alpha v_\alpha w_\alpha$.

We now assume that the electric displacement $D_\alpha(\bm r, \omega)$ and magnetic induction $ B_\alpha (\bm r, \omega)$  are related to the electric and magnetic fields $E_\alpha(\bm r, \omega), H_\alpha(\bm r, \omega)$ by the following linear and local relationships:
\begin{align}
{D}_\alpha(\bm r, \omega) & =   \epsilon_{\alpha \beta }(\bm r, \omega) {E}_{\beta } (\bm r, \omega)~,\label{eq:epsilon}\\
{B}_\alpha(\bm r, \omega)& = \mu_{\alpha \beta }(\bm r, \omega) {H}_{\beta }(\bm r, \omega)~,\label{eq:mu}
\end{align}
$ \epsilon_{\alpha \beta }(\bm r, \omega)$ and $\mu_{\alpha \beta }(\bm r, \omega)$ being position- and frequency-dependent dielectric and permeability tensors.

Taking the divergence of Eq.~\eqref{eq:rotH} and using Eq.~\eqref{eq:divD} yields the continuity equation:
\begin{equation}
\partial_\alpha J_\alpha^{\rm ext}(\bm r, \omega) -i\omega \rho^{\rm ext}(\bm r, \omega)=0~.\label{eq:continuity}
\end{equation}
Inverting Eq.~\eqref{eq:rotE} to get ${H}_{\beta}(\bm r, \omega)$ and substituting the result into Eq.~\eqref{eq:rotE} yields the inhomogeneous Helmoltz equation for the electric field
\begin{align}
&\varepsilon_{\alpha \beta \gamma}\partial_\beta [\mu_{\gamma \delta}^{-1} (\bm r, \omega)\varepsilon_{\delta \zeta \eta }\partial_\zeta E_\eta(\bm r, \omega)]-\frac{\omega^2}{c^2} \epsilon_{\alpha \beta}(\bm r, \omega){ E}_\beta (\bm r, \omega) \nonumber\\& = \frac{4\pi i \omega}{c^2} { J}_\alpha^{\rm ext}(\bm r, \omega). \label{eq:HelE}
\end{align}
Multiplying Eq.~\eqref{eq:rotH} by the inverse of the dielectric permittivity matrix, taking the curl, and making use of Eq.~\eqref{eq:rotE} we find the corresponding equation for the magnetic field:
\begin{align}
&\varepsilon_{\alpha \beta \gamma}\partial_\beta[ \epsilon_{\gamma \beta}^{-1} (\bm r, \omega)\varepsilon_{\delta \zeta \eta }\partial_\zeta H_\eta(\bm r, \omega)]-\frac{\omega^2}{c^2} \mu_{\alpha \beta}(\bm r, \omega){ H}_\beta (\bm r, \omega) \nonumber\\& = \frac{4\pi}{c} \varepsilon_{\alpha \beta \gamma}\partial_\beta  \epsilon_{\gamma \delta}^{-1}(\bm r, \omega) {J}_\delta^{\rm ext}(\bm r, \omega)~.\label{eq:HelH}
\end{align}
Note that taking the divergence of Eq.~\eqref{eq:HelE} and using the continuity equation Eq.~\eqref{eq:continuity} yields Eq.~\eqref{eq:divD}, while taking the divergence of Eq.~\eqref{eq:HelH} gives Eq.~\eqref{eq:divB}.

We introduce the following compact notation for a differential operator ${\cal D}_{\alpha \beta }(T)$ depending on a tensor $T(\bm r, \omega)$:
\begin{equation}
{\cal D}_{\alpha \beta }(T) \equiv
\varepsilon_{\alpha \gamma\delta}\varepsilon_{\zeta \eta \beta}\partial_\gamma[T_{\delta \zeta}(\bm r, \omega)\partial_\eta \dots]~.\label{eq:doperator}
\end{equation}
In the simple case $T_{\alpha \beta}(\bm r, \omega) = \delta_{\alpha \beta}$ this reduces to 
\begin{equation}
{\cal D}_{\alpha \beta}(\bm 1) =\partial_\alpha\partial_\beta-\delta_{\alpha \beta}\partial_\gamma\partial_\gamma~.\label{eq:rotrot}
\end{equation}
With this notation, the Helmholtz equations can be written as
\begin{align}
\label{eq:HelmholtzE}
&{\cal D}_{\alpha \beta}(\mu^{-1})E_\beta(\bm r, \omega)-\frac{\omega^2}{c^2} \epsilon_{\alpha \beta}(\bm r, \omega) E_\beta (\bm r, \omega) =\nonumber\\& = \frac{4\pi i \omega}{c^2}  J_{\alpha}^{\rm ext}(\bm r, \omega)~,\\
&{\cal D}_{\alpha\beta}(\epsilon^{-1})H_\beta (\bm r, \omega)-\frac{\omega^2}{c^2}\mu_{\alpha \beta}(\bm r, \omega) H_\beta (\bm r, \omega)  =\nonumber\\& = \frac{4\pi}{c} \varepsilon_{\alpha \beta \gamma} \partial_\beta\epsilon^{-1}_{\gamma,\delta}(\bm r, \omega)J_\delta^{{\rm ext}}(\bm r, \omega)~.
\end{align}
The Green's function $G_{\alpha \beta}(\bm r,\bm r', \omega)$ associated with the Helmholtz equation for the electric field,  Eq.~\eqref{eq:HelmholtzE}, is known as ``dyadic Green's function'' and obeys the following equation: 
\begin{align}\label{eqn:dyadic1}
&{\cal D}_{\beta \gamma}'({}^T\mu^{-1}) G_{\alpha \gamma}(\bm r,\bm r', \omega)-\frac{\omega^2}{c^2}{}^T\epsilon_{\beta \gamma}(\bm r', \omega)G_{\alpha\gamma}(\bm r, \bm r',\omega)  =\nonumber\\& =4\pi \delta_{\alpha \beta} \delta(\bm r-\bm r')~.
\end{align}
The dyadic Green's function directly connects the total electric field $E_\alpha(\bm r,\omega)$ at point ${\bm r}$ in space with the external source $J_{\beta}^{\rm ext}(\bm r', \omega)$ evaluated at point ${\bm r}^\prime$,
\begin{align}\label{eqn:electric_green_source}
E_\alpha(\bm r,\omega) &= {E_\alpha^{(0)}(\bm r,\omega)} \nonumber\\
&+\frac{i\omega}{c^2}\int d\bm r' G_{\alpha \beta}(\bm r,\bm r',\omega) J_{\beta}^{\rm ext}(\bm r', \omega)~,
\end{align}
${E_\alpha^{(0)}(\bm r,\omega)}$ being a solution of the homogeneous equation associated to Eq.~\eqref{eq:HelmholtzE}.

As usual, we now introduce scalar and vector potentials, $\phi(\bm r, \omega), A_\alpha(\bm r, \omega)$:
\begin{align}
B_\alpha (\bm r,\omega)&=\varepsilon_{\alpha \beta \gamma}\partial_\beta  A_\gamma(\bm r, \omega)~,\\
 E_\alpha (\bm r,\omega)&= -\partial_\alpha \phi(\bm r, \omega)+\frac{i\omega}{c} A_\alpha(\bm r, \omega)~.
 \label{eq:electricfield_interms_of_potentials}
\end{align}
These expressions ensure that  $E_\alpha (\bm r,\omega)$ and $B_\alpha (\bm r,\omega)$ satisfy automatically the two homogeneous equations~\eqref{eq:divB} and~\eqref{eq:rotE}.

Using these definitions into Eq.~\eqref{eq:divD} and Eq.~\eqref{eq:rotH} yields
\begin{align}
\label{eq:poisson0}
&\partial_\alpha \left( \epsilon_{\alpha \beta}(\bm r, \omega)\left[ -\partial_\beta \phi(\bm r, \omega)+\frac{i\omega}{c} A_\beta(\bm r, \omega)\right]\right) \nonumber\\
&= 4\pi \rho^{\rm ext}(\bm r, \omega)~,\\
\label{eq:Helmholtz_for_A}
&{\cal D}_{\alpha \beta}(\mu^{-1})  A_\beta(\bm r, \omega)-\frac{\omega^2}{c^2} \epsilon_{\alpha \beta}(\bm r, \omega) A_\beta(\bm r, \omega) =\nonumber\\& = \frac{4\pi}{c}{J}^{\rm ext}_\alpha (\bm r, \omega)+\frac{i\omega}{c} \epsilon_{\alpha \beta}(\bm r, \omega)
\partial_\beta \phi(\bm r, \omega)~.
\end{align}
We use the gauge freedom to separate out the equation for the scalar potential. We work under the following gauge condition,
\begin{equation}
\label{eq:generalizedC}
\partial_\alpha  [\epsilon_{\alpha\beta}(\bm r, \omega)A_\beta(\bm r, \omega)]=0~.
\end{equation}
This reduces to the standard Coulomb gauge in the limit of a scalar and spatially-uniform dielectric permittivity.

With this gauge-fixing condition, Eq.~\eqref{eq:poisson0} becomes the familiar frequency-dependent Poisson equation
\begin{equation}
\label{eq:poisson}
-\partial_\alpha  [\epsilon_{\alpha\beta}(\bm r, \omega) \partial_\beta  \phi(\bm r, \omega)] = 4\pi \rho^{\rm ext} (\bm r, \omega).
\end{equation}

We now introduce the following ``{\it scalar} (s) - {\it vector} (v)'' decomposition:
\begin{equation}
 J^{\rm ext}_\alpha(\bm r, \omega)= J^{\rm( v)}_\alpha(\bm r, \omega)+ J^{\rm ( s)}_\alpha(\bm r, \omega)~,
\end{equation}
where
\begin{equation}
 J_\alpha^{\rm ( v)}(\bm r, \omega) \equiv { J}^{\rm ext}_\alpha(\bm r, \omega)+\frac{i\omega}{4\pi} \epsilon_{\alpha \beta}(\bm r, \omega)\partial_\beta \phi(\bm r, \omega)
\end{equation}
and
\begin{equation}
J^{\rm ( s)}_\alpha(\bm r, \omega) \equiv -\frac{i\omega}{4\pi} \epsilon_{\alpha \beta}(\bm r, \omega)\partial_\beta \phi(\bm r, \omega)~.
\end{equation}
We stress that, while the vector part ${\bm J}^{\rm ( v)}(\bm r, \omega)$ of the current is clearly solenoidal, i.e. its divergence is zero (a consequence of the continuity equation):
\begin{equation}
\partial_\alpha  J_\alpha^{\rm ( v)}(\bm r, \omega)=0~,
\end{equation}
the scalar counterpart ${\bm J}^{\rm ( s)}(\bm r, \omega)$ is {\it not} irrotational:
\begin{equation}
\varepsilon_{\alpha \beta \gamma}\partial_\beta J_\gamma^{\rm ( s)}(\bm r, \omega) \neq 0~.
\end{equation}
Physically this means that ${\bm J}^{\rm ( v)}(\bm r, \omega)$ is indeed a transverse vector but it does not identify all the transverse current ${\bm J}^{\rm T}(\bm r, \omega)$.

Clearly, when the material reduces to the vacuum, i.e.~$\mu_{\alpha \beta}(\bm{r},\omega)=\epsilon_{\alpha \beta }({\bm r},\omega) = \delta_{\alpha \beta}$, the scalar-vector decomposition proposed above reduces to the standard longitudinal and transverse decomposition enabled by the Coulomb gauge, i.e.~the scalar current reduces to the longitudinal current, ${\bm J}^{\rm ( s)}(\bm r, \omega)\to {\bm J}^{\rm L}(\bm r, \omega)$ such that $\varepsilon_{\alpha \beta \gamma} \partial_j   J^{\rm L}_\gamma(\bm r, \omega) = 0$, while the vector current reduces to the longitudinal one, ${\bm J}^{\rm ( v)} (\bm r, \omega)\to {\bm J}^{\rm T}(\bm r, \omega)$ such that $\varepsilon_{\alpha \beta \gamma} \partial_\beta  J^{\rm T}_\gamma(\bm r, \omega)=0$. In other words, we obtain the standard longitudinal-transverse decoupling of the Maxwell equations. 

We now return to the general case of a linear and local material. The equation for the vector potential reads as following:
\begin{align}
&{\cal D}_{\alpha \beta }(\mu^{-1})  A_\beta(\bm r, \omega)-\frac{\omega^2}{c^2} \epsilon_{\alpha \beta}(\bm r, \omega) A_\beta(\bm r, \omega) \nonumber\\
&= \frac{4\pi}{c}{ J}^{\rm( v)}_\alpha(\bm r, \omega)~,
\end{align}
where on the right-hand side we have introduced the {\it vector} part of the current. 

From the equations for the scalar and vector potentials, it is possible to show that the corresponding scalar $g^{(\rm s)}(\bm r, \bm r', \omega)$ and vector $g_{\alpha \beta}^{(\rm v)}(\bm r, \bm r',\omega)$ Green's functions obey the following equations:
\begin{equation}\label{eqn:scalar_green}
-\partial_\beta' [{}^{\rm T}\epsilon_{\beta\alpha}(\bm r', \omega)\partial_\alpha' g^{(\rm s)}(\bm r, \bm r', \omega)] = 4\pi\delta(\bm r-\bm r')~,
\end{equation}
\begin{equation}\label{eqn:vector_green}
\begin{split}
&\varepsilon_{\beta \delta \zeta}\varepsilon_{\eta \lambda \gamma}\partial^{\prime}_\delta[{}^{\rm T}\mu_{\zeta \eta}^{-1}(\bm r',\omega) \partial^{\prime}_\lambda g_{\alpha \gamma}^{(\rm v)}(\bm r, \bm r',\omega)]  -\\
&- \frac{\omega^2}{c^2}{}^{\rm T}\epsilon_{\beta \gamma}(\bm r',\omega)g_{\alpha \gamma}^{(\rm v)}(\bm r, \bm r',\omega)=\\
&=
{4\pi}\left[\delta_{\alpha \beta}\delta(\bm r-\bm r') - \frac{1}{4\pi}\epsilon_{\gamma \beta}({\bm r}^\prime,\omega)\partial_\gamma^\prime\partial_\alpha g^{(\rm s)}({\bm r}, {\bm r}^\prime,\omega)\right]~,
\end{split}
\end{equation}
together with the gauge-fixing condition \eqref{eq:generalizedC} on the vector Green's function:
\begin{equation}
\label{eq:generalizedC_forG}
\partial'_\beta[{}^{\rm T}\epsilon_{\beta\gamma}(\bm r',\omega)g_{\alpha \gamma}^{(\rm v)}(\bm r, \bm r',\omega)]=0~.
\end{equation}
Here, ${}^{\rm T}O_{\beta\gamma}(\bm r',\omega)$ denotes the transpose of the tensor $O(\bm r',\omega)$.
Once these equations are solved for $g^{(\rm s)}(\bm r, \bm r', \omega)$ and $g_{\alpha \beta}^{(\rm v)}(\bm r, \bm r',\omega)$, the potentials are found by carrying out the following convolutions with the sources:
\begin{align}\label{eqn:scalar_green_source}
\phi(\bm r,\omega) = { \phi_0(\bm r, \omega)} + \int d \bm r' g^{(\rm s)}(\bm r, \bm r',\omega) \rho_{\rm ext}(\bm r', \omega)
\end{align}
and
\begin{align}\label{eqn:vector_green_source}
A_\alpha(\bm r,\omega) = {A_{0,\alpha}(\bm r, \omega)}+\frac{1}{c}\int d\bm r' g_{\alpha \beta}^{(\rm v)}(\bm r, \bm r',\omega) J_\beta^{(\rm v)}(\bm r',\omega)~,
\end{align}
where, as usual, $\phi_0(\bm r, \omega)$ and $A_{0,\alpha}(\bm r, \omega)$ are solutions of the homogeneous equations associated to Eqs.~(\ref{eq:poisson}) and~(\ref{eq:Helmholtz_for_A}), respectively. Eq.~(\ref{eqn:scalar_green_source}) is obtained by multiplying Eq.~\eqref{eqn:scalar_green} [Eq.~\eqref{eqn:vector_green}] by $\phi({\bm r}^\prime,\omega)$ [$A_\alpha({\bm r}^\prime,\omega)$] and integrating over $\bm{r}^\prime$. 

We now express the source terms for the scalar and vector potentials only in terms of the external current ${\bm J}^{\rm ext}(\bm{r},\omega)$. In Eq.~\eqref{eqn:scalar_green_source} we use the continuity equation finding
\begin{align}\label{eqn:scalar_green_source2}
\phi(\bm r,\omega) =  \phi_0(\bm r, \omega) - \frac{1}{i\omega}\int d \bm r' \partial_\alpha^{\prime}g^{(\rm s)}(\bm r, \bm r',\omega) J^{\rm ext}_{\alpha}(\bm r', \omega)~.
\end{align}

The vector potential expressed in terms of the source and the vector Green's function can be further manipulated by introducing the total current ${J}_\alpha^{\rm ext}(\bm{r},\omega)$. We rewrite the first term on the right-hand side of Eq.~\eqref{eqn:vector_green_source} as
\begin{align}
&\frac{1}{c}\int d\bm r' g_{\alpha \beta}^{(\rm v)}(\bm r, \bm r',\omega) J_\beta^{(\rm v)}(\bm r',\omega) =
    \nonumber\\
    &= \frac{1}{c}\int d\bm r' g_{\alpha \beta}^{(\rm v)}(\bm r, \bm r',\omega) J^{{\rm ext}}_{ \beta}(\bm r',\omega)+ \nonumber\\ 
    & - \frac{i\omega}{4\pi c} \int d\bm r' \partial_{\gamma}^\prime \epsilon_{\beta \gamma}({\bm r}^\prime,\omega) g_{\alpha \beta}^{(\rm v)}(\bm r, \bm r',\omega)\phi({\bm r}^\prime,\omega)~.
\end{align}
The third line of the previous equation vanishes due to the gauge-fixing condition \eqref{eq:generalizedC_forG} yielding
\begin{align}
& \frac{1}{c}\int d\bm r' g_{\alpha \beta}^{(\rm v)}(\bm r, \bm r',\omega) J_\beta^{(\rm v)}(\bm r',\omega)= \nonumber\\ &= \frac{1}{c}\int d\bm r' g_{\alpha \beta}^{(\rm v)}(\bm r, \bm r',\omega) J^{{\rm ext}}_{ \beta}(\bm r',\omega) ~.
\end{align}
Eq.~\eqref{eqn:vector_green_source} can be finally rewritten by means of the previous equation as
\begin{align}\label{eqn:vector_green_source2}
A_\alpha (\bm r,\omega) =  A_{0,\alpha }(\bm r, \omega)+\frac{1}{c}\int d\bm r' g_{\alpha \beta}^{(\rm v)}(\bm r, \bm r',\omega) J^{\rm ext}_{\beta}(\bm r',\omega)~.
\end{align}
The vectorial Green's function links $A_\alpha (\bm r,\omega)$ with the external source $J^{\rm ext}_{\beta}(\bm r',\omega)$.
By comparing Eq.~\eqref{eq:electricfield_interms_of_potentials}  with Eqs.~\eqref{eqn:scalar_green_source} and~\eqref{eqn:vector_green_source2} we conclude that the total dyadic Green's function can be decomposed into scalar and vectorial contributions
\begin{equation}
\label{eq:decomposition}
G_{\alpha \beta}(\bm r, \bm r',\omega)= -\frac{c^2}{\omega^2} \partial_\alpha \partial_\beta' g^{(\rm s)}(\bm r, \bm r', \omega) + g_{\alpha \beta}^{(\rm v)}(\bm r, \bm r',\omega)~.
\end{equation}
This is the most important technical result of this Section.

The decomposition (\ref{eq:decomposition}) allows us to prove the most important result of this Section from the Physics point of view: the Green's function of a sub-wavelength nanoplasmonic cavity is dominated by the scalar component $g^{(\rm s)}(\bm r, \bm r', \omega)$, which solves the frequency-dependent Poisson equation~\eqref{eq:poisson}.
As previously discussed, indeed, a sub-wavelength nanoplasmonic cavity confines the field in a mode volume $V_{\rm mode}=\lambda_{\rm p}^3$, significantly smaller than the volume constrained by the free-space diffraction limit $\lambda^3_{\rm d}$. This corresponds to the non-retarded limit $\lambda_{\rm d} \gg \lambda_{\rm p}$.
The free-space diffraction limit length can be related to the frequency  $\omega$ as $\lambda_{\rm d} = \pi (c/\omega)$. To estimate the spatial variation of the field in a sub-wavelength nanoplasmonic cavity, we can set  $\partial_\alpha, \partial^\prime_\beta \approx \lambda^{-1}_{\rm p}$ in Eq.~\eqref{eq:decomposition}. As a result, in sub-wavelength nanoplasmonic cavity the first term on the right-hand side of Eq.~\eqref{eq:decomposition} is amplified by a factor of  $(\lambda_{\rm d}/\lambda_{\rm p})^2\gg 1$ with respect to the second term. Thus, these cavities can be essentially characterized by the scalar Green's function $g^{(\rm s)}(\bm r, \bm r', \omega)$ alone, which yields the cavity's quasi-static electric field:
\begin{eqnarray}
\label{eq:longGreen}
G^{\rm nano}_{\alpha \beta}(\bm r, \bm r', \omega) \approx -\left(\frac{c}{\omega}\right)^2\partial_\alpha \partial_\beta^\prime g^{(\rm s)}(\bm r, \bm r',\omega)~,
\end{eqnarray}
where the non-retarded scalar Green's function $g^{(\rm s)}(\bm r, \bm r',\omega)$ satisfies the Poisson-like equation \eqref{eqn:scalar_green}.

We now show that the results reported in Sect.~\ref{sec:effectiveEEinteractions}, particularly Eqs.~(\ref{eq:PQ5})-(\ref{eq:Ftensor}), are rather dominated by the vectorial contribution $g_{\alpha \beta}^{(\rm v)}(\bm r, \bm r',\omega)$. We start by calculating the vectorial Green's function $g^{(\rm v, FB)}_{\alpha,\beta}(\bm{r}^\prime,\bm{r},\omega)$ for a Fabry-P\'erot cavity. This is conveniently accomplished by using the correlator~\cite{Giuliani_and_Vignale}
\begin{align}\label{eq:chi}
&g^{(\rm v, FB)}_{\alpha,\beta}(\bm{r},\bm{r}^\prime,\tau)= \nonumber\\
&=\frac{i}{\hbar}\Theta(\tau)\tensor[_{\rm ph }]{\langle}{} {0}|\Big[\hat{A}_{{\rm cav},\alpha}(\bm{r}_{\parallel}, z),\hat{A}^{\rm I}_{{\rm cav},\beta}(\bm{r}^{\prime} _{\parallel}, z^{\prime},-\tau)\Big]|0\tensor[]{\rangle}{_{\rm ph}}~,  
\end{align}
where $\Theta(\tau)$ is the usual Heaviside function and $[\hat{X},\hat{Y}]$ denotes the commutator between the operators $\hat{X}$ and $\hat{Y}$. The quantity $\hat{A}_{{\rm cav},\alpha}(\bm{r}_{\parallel}, z)$  denotes the quantized vector potential in a Fabry-P\'erot cavity and has been given above in Eq.~\eqref{vectorpot_2}. The quantity
$\hat{A}^{\rm I}_{{\rm cav},\beta}(\bm{r}_{\parallel}, z,t)$ refers to the same quantity but in the interaction picture. Specifically, $\hat{A}^{\rm I}_{{\rm cav},\beta}(\bm{r}_{\parallel}, z,t)\equiv U_0^\dagger(t)\hat{A}_{{\rm cav},\beta}(\bm{r}_{\parallel}, z)U_0(t)$
where $U_0(t)\equiv \exp(-i\hat{\mathcal{H}}_{\rm ph}t)$.

The Fourier transform with respect to time of $g^{(\rm v, FB)}_{\alpha,\beta}(\bm{r},\bm{r}^\prime,\tau)$ is taken with the addition of the usual small $\eta=0^+$ factor~\cite{Giuliani_and_Vignale}:
\begin{equation}
g^{(\rm v, FB)}_{\alpha,\beta}(\bm{r},\bm{r}^\prime, \omega)= \lim\limits_{\eta \to 0} \int_{-\infty}^\infty d\tau g^{(\rm v, FB)}_{\alpha,\beta}(\bm{r},\bm{r}^\prime,\tau) e^{i(\omega+i\eta)\tau}~.
\end{equation}
Expanding the vector potential using the normal modes described in Eq.~\eqref{vectorpot_2}, we arrive at the following result~\cite{Charlie_Thesis}:
\begin{align}
\label{eq:GTFB}
&g^{(\rm v, FB)}_{\alpha,\beta}(\bm{r},\bm{r}^\prime, \omega)=\sum_{\bm{q}_{\parallel},s,n_z}[A^{\rm(2D)}_{\bm{q}_{\parallel} ,n_z}]^2 e^{i\bm{q}_{\parallel}\cdot(\bm{r}_{\parallel}-\bm{r}^\prime_{\parallel})} 
 \times\nonumber \\&\times \left[
   \frac{{\bm e}^{(\beta)}_{-\bm{q}_{\parallel},s,n_z }(z^\prime)~{\bm e}^{(\alpha)*}_{-\bm{q}_{\parallel},s,n_z}(z)}{\hbar(\omega+i\eta+\omega_{-\bm{q}_{\parallel},n_z })} -  \frac{{\bm e}^{(\alpha)}_{\bm{q}_{\parallel},s,n_z }(z)~{\bm e}^{(\beta)*}_{\bm{q}_{\parallel},s,n_z}(z^\prime)}{\hbar(\omega+i\eta-\omega_{\bm{q}_{\parallel},n_z })}\right]~, 
\end{align}
where ${\bm e}^{(\alpha)}_{\bm{q}_{\parallel},s,n_z }(z)$ denotes $\alpha$-th component of the polarization vector ${\bm e}_{\bm{q}_{\parallel},s,n_z }(z)$.

In order to link this vectorial Green's function with the results discussed in Sect.~\ref{sec:effectiveEEinteractions} we need to take the afore mentioned $\omega\to 0$  adiabatic limit and the limit $\eta\to 0$. Additionally, we are specifically interested in fields evaluated at the graphene plane and we therefore need to set $z = z' = 0$ in Eq.~(\ref{eq:GTFB}). By noticing that ${\bm e}^{(\beta)}_{-\bm{q}_{\parallel},s,n_z }(0)~{\bm e}^{(\alpha)*}_{-\bm{q}_{\parallel},s,n_z}(0)={\bm e}^{(\beta)}_{\bm{q}_{\parallel},s,n_z }(0)~{\bm e}^{(\alpha)*}_{\bm{q}_{\parallel},s,n_z}(0)$ and $\omega_{-\bm{q}_{\parallel},n_z }=\omega_{\bm{q}_{\parallel},n_z }$, we can simplify Eq.~\eqref{eq:GTFB} as follows:
\begin{align}
\label{eq:GTFB1}
&g^{(\rm v, FB)}_{\alpha,\beta}(\bm{r}_\parallel,z=0,\bm{r}_\parallel^\prime, z^\prime=0,\omega=0)=
\nonumber \\ &= 2 \sum_{\bm{q}_{\parallel},s,n_z}\left\{   \frac{ e^{i\bm{q}_{\parallel}\cdot(\bm{r}_{\parallel}-\bm{r}^\prime_{\parallel})} [A^{\rm(2D)}_{\bm{q}_{\parallel} ,n_z}]^2{\bm e}^{(\alpha)}_{\bm{q}_{\parallel},s,n_z }(0)~{\bm e}^{(\beta)*}_{\bm{q}_{\parallel},s,n_z}(0)}{\hbar\omega_{\bm{q}_{\parallel},n_z }}\right\}~.
\end{align}
By comparing the $\alpha,\beta$ component of the tensor $\mathcal{F}_{\ell,{\ell}^\prime,\bm{q}_{\parallel}}$ introduced in Eq.~\eqref{eq:Ftensor} with the previous expression for the adiabatic vectorial Green's function $g^{(\rm v, FB)}_{\alpha,\beta}(\bm{r}_\parallel,0,\bm{r}_\parallel^\prime,0,0)$ we recognize that they are the same object, modulo a spatial Fourier transform:
\begin{align}
\label{eq:GTFB1}
\mathcal{F}^{(\alpha,\beta)}_{\ell,{\ell}^\prime,\bm{q}_{\parallel}} &=\frac{1}{2N^2_{\rm cell}}\sum_{\bm{q}^\prime_\parallel}\sum_{i,i^\prime} e^{-i(\bm{q}_{\parallel}\cdot\bm{R}_{i}+\bm{q}^{ \prime}_{\parallel}\cdot\bm{R}_{i^\prime})} \nonumber\\
&\times g^{(\rm v, FB)}_{\alpha,\beta}(\bm{R}_{i} + \bm{\delta}_\ell/2,0, \bm{R}_{i^\prime} + \bm{\delta}_{\ell^\prime}/2,0,0)~.
\end{align}
We conclude that the tensor $\mathcal{F}_{\ell,{\ell}^\prime,\bm{q}_{\parallel}}$ in Eq.~\eqref{eq:Ftensor} shares the same ``vectorial nature'' of the vectorial Green's function $g^{(\rm v, FB)}_{\alpha,\beta}$. We now observe that, in the $\omega\to$ adiabatic limit, the second term in Eq.~\eqref{eq:Helmholtz_for_A}, which is proportional to ${\omega^2}/{c^2}$, can be neglected, and one finds Ampere's law of magnetostatics:
\begin{equation}
\label{eq:HelmholtzMagnestostatic_for_A}
{\cal D}_{\alpha \beta}(\mu^{-1})  A_\beta(\bm r, 0) = \frac{4\pi}{c}{J}^{(\rm v)}_\alpha (\bm r, 0)~.
\end{equation}
The adiabatic vectorial Green's function $g_{\alpha \beta}^{(\rm v)}(\bm{r},\bm{r}^\prime, \omega=0)$ therefore shares the same Physics, i.e.~that of classical magnetostatics. In the same spirit, the current-current EEEI in Eq.~\eqref{eq:PQ5} are equivalent to magnetostatic Amperean interactions. 

Before concluding, we proceed to illustrate that deep sub-wavelength nanoplasmonic cavities are primarily governed by the scalar Green's function $g^{(\rm s)}(\bm r, \bm r',\omega)$. In essence, this quantity describes how instantaneous Coulomb interactions in vacuum are modified by the presence of the cavity. Hence, it is not possible to extrapolate the formula for the critical temperature $T_{\rm c}$ given in Eq.~\eqref{eq:nu2}, which was calculated on the sole basis of $g^{(\rm v, FB)}_{\alpha,\beta}(\bm{r},\bm{r}^\prime, \omega=0)$, to a sub-wavelength nanoplasmonic cavity. Once again, no ``mode volume confinement'' argument can be used for magnetostatic Amperean interactions. 

In a deep sub-wavelength nanoplasmonic cavity, the electronic system predominantly interacts with the cavity's electrical potential $\hat{\phi}_{\rm cav}(\bm{r})$. This interaction is of the usual electrostatic type~\cite{Giuliani_and_Vignale}, i.e.
\begin{align}
{\cal \hat{H}}_{\rm int} = -e \int d\bm{r} \hat{n}(\bm{r})\hat{\phi}_{\rm cav}(\bm{r})~,
\end{align}
where $\hat{n}(\bm{r})$ is the density operator of the electronic system. By carrying out calculations analogous to those detailed in Sect.~\ref{sec:effectiveEEinteractions}, we can trace out the cavity field to obtain the dominant EEEI in a sub-wavelength nanoplasmonic cavity. The result is:
\begin{eqnarray}
\label{eq:g_coulomb}
\hat{\mathcal{H}}_{\rm EEEI}= \frac{e^2}{2}\int d\bm{r}  d\bm{r}^\prime\hat{n}(\bm{r})  g^{(\rm s)}(\bm r, \bm r',0)\hat{n}(\bm{r}^\prime)~.
\end{eqnarray}
This formula shows that the main effect of a sub-wavelength nanoplasmonic cavity is to strongly alter the usual Coulomb interaction in vacuum. Indeed, in free-space, ${ \epsilon}_{\alpha\beta}(\bm r, \omega)=\delta_{\alpha\beta}$ and thus the scalar Green's function reduces to the well-known Coulomb interaction, i.e.~$ g^{(\rm s)}(\bm r, \bm r', \omega)  \to 1/|\bm r- \bm r'|$.    
On the other hand, the current-current EEEIs (which are responsible for Amperean superconductivity) discussed in Sect.~\ref{sec:effectiveEEinteractions} are controlled by Ampere's law and unaltered by such cavities. In summary, we conclude that Amperean superconductivity, which is the main focus of this Article, cannot be induced in sub-wavelength cavities.

Before concluding, however, we would like to comment on the possibility of achieving superconductivity in the ordinary density-density channel (rather than Amperean superconductivity in the current-current channel). To this end, one needs to transcend the adiabatic $\omega=0$ approximation and keep track of the frequency dependence of $g^{(\rm s)}$, which physically describes {\it retardation}. Due to the latter, an attractive EEEI can emerge at certain frequencies, possibly leading to superconductivity in the density channel. The density-density interaction in Eq.~\eqref{eq:g_coulomb}, which is evaluated at $\omega=0$, encodes only the physics of static screening~\cite{Giuliani_and_Vignale} due to polarization charges in the metallo-dielectric environment surrounding the electron system in graphene. As such, $g^{(\rm s)}(\bm r, \bm r',0)$ is always repulsive. Going beyond the $\omega=0$ adiabatic approximation in deriving EEEIs is well beyond the scope of the present Article.

\section{Summary and conclusions}
\label{sec:conclusions}

In summary, we have derived effective electron-electron interactions for the case of graphene placed inside a planar Fabry-P\'erot cavity. Integrating out the cavity degrees of freedom, we have obtained an effective current-current interaction, which is reported in Eq.~(\ref{eq:PQ6}). Using this interaction inside the linearized gap equation (\ref{eq:geq1}) for Amperean superconductivity results in a critical temperature which is $\sim 10^{-6}~{\rm K}$. According to our calculations, cavity-induced Amperean superconductivity occurs at such low temperatures that it is impossible to be measured. Finally, we have argued that any attempt to boost these ultra-low critical temperatures to measurable values via ``mode volume confinement'' arguments is unjustified. 

Concluding with a note of positivity, we hasten to emphasize that our conclusions have no implications at all on the exciting possibility of stabilizing exotic states of matter (other than Amperean superconductors) via engineered nanoplasmonic cavities. A Green's function approach to effective electron-electron interactions in these sub-wavelength cavities is reported in Sect.~\ref{sec:TransversVsLong} and will be the subject of forthcoming studies.

\section*{Acknowledgements} This work was supported by: i) the European Union's Horizon 2020 research and innovation programme under grant agreement no.~881603 - GrapheneCore3; ii) the University of Pisa under the ``PRA - Progetti di Ricerca di Ateneo" (Institutional Research Grants) -  Project No.~PRA\_2020-2021\_92 ``Quantum Computing, Technologies and Applications''; iii) the Italian Minister of University and Research (MUR) under the ``Research projects of relevant national interest  - PRIN 2020''  - Project no.~2020JLZ52N, title ``Light-matter interactions and the collective behavior of quantum 2D materials (q-LIMA)'';  iv) CEX2019-000910-S [MCIN/ AEI/10.13039/501100011033]; v) Fundaci\'{o} Cellex; vi) Fundaci\'{o} Mir-Puig; and vii) Generalitat de Catalunya through CERCA. F.M.D.P. was supported by the ``Piano di Incentivi per la Ricerca di Ateneo 2020/2022'' of the University of Catania, through the projects ``QUAPHENE'' and ``Q-ICT''. 

It is a great pleasure to thank A. H. MacDonald, A. Imamo\u{g}lu, D. E. Chang, A. Cavalleri, D. Miserev, V. Kozin, D. Loss, J. Klinovaja, F. Piazza, M. A. Sentef, C. Eckhardt, G. Menichetti, G. Citeroni, and I. Gianardi for useful discussions.

\newpage
\appendix
\section{Details on the tight binding Hamiltonian and the current operator}\label{sec:GrapheneH}

In this Appendix we give further details on the diagonalization of the free electron Hamiltonian reported in Eq.~\eqref{eq:electronic_Hamiltonian_graphene_real_space}.

The relation between the field operators $\hat c_{{\bm k},\sigma,\alpha}$, which destroy an electron with a given momentum ${\bm{k}}$,  spin $\sigma$, and sub-lattice index $\alpha=A,B$, and the equivalent operators in the real-space representation is:
\begin{align}\label{eq:j_to_k}
&\hat c_{{\bm R}_j,\sigma,A} = \frac{1}{\sqrt{N_{\rm cell}}} \sum_{{\bm k} \in {\rm BZ}} \hat c_{{\bm k},\sigma,A}e^{i {\bm k}\cdot  \bm{R}_{i} }~,\nonumber\\
&\hat c_{{\bm R}_{(i,\ell)},\sigma,B} = \frac{1}{\sqrt{N_{\rm cell}}} \sum_{{\bm k} \in {\rm BZ}} \hat c_{{\bm k},\sigma,B} e^{i {\bm k}\cdot  (\bm{R}_{i}+{\bm \delta}_\ell) }~.
\end{align}
The following identity holds true:
\begin{equation}
\label{eq:identity}
 \frac{1}{N_{\rm cell}} \sum^{N_{\rm cell}}_{i=1} e^{i ({\bm k}-{\bm k}')\cdot  \bm{R}_{i} }=\delta_{{\bm k},{\bm k}'+{\bm G}} 
 ~,
\end{equation}
where ${\bm G} = n_1 {\bm b}_1 + n_2 {\bm b}_2$ is an arbitrary vector of the reciprocal space. Here, $n_1, n_2$ are integers and ${\bm b}_1=2 \pi/(3a)(1, \sqrt{3})$, $\bm{b}_2=2 \pi/(3 a)(1,- \sqrt{3})$ are primitive reciprocal lattice vectors.

Inserting Eq.~(\ref{eq:j_to_k}) inside Eq.~(\ref{eq:electronic_Hamiltonian_graphene_real_space}) and using the identity (\ref{eq:identity}), we rewrite ${\cal \hat{H}}_{\rm e}$ as following:
\begin{eqnarray}
\hat{\cal H}_{\rm e}&=& \sum_{{\bm k} \in { {\rm BZ}}}\sum_{\sigma=\uparrow,\downarrow}  \left[  {f}{( \bm k)} \, \hat c_{{\bm k},\sigma,A}^{\dagger} \hat c_{{\bm k }, \sigma,B} + {\rm H.c.} \right]+\nonumber \\ &-&\mu \sum_{{\bm k} \in { {\rm BZ}}}  \sum_{\sigma=\uparrow,\downarrow}  \sum_{\alpha=A,B}  \hat c_{\bm{k},\sigma,\alpha}^{\dagger}  \hat c_{\bm{k}, \sigma, \alpha}~,
\end{eqnarray}
where $f( \bm k)  =- t \sum_{\ell=1}^3 e^{i {\bm k } \cdot  {\bm \delta}_\ell}$.

Finally, we introduce the operators $\hat d_{{\bm k }, \sigma,\nu}$ , $\hat d^\dagger_{{\bm k }, \sigma,\nu}$ with $\nu=\pm$, which allow us to write the free-electron Hamiltonian $\hat{\cal H}_{\rm e}$ in the previous equation in the diagonal form \eqref{eq:free_electron} reported in the main text:
\begin{align}\label{eq:c_vs_d}
&\hat c_{{\bm k},\sigma, A}= \frac{1}{\sqrt 2} \left( \hat d_{{\bm k}, \sigma, +} + \hat d_{{\bm k},\sigma, -} \right)~,\nonumber\\
&\hat c_{{\bm k},\sigma, B}
= \frac{1}{\sqrt 2}\varphi^\ast_{\bm{k}}\left( \hat d_{{\bm k},\sigma, +} - \hat d_{{\bm k},\sigma, -} \right)~,
\end{align}
where $\varphi_{\bm{k}}={{f}{( \bm k)}}/{\big |{f}{( \bm k)}\big|}$.

\begin{widetext}
We now proceed to give details on how the current is calculated.
The paramagnetic current $\hat{\bm J}^{(j,\ell)}$ in Eq.~\eqref{eq:current} can be expressed in terms of the momentum-space creation and annihilation operators by employing Eq.~\eqref{eq:j_to_k}:
\begin{align}
\hat{\bm J}^{(j,\ell)}&=   \left(\frac{i t}{\hbar}\right) \frac{1}{N_{\rm cell}}\sum_{{\bm k},{\bm k}' \in {\rm BZ}}   \sum_{\sigma=\uparrow, \downarrow} \bm{\delta}_\ell  \left[e^{-i {(\bm k - \bm k^\prime)}\cdot \bm{R}_{j}} e^{i {\bm k}^\prime\cdot {\bm \delta}_\ell}   \hat c_{{\bm k},\sigma,A}^{\dagger} \hat c_{{\bm k}', \sigma,B}   - \rm{\rm{H.c.}}\right]~.
\end{align}
We then express the current in terms of the operators $\hat d_{{\bm k }, \sigma,\nu}$ defined in Eq.~\eqref{eq:c_vs_d}:
\begin{eqnarray}
\label{eq:Jd}
\hat{\bm J}^{(i,\ell)}&=&  \left(\frac{i t}{\hbar}\right) \frac{1}{N_{\rm cell}}\sum_{{\bm k},{\bm k}' \in {\rm BZ}}  \sum_{\sigma=\uparrow, \downarrow} \bm{\delta}_\ell  \left[e^{-i {(\bm k - \bm k^\prime)}\cdot \bm{R}_{i} } e^{i {\bm k}'\cdot {\bm \delta}_\ell} \frac{\varphi^\ast_{\bm{k}}}{2}   \sum_{\nu,\nu^\prime} \nu^\prime \hat d_{{\bm k},\sigma,\nu}^{\dagger} \hat d_{{\bm k^\prime},\sigma,\nu^\prime} - \rm{H.c.}\right]~.
\end{eqnarray}
The previous equation can be written as
\begin{eqnarray}
\label{eq:currentFourierApp}
\hat{\bm J}^{(i,\ell)}&=& 
\frac{1}{N_{\rm cell}}\sum_{{\bm k},{\bm k}' \in {\rm BZ}}\left[    e^{-i {(\bm k - \bm k^\prime)}\cdot \bm{R}_{i}} \hat{\bm \jmath}_{\ell,\bm k, \bm k^\prime} +  e^{i {(\bm k - \bm k^\prime)}\cdot \bm{R}_{i}} \hat{\bm \jmath}^\dagger_{\ell,\bm k, \bm k^\prime}\right]~,
\end{eqnarray}
where
\begin{equation}
\label{eq:componentsJApp}
\hat{\bm \jmath}_{\ell,\bm k, \bm k^\prime}  \equiv   \left(\frac{i t}{\hbar}\right) \sum_{\sigma=\uparrow, \downarrow} \bm{\delta}_\ell e^{i {\bm k}'\cdot {\bm \delta}_\ell}   \frac{\varphi^\ast_{\bm{k}}}{2}  \sum_{\nu,\nu^\prime} \nu^\prime \hat d_{{\bm k},\sigma,\nu}^{\dagger} \hat d_{{\bm k^\prime},\sigma,\nu^\prime}~.
\end{equation}
By swapping $\bm{k}$ and $\bm{k}^\prime$ in the first term of Eq.~\eqref{eq:currentFourierApp}
we finally get
\begin{equation}
\label{eq:currentFourierApp1}
\hat{\bm J}^{(i,\ell)}= 
\frac{1}{N_{\rm cell}}\sum_{{\bm k},{\bm k}' \in {\rm BZ}}    e^{-i {(\bm{k}^\prime - \bm{k})}\cdot \bm{R}_{i}} \left(\hat{\bm \jmath}_{\ell,\bm k^\prime, \bm k} +  \hat{\bm \jmath}^\dagger_{\ell,\bm k, \bm k^\prime}\right)~.
\end{equation}
By defining
\begin{align}\label{eq:Fourier_link_currentApp}
 \hat{\bm{J}}_{\ell,\bm{q}_{\parallel}} \equiv \sum_{{\bm k},\bm{k}^\prime \in {\rm BZ}}   \left(
 \hat{\bm \jmath}_{\ell, \bm{k}^\prime, \bm k} + 
 \hat{\bm \jmath}^\dagger_{\ell,\bm k, \bm{k}^\prime} 
 \right)\delta_{\bm{k}-\bm{k}^\prime,\bm{q}_{\parallel}}~,
\end{align}
we can rewrite Eq.~\eqref{eq:currentFourierApp1} in terms of this new quantity as
\begin{align}\label{eq:currentFourierApp}
\hat{\bm J}^{(i,\ell)}= 
\frac{1}{N_{\rm cell}}\sum_{\bm{q}_{\parallel}}   e^{i \bm{q}_{\parallel}\cdot \bm{R}_{i}}  
\hat{\bm{J}}_{\ell,\bm{q}_{\parallel}}~,
\end{align}
\end{widetext}
which corresponds to Eq.~\eqref{eq:currentFourier} of the main text. 

\section{Diamagnetic term}\label{sec:Diamagnetic}

The diamagnetic term $\hat{\cal H}_{\rm diam}$ reads as following:
\begin{widetext}
\begin{equation}
\hat{\cal H}_{\rm diam}=  \frac{1}{2}\left(\frac{e}{c}\right)^2  \sum^{N_{\rm cell}}_{j=1}  \sum_{\ell=1}^3  \hat{\bm A}_{\rm cav}\big(\bm{R}_{j}  + \frac{\bm{\delta}_\ell}{2},z=0\big) \cdot \mathcal{T}^{(j,\ell)}\cdot  \hat{\bm A}_{\rm cav}\big(\bm{R}_{j}  + \frac{\bm{\delta}_\ell}{2},z=0\big)~,
\end{equation}
where $\mathcal{T}^{(j,\ell)}$ is the diamagnetic operator:
\begin{align}
\mathcal{T}^{(j,\ell)} &\equiv \left(\frac{c}{e}\right)^2 \frac{\partial^2 \hat {\cal H}}{\partial\hat {\bm A}_{\rm cav} \left(\bm{R}_{j}  + \frac{\bm{\delta}_\ell}{2},z=0\right)\partial\hat {\bm A}_{\rm cav} \left(\bm{R}_{j}  + \frac{\bm{\delta}_\ell}{2},z=0\right)} \Bigg|_{\hat{\bm A}_{\rm cav} = {\bm 0}} \nonumber\\
&=  \sum_{\sigma=\uparrow, \downarrow} {\frac{t}{\hbar^2}}\; \bm{\delta}_\ell \bm{\delta}_\ell  \;  \left[\hat c_{{\bm R}_j,\sigma,A}^{\dagger} \hat c_{{\bm R}_{(j,\ell)},\sigma, B} + \hat c_{{\bm R}_{(j,\ell)},\sigma, B} ^{\dagger}  \hat c_{{\bm R}_j,\sigma,A}\right]~.
\end{align}
We can therefore estimate the magnitude of the diamagnetic contribution as following:
\begin{align}\label{eq:order-of-magnitude-diamagnetic-term}
\hat{\cal H}_{\rm diam} &\sim \left(\frac{e}{c}\right)^2 \times 
\underbrace{\left(\epsilon_{\rm r}^{-1/4} \hbar c \frac{1}{\sqrt{V}} \frac{1}{\sqrt{\hbar\omega_{\rm D}}}\right)^2}_{\rm from~the~vector~potential}
\times
\underbrace{t \left(\frac{a}{\hbar}\right)^2}_{{\rm from~the~diamagnetic~operator}}=\nonumber\\
&= \frac{1}{t} \left(\epsilon_{\rm r}^{-1/4} \frac{a}{\sqrt{V}} \frac{1}{\sqrt{\hbar\omega_{\rm D}}} t e\right)^2=\frac{g^2}{t}~,
\end{align}
\end{widetext} 
where the light-matter coupling constant $g$ has been introduced in Eq.~\eqref{eq:coupling_constant_of_the_whole_theory}.


\begin{thebibliography}{77}
\bibitem{footnote1}
This approximation is justified since we are interested in small photonic momenta ${q}_\parallel \ll 1/a$ and later we will evaluate both $\bm{k}$ and $\bm{k}^\prime$ on the Fermi surface, away from the corner of the Brillouin zone, which is the only region where Umklapp processes can occur for small ${q}_\parallel$.
\end{thebibliography}
\end{document}